\documentclass{elsarticle}

\usepackage[ margin=1.0in]{geometry}
\usepackage{etoolbox}
\usepackage{subfiles}

\usepackage{hyperref}
\usepackage{xr-hyper}
\usepackage{nicefrac}

  \biboptions{sort&compress}

\newlength{\fchartw}  
\newlength{\fdecw}  
\newlength{\gfchartw}  
\newlength{\gfdecw}  
\usepackage{tikz}
\usetikzlibrary{shapes.geometric, arrows}
\tikzstyle{startstop} = [rectangle, minimum width=3cm, minimum height=1cm,text centered, text width=\fchartw, draw=black, fill=red!30]
\tikzstyle{io} = [rectangle, minimum width=3cm, minimum height=1cm,text centered, text width=\fchartw, draw=black, fill=orange!30]
\tikzstyle{process} = [rectangle, minimum width=3cm, minimum height=1cm,text centered, text width=\fchartw, draw=black,fill=orange!30]
\tikzstyle{decision} = [rectangle, minimum width=3cm, minimum height=1cm,text centered, text width=\fchartw, draw=black, fill=green!30]
\tikzstyle{arrow} = [thick,->,>=stealth]

\usepackage{xcolor}
\usepackage{amsmath}
\usepackage{amsfonts}   
\usepackage{amssymb}    
\usepackage{graphicx}
\usepackage{subfigure}

\usepackage{titlesec}
\usepackage{sectsty}
\usepackage{lipsum}

\newcommand\nc{\newcommand}

\nc\myfrac[2]{\frac{#1}{#2}}

\nc\cB{{\mathcal{B}}}
\nc\cC{{\mathcal{C}}}
\nc\cN{{\mathcal{N}}}
\nc\tcN{\tilde{{\cal{N}}}}
\nc\cG{{\mathcal{G}}}
\nc\cD{{\mathcal{D}}}
\nc\cU{{\mathcal{U}}}
\nc\cL{{\mathcal{L}}}
\nc\cT{{\mathcal{T}}}
\nc\cF{{\mathcal{F}}}
\nc\cK{{\mathcal{K}}}
\nc\Grad{\nabla}
\nc\Lapcian{\nabla^2}
\nc\hGrad{\hat{\nabla}}
\nc\tGrad{\tilde{\nabla}}
\nc\Div{\nabla \cdot}
\nc{\curl}[1]{ \nabla \times { #1 } }

\nc\PiS{{\Pi_{\textrm{S}}}}
\nc\dPiS{{\Pi_{\textrm{S}}'}}

\nc{\nsall}[5]{#1^{#2}_{(#3),(#4,#5)}}
\nc{\nsx}[3]{#1_{(#2,#3)}}
\nc{\nstn}[3]{#1^{#2}_{(#3)}}

\nc{\ep}{\epsilon}
\nc{\pa}{\partial}
\nc{\pad}[2]{\frac{\partial #1}{\partial #2}}
\nc{\padn}[3]{\frac{\partial^{#3} #1}{\partial #2^{#3}}}
\nc{\vad}[2]{\frac{\delta #1}{\delta #2}}
\nc{\myTitle}{My Title}

\nc{\markchange}[1]{{\color{red}#1}}

\nc{\pref}[1]{\protect\ref{#1}}

\numberwithin{equation}{section}

\usepackage[acronym]{glossaries}

\newacronym{NLC}{NLC}{Nematic Liquid Crystal}

\newacronym{GPU}{GPU}{Graphics Processing Unit}
\newacronym{CPU}{CPU}{Central Processing Unit}
\newacronym{CUDA}{CUDA}{Compute Unified Device Architecture}
\newacronym{ADI}{ADI}{Alternating Direction Implicit}
\newacronym{SIMD}{SIMD}{Single Instruction, Multiple Data}
\newacronym{LSA}{LSA}{Linear Stability Analysis}
\newacronym{API}{API}{Application Programming Interface}
\makeglossaries

\begin{document}

\title{Computing dynamics of thin films via large scale GPU-based simulations}
\author[mymainaddress]{Michael-Angelo Y.-H. Lam}
\author[mymainaddress]{Linda J. Cummings}
\author[mymainaddress]{Lou Kondic\corref{cor1}}
\ead{kondic@njit.edu}
\cortext[cor1]{Corresponding author}
\fntext[fn1]{Research supported by the National Science Foundation under grant  DMS-121171.}
\fntext[fn2]{Research supported by the National Science Foundation under grant  CBET-1604351.}
\fntext[fn3]{Research supported by the National Aeronautics and Space Administration under grant No. NNX16AQ79G.}
\address[mymainaddress]{Department of Mathematical Sciences, New Jersey Institute of Technology, Newark, NJ, 07102, USA}

\begin{abstract}
We present the results of large scale simulations of $4$th order nonlinear partial 
differential equations of diffusion type that are typically encountered when modeling 
dynamics of thin fluid films on substrates.  The simulations are based on the alternate
direction implicit (ADI) method, with the main part of the computational work carried out in 
the GPU computing environment.  Efficient and accurate computations 
allow for simulations on large computational domains in three spatial
dimensions (3D) and for long computational times.   We apply the methods developed to the 
particular problem of instabilities of thin fluid films of nanoscale thickness. The large scale of 
the simulations minimizes the effects of boundaries, and also allows for simulating domains 
of the size encountered in published experiments.   As an outcome, we can analyze the development
of instabilities with an unprecedented level of detail.   A particular focus
is on analyzing the manner in which instability develops, in particular regarding differences
between spinodal and nucleation types of dewetting for linearly unstable films, as well as 
instabilities of metastable films.   Simulations in 3D allow for consideration of some recent
results that were previously obtained in the 2D geometry ({\it J. Fluid Mech.} {\bf 841}, 925 (2018)).  
Some of the new results include using Fourier transforms as well as topological invariants (Betti numbers)
to distinguish the outcomes of spinodal and nucleation types of instabilities, describing 
in precise terms the complex processes that lead to the formation of satellite drops, as well
as distinguishing the shape of the evolving film front in linearly unstable and metastable
regimes.   We also discuss direct comparison between simulations and available experimental
results for nematic liquid crystal and polymer films.
\end{abstract}
\maketitle


Simulating the dynamics of thin fluid films on substrates presents significant computational challenges, 
requiring consideration of complex setups that include evolving free surfaces and computational 
domains.  Solving such problems using
full numerical simulations of Navier-Stokes equations coupled with complex boundary conditions is
computationally very expensive.   While there has been significant progress based on Volume of 
Fluid, phase field and related methods~\cite{Jacqmin1999,Jacqmin2000,JCP2015,Renardy2001,seric2017,Yue2010,
ZHOU2010498}, it is desirable 
to be able to consider evolving thin films in a simpler and cheaper manner. 

Motivated by the desire to reduce computational cost, and even to gain some insight based on analytical 
methods, thin fluid films are often considered using asymptotic approaches, in particular based on the long wave
approximation.  Such an approach, in its simplest form, leads to a highly nonlinear 4th order partial 
differential equation (PDE) of diffusion type, which must then be solved numerically.  While this is a much 
simpler task than that presented by the original problem, it still leads to significant computational 
challenges, particularly
in configurations involving contact lines.  Contact lines introduce short length scales in the problem,
that result from the need to include additional physics resulting from liquid-solid interaction forces.   
These short length scales need to be resolved for the purpose of producing accurate results, leading 
to (in a finite difference setting) a large number of grid points.  For evolving problems, the resulting outcome
is still computationally expensive, typically limiting the simulations to relatively small computational 
domains and short times.  Particularly when film instabilities are considered, simulating small computational
domains is a significant restriction, since it is not obvious to which degree computational domain boundaries
influence the results.   

Numerical approaches that have been applied to the 4th order nonlinear partial (PDE) of diffusion type are numerous
and we will not attempt to provide a comprehensive review.   Only some examples are mentioned, with focus on the three
dimensional setting that is of interest here.  There is a number of
finite difference based methods~\cite{becker_nat03,dk_jcp02,fetzer_05,lin_pof12,sharma_epje03}, 
pseudo-spectral approaches~\cite{TBBB2003epje},  and finite element type of methods~\cite{EWGT2016prf}.   
Within finite difference based computations, significant progress has been achieved using the ADI 
(Alternate Direction Implicit) approach, which allows for efficient and accurate simulations, with good
stability properties.   Such an approach was discussed within the context of thin films~\cite{Witelski2003}
and later applied to a number of different problems, e.g.~\cite{kd_pre09,lin_pof12,mk_jfm09}.   The main computational expense
in this approach consists of solving a large number of linear algebra problems, involving in particular large
sparse (pentadiagonal) matrices.   

In this paper, we present a fast numerical method using a \gls{GPU} and 
the \gls{CUDA} \gls{API}.  Using this approach, the most time consuming 
parts of the simulations are carried out on a \gls{GPU}, producing as an outcome
simulations that are more than an order of magnitude faster than a similarly 
coded serial \gls{CPU} method.   

The newly developed computational methods allow for consideration of problems that until now have 
been out of reach using reasonable computational resources.   In this work we will focus on
problems involving instabilities of thin films on the nanoscale, but it should be noted that the presented
computational approach is quite general, and could be applied to a number of problems within the context
of thin films, and also other physical problems that reduce to similar mathematical formulations that lead 
to the Cahn-Hilliard and Kuramoto-Sivashinsky type of equations.

Returning to the discussion of thin films, we note that such films are often unstable due to 
destabilizing liquid-solid interaction forces.   The nature of these instabilities has been considered
extensively in a variety of settings including polymer films~\cite{becker_nat03,Jacobs2008,seeman_prl01}, 
liquid crystal films~\cite{Herminghaus1998,poulard2005,Schlagowski2002}, as well
as liquid metals~\cite{fowlkes_nano11,gonzalez2013}  (see also~\cite{cm_rmp09,oron_rmp97} for excellent reviews of this topic).    
The instability development leads to many questions even in relatively simple settings.   One question
involves differences between the instabilities due to the presence of a free surface, and those due to localized 
perturbations; the latter
could be present due to some imperfection of either the fluid, or the substrate, or 
both.  
Free surface perturbations are typically global in character, random, and small in size (compared
to the film thickness); the instabilities due to their presence are typically classified as spinodal dewetting, 
with the name evolving from mathematically similar spinodal decomposition.  Localized perturbations, 
on the other hand, are typically large and lead to so-called nucleation type instability.   One question is 
how to compare the instabilities resulting from these two mechanisms.  While elaborate 
approaches based on Minkowski functionals have proved useful~\cite{becker_nat03}, one wonders whether, 
given a sufficiently large computational domain, one could analyze the results using some more 
straightforward approach.   We will discuss one possible approach in this work. 

There are additional aspects of thin film instabilities that require large scale simulations to resolve.  One could
ask, for example, whether the drop size resulting from film breakup due to spinodal and nucleation type instabilities are
comparable.  This question is of relevance to applications where instability is harnessed to produce drops of 
a specific size for use in some application,
such as (for example) plasmonic resonance in the context of metal films~\cite{atwater_natmat10}.  
Turning this around, if one is interested in the source of instability, one could analyze the resulting drop 
sizes, and use this information to infer the source.  Furthermore, one could ask whether instability and resulting 
film breakup lead to the formation of small drops, often called satellite drops~\cite{Lam2018,Seric2014}.    We will see that some of our
findings could serve as a basis for answering these questions. 

Clearly, a number of questions could be asked.  We will focus on particular 
examples of nematic liquid crystal films and, to a smaller extent, on polymer films;
however, the main features of our results are general, and will be of relevance to a number of problems 
involving thin films.  What distinguishes different films, substrates and film thickness regimes, at least for 
slow evolution where inertial effects are not significant, is essentially the form of (effective) disjoining pressure that, 
in addition to liquid/solid interaction, may include the effects of anchoring in the context of liquid crystals; interactions of electric
or magnetic type in the context of ferrofluids~\cite{Seric2014}; or composite substrates in the case of polymer 
films~\cite{Jacobs2008}.  The influence of the functional form of such effective disjoining pressure on film 
stability was discussed in two spatial dimensions (2D) recently~\cite{Lam2018}.  In that work, we presented 
numerical evidence for the formation of secondary (satellite) drops for positive values of the effective disjoining 
pressure, and discussed various regimes of instability development within
linearly unstable as well as metastable regimes.   The present paper will discuss some of these results in the 3D geometry.

The rest of this paper is organized as follows. Section~\ref{sec:GovEqn} provides the basic mathematical 
formulation of the problems to be considered. The numerical methods are discussed in \S~\ref{sec:Numerical_Method}, 
and the performance of the presented method in \S~\ref{sec:Num_Perf}.  The aspects related to implementation on 
a \gls{GPU} are relegated to \ref{sec:GPUImplement}.   An application specific to nematic liquid crystal (NLC) films 
is discussed in \S~\ref{sec:NLC}, and comparison to experimental results for both NLC and polymer films 
in \S~\ref{sec:ExperimentalResults}.  We present our conclusions in \S~\ref{sec:Conclusions}.

\section{Governing Equation} \label{sec:GovEqn}

We consider equations describing nonlinear diffusion of the following general form  
\begin{equation} \label{eq:GoverningEquation}
 {u}_t + \nabla \cdot \left[  {f}_0 ({u}) \nabla \nabla^2 u +  {f}_1 ({u}) \nabla {u} \right] =0 \;,
\end{equation}         
where ${u}(x,y,t)$ is the evolving quantity of interest; and ${f}_0({u})$, ${f}_1({u})$ are some smooth nonlinear functions of $u$.  
The square bracketed term may be interpreted as the flux that governs the diffusion process.   Governing equations of the form stated in (\ref{eq:GoverningEquation}) 
appear in a variety of models, such as those leading to Cahn-Hilliard, Kuramoto-Sivashinsky and related equations.  
\subsection{Description of model problems}

Equations of the form~(\ref{eq:GoverningEquation}) commonly appear in the context of thin fluid films.   In this context, the variable $u$
describes the film thickness, $h$.   For future reference, we specify here the governing equation for thin films in terms of (hatted)  dimensional
variables
\begin{equation} \label{eq:DimGoverningEquation}
\mu \hat{h}_t + \hat\nabla \cdot \left[  \hat{f}_0 (\hat{h}) \hat\nabla \hat\nabla^2 h +  \hat{f}_1 (\hat{h}) \hat\nabla \hat{h} \right] =0 \;,
\end{equation} 
where $\hat h(\hat x,\hat y,\hat t)$ is the fluid film thickness and $\mu$ is the viscosity.  
To nondimensonalize (\ref{eq:DimGoverningEquation}), four scaling factors are defined: 
${H}$, a representative film thickness scale,
${L}$, the typical lengthscale of variations in the plane of the film, $(\hat x, \hat y)$;
$\delta={H}/{L}\ll1$, the small aspect ratio; and
${T}$, the timescale of fluid flow.
Scaling $(\hat x, \hat y)$, $\hat{t}$ and $\hat{h}$ in the obvious way,
(\ref{eq:DimGoverningEquation}) becomes
\begin{equation} \label{eq:NonDimGoverningEquation}
 h_t + \nabla \cdot \left[ f_0 (h) \nabla \nabla^2 u + f_1 (h) \nabla h \right] =0 \;, \quad \textrm{with} \quad 
 f_0 =\frac{ T {F}_0 }{\mu L^4} \tilde{f}_0 (h) \quad \textrm{and} \quad
 f_1 =\frac{ T  {F}_1}{\mu   L^2} \tilde{f}_1 (h) \;,
\end{equation}
where ${F}_0$ and ${F}_1$ are some positive dimensional prefactors associated with $\hat{f}_0$ and $\hat{f}_1$, respectively, such that $\tilde{f}_0$ and $\tilde{f}_1$ are nondimensional functions of the dimensionless film height $h$, for example, $ \hat{f}_0(\hat{h}) =F_0 \tilde{f}_0(h)$.   The prefactors of $\tilde{f}_0$ and $\tilde{f}_1$ in (\ref{eq:NonDimGoverningEquation}) are nondimensional; therefore, for simplicity, we absorb these prefactors into the definitions of the relevant functions.   

\subsection{Linear Stability Analysis (LSA)} \label{sec:LSA}

We now present a brief overview of the \gls{LSA} of a flat film of thickness $H_0$ in 2D (two dimensions), which will be used to validate our numerical code.   To derive the dispersion relation, the solution is assumed to be of the form $h(x,t) = H_0(1 + \ep e^{iqx + i\omega t}$), where $\ep \ll 1$. Substituting this ansatz
 into (\ref{eq:GoverningEquation}) yields 
\begin{equation} \label{eq:DispersionRelation}
\omega = i\left[ f_1(H_0) q^2 - f_2(H_0) \right]q^2 \;. 
\end{equation}
A film of thickness $H_0$ is linearly unstable if $f_2(H_0)>0$, and in the unstable flat film thickness regime, the critical wavenumber (below which films are unstable), the most unstable mode, and the maximum growth rate are given by
\begin{equation} \label{eq:unstableLSAmax}
q_c=\sqrt{\frac{f_2(H_0)}{f_1(H_0)}} \; , \quad
q_m=\sqrt{\frac{f_2(H_0)}{2 f_1(H_0)}} \quad \textrm{and} \quad 
\omega_m=   \frac{[f_1(H_0) ]^2}{4 f_1(H_0)}\;, 
\end{equation}
respectively.

\subsection{Thin Film Models} \label{sec:ThinFilmModels}

We consider in this paper three different models i.e., three different sets of $f_0(h)$ and $f_1(h)$ in (\ref{eq:NonDimGoverningEquation}).  The first model considered is a test case leading to a simple linear partial differential equation, i.e,
\begin{equation}
 f_0(h) = c_0 \quad \textrm{and} \quad f_1(h) = c_1\;,
\end{equation}
where the sign of $c_1$ is chosen so that a flat film is linearly unstable. For the remaining two models, we assume ${f}_0$ and ${f}_1$ are of the form 
\begin{equation} \label{eq:GeneralDisPreForm}
 {f}_0( h) =  \cC  h^3 \quad \textrm{and} \quad f_1(h) =  \frac{h^3 \Pi'( h) }{\cC}\;,
\end{equation}
where $\cC$ is the inverse Capillary number (ratio of surface tension forces to viscous force) and $\Pi( h)$ is the disjoining pressure, typically describing 
the strength of fluid/solid interaction.   In our second model, the disjoining pressure is specified as an
`effective' disjoining pressure derived in the context of nematic liquid crystal films~\cite{Lin2013,Lam2018}.   For this model, 
$\Pi(h) = \Pi_{\textrm {NLC}}(h)$, with 
\begin{equation} \label{eq:DisjPressNLC}
 \Pi_{\textrm {NLC}}(h) = \cK\left[ \left(\frac{b}{h}\right)^3-\left(\frac{b}{h}\right)^2\right] + \frac{\cN}{2}\left[\frac{m(h)}{h}\right]^2 \;, 
\end{equation}
where 
\begin{equation} \label{eq:mh_model}
m(h) = g(h) \frac{h^2}{h^2+\beta^2} \; ; \quad g(h)= \frac{1}{2}\left[ 1 + \tanh \left( \frac{h-2b}{w}\right) \right]\; .
\end{equation}
The scales are chosen based on the experiments of Herminghaus {\it et al.} and Cazabat {\it et al.} for thin films of NLC~\cite{Cazabat2011,Herminghaus1998}
\begin{equation} \label{eq:NLC_scales}
	H = 100 \; \textrm{nm} \;, \quad
	L = 10 \;  \mu\textrm{m} \;, \quad
	T = 1 \;  \textrm{s} \; .
\end{equation}
The values of the nondimensional parameters are also based on these experiments, as discussed in 
some detail in our earlier work~\cite{Lam2018}, and are set to
\begin{equation} \label{eq:NLC_paras}
	\cC = 0.0857 \;, \quad
	\cK = 36.0 \; ,\quad
	\cN = 1.67 \; ,\quad
	\beta = 1 \; ,\quad
	w = 0.05 \; ,\quad
	b = 0.01 \; .
\end{equation}
These parameters are used throughout the paper except if specified differently.   We leave 
the discussion of the details of this model to $\S$~\ref{sec:NLC}, where we focus on extending previous results for 2D~\cite{Lam2018} to 3D films; however, for now, we note that the term with prefactor $\cK$ in the disjoining pressure defined in (\ref{eq:DisjPressNLC}) is the power-law form of disjoining pressure consisting of Born repulsion and the van der Waals force. The power-law form of the disjoining pressure is commonly used in the literature; see e.g., the review~\cite{cm_rmp09} for detailed discussion. The term with a prefactor $\cN$ in (\ref{eq:DisjPressNLC}) is due to the elastic response of \gls{NLC}, further discussed in $\S$~\ref{sec:NLC}.

The third model describes polymeric films~\cite{Jacobs2008}, where $\Pi (h)= \Pi_{\textrm {POL}} (h)$, with 
\begin{equation} \label{eq:DisjPressPolymer}
	\Pi_{\textrm {POL}} ( h) = - \frac{\partial \psi_{\textrm {POL}} }{\partial h} \;,  \quad \textrm{where} \quad
	 \psi_\textrm{POL}(  h) = 
	\frac{ C}{ h^8} - \frac{ A_\textrm{SiOx}}{12\pi  h^2} + \frac{ A_\textrm{SiOx}- A_\textrm{Si}}{12\pi ( h+ d)^2} \;,
\end{equation}
and the coefficients in (\ref{eq:GeneralDisPreForm}) and (\ref{eq:DisjPressPolymer}) are given by
\begin{equation} \label{eq:Polymer_scales}
	\cC = 0.00581 \;, \quad
	C = 1.181 \; ,\quad
	A_\textrm{SiOx}  = 41.25 \; ,\quad
	A_\textrm{Si} = -243.75 \; ,\quad
	d = 191 \; ,\quad
\end{equation}
with the scalings
\begin{equation} \label{eq:Polymer_paras}
	H = 1 \; \textrm{nm} \;, \quad
	L = 100 \;  \textrm{nm} \;, \quad
	T = 60 \; \textrm{s} \; ,
\end{equation}
(here the physical parameters are taken from Seemann et al.~\cite{Seemann2001b}).   The first term in (\ref{eq:DisjPressPolymer}) is due to steric effects (non-bonding intermolecular interactions), and the last two terms are van der Waals forces in a thin film of a polymer deposited on a silicon substrate (Si), coated with a silicon oxide (SiOx) layer of thickness ${d}$. 

\section{Numerical Method} \label{sec:Numerical_Method}

The numerical approach that we employ is based on the Alternate Direction Implicit method, discussed in the context of thin film flows~\cite{Witelski2003} and 
implemented in recent works, such as~\cite{lin_pof12,Lin2013}.  
In the present paper we provide a review of the method, both for completeness, and for the purpose of discussing
particular issues involving implementation in a GPU computing environment.  More precisely, in $\S$~\ref{sec:ConLaw} we discuss the spatial dicretization
 scheme in terms of fluxes; temporal 
discretization is described in $\S$~\ref{sec:TemporalDiscretization}; discretization of fluxes in $\S$~\ref{sec:FluxDiscretization}; and boundary conditions are discussed in $\S$~\ref{sec:BoundaryCondition}.

\subsection{Conservation Law} \label{sec:ConLaw}

In terms of a conservation law, the governing equation may be expressed as
\begin{equation}  \label{eq:ConservationLaw}
  u_t  + \nabla \cdot \mathbf{F}(u) = 0\;,
  \quad \textrm{where} \quad  
  \mathbf{F}(u) =   f_0 (u) \nabla \nabla^2 u    +  f_1 (u) \nabla u   
\end{equation}
is the flux vector, with two components, i.e., $\mathbf{F}(u)=\{ F_x(u),F_y(u) \}$.   To simplify the results, we generalize the flux to linear combinations of terms of the form
\begin{equation} \label{eq:GenFluxForm}
  \mathbf{F}(u) =   f (u) \mathbf{L} [u],
\end{equation}
where $f(u)$ is some smooth function of $u$ and $\mathbf{L}$ is some linear differential operator with two components, i.e., $\mathbf{L}[u]=\{ L_x[u],L_y[u] \}$.  The results to be shown can be easily extended to the flux in our original governing equation (\ref{eq:ConservationLaw}). 

The governing equation for $\bar{u}$, the average value of $u$ on some sub-domain $\Omega$, is given by
\begin{equation} 
  \bar{u}_t =  -\frac{1}{A} \oint_{\partial \Omega} \mathbf{F}(u) \cdot \mathbf{n} \; d s \;,
\end{equation}
where $A$ is the area of the sub-domain $\Omega$, $\partial \Omega$ denotes its boundary, and $\mathbf{n}$ is the outward-pointing normal to $\partial \Omega$.   To subdivide the entire domain, the solution is discretized on an equipartitioned grid; specifically, the grid points in the $x$ and $y$ directions are defined as
\begin{equation} 
\begin{split}
        x_i &= X_0 +  i \Delta s, \, \quad 0\le i\le I\; \quad \textrm{and} \\
        y_j &= Y_0 +  j \Delta s, \, \quad 0\le j\le J\; ,
\end{split}
\end{equation}
where $\Delta s$ is the grid spacing; $I+1$ and $J+1$ are the numbers of points in the $x$ and $y$ directions, respectively; and $X_0$ and $Y_0$ are the initial points in $x$ and $y$, respectively.  Furthermore, the cell average points are given by
\begin{equation} 
  \bar{u}_{(i,j)} =  \frac{1}{A} \oint_{ \Omega_{(i,j)} } u(x,y) \; dA , 
  \quad \textrm{where} \quad \Omega_{(i,j)}=[x_i,x_{i+1}]\times[y_{j},y_{j+1}]
  \quad \textrm{and} \quad A_{(i,j)}=\Delta s^2 \;,
\end{equation}
for $0\le i < I$ and $0\le j < J$; and their respective governing equations are
\begin{equation} \label{eq:GovEqnCell}
  \bar{u}_{t,(i,j)} = -
	\frac{1}{\Delta s^2 }
		\left[
		  \int_{x_{i}}^{x_{i+1}}  F_{y,(,j)}  \; dx
		+ \int_{y_{j+1}}^{y_{j}} F_{x,(i+1,)} \; dy 
		- \int_{x_{j+1}}^{x_{j}} F_{y,(,j+1)} \; dx
		- \int_{y_{j}}^{y_{j+1}} F_{x,(i,)}  \; dy  		
		\right] \;,
\end{equation}
where
\begin{equation} \label{eq:CellVolumeFlux}
  F_{x,(i,j)} =  F_x(u(x_{i},y_j)) \;, \quad
  F_{x,(,j)} =  F_x(u(x,y_j)) \;, \quad 
  F_{x,(i,)} =  F_x(u(x_{i},y)) \;, 
\end{equation}
and similar notation is used for $F_y$ and $\bar{u}_t$.   To solve (\ref{eq:GovEqnCell}), two second-order accurate approximations are implemented: 1) the integrals on the right-hand side of (\ref{eq:GovEqnCell}) are evaluated using the midpoint rule, e.g.,
\begin{equation} \label{eq:FluxApprox}
		 \int_{x_{i}}^{x_{i+1}} F_{y,(,j)} \; dx = 
		 \Delta s \left[ F_{y,(i+\myfrac{1}{2},j)}  + O(\Delta s^2) \right] \; ,
\end{equation}
where
\begin{equation} 
\begin{split}
        x_{i+\myfrac{1}{2}} &= X_0 + \left( \myfrac{1}{2} + i \right)\Delta s \, \quad 0\le i<I\; , \quad \textrm{and} \\
        y_{j+\myfrac{1}{2}} &= Y_0 + \left( \myfrac{1}{2} + j \right)\Delta s \, \quad 0\le j<J\; ,
\end{split}
\end{equation}
are the center points of a cell; and 2) the cell averaged value is approximated by the cell-centered value, i.e., 
\begin{equation} \label{eq:CellAvgApprox}
 \bar{u}_{(i,j)}  = u_{(i+\myfrac{1}{2},j+\myfrac{1}{2})}   + O(\Delta s^2)  = u ( x_{i+\myfrac{1}{2}}, y_{j+\myfrac{1}{2}} )   + O(\Delta s^2) \;.
\end{equation}
Substituting (\ref{eq:FluxApprox}) and (\ref{eq:CellAvgApprox}) into (\ref{eq:GovEqnCell}) yields
\begin{equation} \label{eq:GovEqnCell2}
	u_{t,(i+\myfrac{1}{2},j+\myfrac{1}{2})} = 
		- \frac{1}{\Delta s}
			\left[ 
				  F_{x,(i+1,j+\myfrac{1}{2})}
				- F_{x,(i,j+\myfrac{1}{2})}
				+ F_{y,(i+\myfrac{1}{2},j+1)}
				- F_{y,(i+\myfrac{1}{2},j)}
			\right]
		+ O(\Delta s^2)\; .
\end{equation}

\subsection{Temporal Discretization} \label{sec:TemporalDiscretization}

Let us write ($\ref{eq:GovEqnCell2}$) in the following general form 
\begin{equation} \label{eq:NumericGeneralForm}
	\mathbf{u}_t =-  \mathbf{D} \mathbf{u}\; ,
\end{equation}
where $\mathbf{D}$ is a nonlinear matrix differential operator, representing the fluxes on the right-hand side of ($\ref{eq:GovEqnCell2}$); and $\mathbf{u}=\{u_p\}$ is the collection of grid point values $u_{(i,j)}$.   The grid points in $\mathbf{u}$ are ordered lexicographically and may be in row-major form, $p=Ji+j$; or column-major form, $p=Ij+i$.   Note that vector notation will be used to denote vectors associated with cell centered quantities.

To begin, a central difference discretization in time is applied to (\ref{eq:NumericGeneralForm}),
\begin{equation} \label{eq:GeneralTimeDiscit}
  \mathbf{u}^{n+1} - \mathbf{u}^{n} = -\Delta t \mathbf{D} \mathbf{u}^{n+1/2} 
\end{equation}
where $\Delta t$ is the time step, and $n$ superscripts denote the value at the current time.   Using the trapezoidal rule to evaluate the non-linear term at the half step, $\mathbf{D} \mathbf{u}^{n+1/2}=\left[ \mathbf{D} \mathbf{u}^{n+1} + \mathbf{D} \mathbf{u}^{n} \right]/2 + O(\Delta t^2)$, (\ref{eq:GeneralTimeDiscit}) may be expressed as
\begin{equation} \label{eq:TrapRule}
  \left(\mathbf{I} + \frac{\Delta t}{2} \mathbf{D} \right)\mathbf{u}^{n+1}  =  \left(\mathbf{I} - \frac{\Delta t}{2} \mathbf{D} \right)\mathbf{u}^{n}  \;,
\end{equation}
where $\mathbf{I}$ is the identity matrix.   Equation (\ref{eq:TrapRule}) contains a nonlinear implicit term, i.e., nonlinear dependence on the unknown $\mathbf{u}^{n+1}$; therefore the above equation is solved numerically using a Newton iterative scheme.    

\subsubsection{Newton Iterative Scheme}

To rewrite (\ref{eq:TrapRule}) as a root-solving problem, we define
\begin{equation}
 \mathbf{G}(\mathbf{u}^{n+1}) = \left[  \left(\mathbf{I} + \frac{\Delta t}{2} \mathbf{D} \right)  \mathbf{u}^{n+1} - \left(\mathbf{I} - \frac{\Delta t}{2} \mathbf{D} \right)  \mathbf{u}^{n} \right] = 0 \;,
\end{equation}
and find $\mathbf{u}^{n+1}$ satisfying this equation using Newton's iterative scheme.  First, we compute the Jacobian of $G$ with respect to $\mathbf{u}^{n+1}$, yielding
\begin{equation} \label{eq:JacobianDerivativeOperator}
\mathbf{J}_{\mathbf{G}} = \mathbf{I} + \frac{\Delta t}{2} \mathbf{J}_{\mathbf{D}} \;, \\
\end{equation}
where $\mathbf{J}_\mathbf{D}$ is the Jacobian of $\mathbf{D} \mathbf{u}^{n+1}$.   The Newton iterative scheme may be expressed as
\begin{equation} \label{eq:NewtonIterativeForm}
 \begin{split} 
  \mathbf{J}_{\mathbf{G},(k)}  \mathbf{w} & = \left( \mathbf{I} + \frac{\Delta t}{2} \mathbf{J}_{\mathbf{D}} \right) \mathbf{w} = -\mathbf{G}(\mathbf{u}^{n+1}_{(k)}) \\ 
	\mathbf{u}^{n+1}_{(k+1)} & = \mathbf{u}^{n+1}_{(k)} + \mathbf{w} \;,
 \end{split}
\end{equation}
where $\mathbf{u}_{(k)}$ corresponds to the values at the current iterative step, $\mathbf{w}$ is an intermediate variable, and the $(k)$ subscript in $\mathbf{J}_{\mathbf{G},(k)}$ indicates that the nonlinear $\mathbf{u}^{n+1}$ terms in $\mathbf{J}_{\mathbf{G},(k)}$ are evaluated using $\mathbf{u}^{n+1}_{(k)}$.   The iterative scheme is initialized by 
setting $\mathbf{u}^{n+1}_{(0)}=\mathbf{u}^{n}$.   To reduce computational complexity, we apply an \gls{ADI} type scheme, discussed next.  

\subsubsection{Alternating Direction Implicit (ADI) Method}

The \gls{ADI} method splits the implicit differential operator (\ref{eq:JacobianDerivativeOperator}) in two parts (implicit $x$ derivatives for a fixed $y$ and vice versa). Specifically, the right-hand side of (\ref{eq:JacobianDerivativeOperator}) may be expressed as 
\begin{equation} \label{eq:DirectionSplit}
  \mathbf{I} + \frac{ \Delta t}{2} \mathbf{J}_\mathbf{D} =
\left( \mathbf{I} + \frac{ \Delta t}{2} \mathbf{J}_y\right) 
  \left( \mathbf{I} + \frac{ \Delta t}{2} \mathbf{J}_x\right)
 + \frac{\Delta t}{2} \mathbf{J}_{xy} - \frac{\Delta t^2}{4} \mathbf{J}_x \mathbf{J}_y \;, 
\end{equation}
where $\mathbf{J}_x$ and $\mathbf{J}_y$ are the pure $x$ and $y$ derivative terms in $\mathbf{J}_\mathbf{D}$, respectively; and $\mathbf{J}_{xy}$ represents the remaining mixed derivative terms, i.e. $\mathbf{J}_{xy}=\mathbf{J}_\mathbf{D}-\mathbf{J}_x-\mathbf{J}_y$.  

Neglecting the implicit mixed derivatives, $\mathbf{J}_{xy}$, we may substitute (\ref{eq:DirectionSplit}) into (\ref{eq:NewtonIterativeForm}) reducing the method to first-order accuracy; however, as noted and 
confirmed numerically~\cite{Witelski2003}, implementing an appropriate iterative method, the \gls{ADI} scheme will converge to second order accuracy in time. We will also show second order convergence for our implementation.   The Newton iterative scheme in (\ref{eq:NewtonIterativeForm}) becomes 
\begin{eqnarray} 
\left( \mathbf{I} + \frac{ \Delta t}{2} \mathbf{J}_{y,(k)} \right) 
\mathbf{w}   & = & -
\left[  \left(\mathbf{I} + \frac{\Delta t}{2} \mathbf{D} \right)  \mathbf{u}^{n+1}_{(k)} - \left(\mathbf{I} - \frac{\Delta t}{2} \mathbf{D} \right)  \mathbf{u}^{n} \right]  \;, \label{eq:JacobianSystem1} \\
  \left( \mathbf{I} + \frac{ \Delta t}{2} \mathbf{J}_{x,(k)}\right)
\mathbf{v}  & = & \mathbf{w}  \; , \label{eq:JacobianSystem2} \\
\mathbf{u}^{n+1}_{(k+1)} &= & \mathbf{u}^{n+1}_{(k)} + \mathbf{v} \;,  \label{eq:JacobianSystem3}
\end{eqnarray}
where $\mathbf{w}$ and $\mathbf{v}$ are intermediate variables.
This pseudo-Newton iterative method reduces (by an approximation) the left hand side of (\ref{eq:DirectionSplit}), from a large sparse matrix of size $IJ\times IJ$, to two block diagonal matrices (right hand side of (\ref{eq:DirectionSplit})).   In the implicit $y$ step, using column-major ordering, we form a block diagonal matrix, $\mathbf{J}_y$, with $I$ blocks of size $J\times J$. Each block can be inverted independently of others, thus reducing the computational complexity. Furthermore, each block is $n$-diagonal, where the width of the diagonal band, $n$, depends on the stencil used to  evaluate spatial derivatives.  For our implementation, $n=5$ (penta-diagonal system). The $n$-diagonal matrices can be inverted with linear complexity, further reducing computational 
cost, for example by using an extension of the Thomas algorithm for an $n$-diagonal matrix. Similarly, in the implicit $x$ step, using row-major ordering, we form a block diagonal matrix, $\mathbf{J}_x$, with $J$ blocks of size $I\times I$. 

\subsubsection{Adaptive Time Stepping} \label{sec:adaptive_time_step}

\begin{figure}
	\begin{tikzpicture}[node distance=1.7cm]
		\node (start) [startstop] {Start};
		\node (in1) [io, below of=start] {Load initial condition and set initial time step, $\Delta t$.};
		\node (pro1) [process, below of=in1] {Initialize iterative scheme $u^{n+1}_{(0)}=u^{n}$.};
		\node (pro2) [process, below of=pro1] {Solve (\ref{eq:JacobianSystem1})--(\ref{eq:JacobianSystem3}).};
		\node (dec1) [decision, below of=pro2] {Does next iteration satisfy solution criteria? };
		\node (pro3) [process, below of=dec1] {Accept solution as next time step, $u^{n+1}=u^{n+1}_{(k+1)}$. };
		\node (dec5) [decision, below of=pro3] {Final time step reached? };
		\node (end) [startstop, below of=dec5] {End};

		\node (dec2) [decision, right of=dec1, xshift=4.5cm] {Has a maximum iteration count been reached? };
		\node (pro2b) [process, right of=pro2, xshift=4.5cm] {Reject current solution at $n+1$ and reduce $\Delta t$.};
		\node (dec2b) [decision, right of=pro1, xshift=4.5cm] {Is $\Delta t$ below some minimum threshold?};
		\node (endb) [startstop, right of=in1, xshift=4.5cm] {Failed end state, $\Delta t$ too small.};

		\node (dec4c) [decision, left of=dec1, xshift=-4.5cm] {Has a minimum number of successful time steps been computed?};
		\node (pro5c) [process, left of=pro2, xshift=-4.5cm] {Increase $\Delta t$.};
		\node (dec6c) [decision, left of=pro1, xshift=-4.5cm] {Is $\Delta t$ above some maximum threshold?};
		\node (proc7c) [process, left of=in1, xshift=-4.5cm] {Set $\Delta t$ to maximum threshold value.};

		\draw [arrow] (start) -- (in1);
		\draw [arrow] (in1) -- (pro1);
		\draw [arrow] (pro1) -- (pro2);
		\draw [arrow] (pro2) -- (dec1);
		\draw [arrow] (dec1) -- node[anchor=east] {yes} (pro3);
		\draw [arrow] (pro3) -- (dec5);
		\draw [arrow] (dec5) -- node[anchor=east] {yes} (end);

		\draw [arrow] (dec1) -- node[anchor=north] {no} (dec2);
		\draw [arrow] (dec2) -- node[anchor=north] {no} (pro2);
		\draw [arrow] (dec2) -- node[anchor=east] {yes} (pro2b);
		\draw [arrow] (pro2b) -- (dec2b);
		\draw [arrow] (dec2b) -- node[anchor=north] {no} (pro1);
		\draw [arrow] (dec2b) -- node[anchor=east] {yes} (endb);

		\draw [arrow] (dec5) -| node[anchor=north] {no} (dec4c);
		\draw [arrow] (dec4c) -- node[anchor=east] {no} (pro1);
		\draw [arrow] (dec4c) -- node[anchor=east] {yes} (pro5c);
		\draw [arrow] (pro5c) -- (dec6c);
		\draw [arrow] (dec6c) -- node[anchor=north] {no} (pro1);
		\draw [arrow] (dec6c) -- node[anchor=east] {yes} (proc7c);
		\draw [arrow] (proc7c) -- (pro1);

	\end{tikzpicture}
	\caption{Flow chart of the adaptive time stepping with a nested pseudo-Newton
iterative scheme. Note that red rectangles correspond to the start or end
of the flow chart, orange boxes designate processes, and green boxes denote decisions. }
	\label{fig:FlowChartTime}
\end{figure}
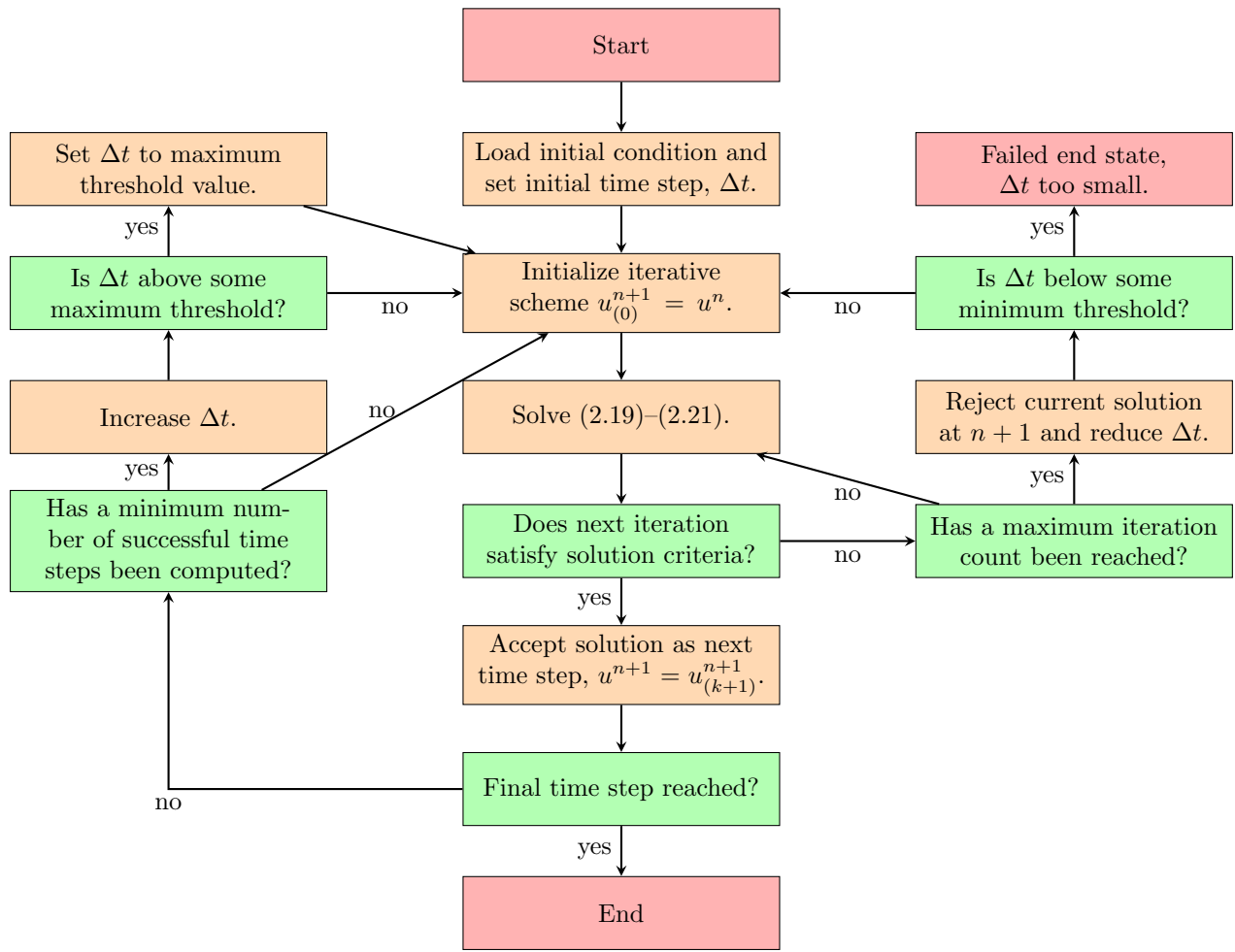

To control the time step, we first specify the convergence conditions. Specifically, the solution at the next time step is accepted (i.e. $\mathbf{u}^{n+1}=\mathbf{u}^{n+1}_{(k+1)}$) if
\begin{equation} \label{eq:NewtonError}
 \max_{0\le i < I,0\le j < J} \left| \frac{v_{(i+\myfrac{1}{2},i+\myfrac{1}{2})}}{u_{(k),(i+\myfrac{1}{2},i+\myfrac{1}{2})}} \right| < \ep_\textrm{Tol}
\end{equation}
where $\ep_\textrm{Tol}$ is the error tolerance. In addition, if the error tolerance is not satisfied in a specified maximum number of iterative steps, the scheme is assumed to have failed, triggering time step decrease.

\textbf{Note 1:} Our simulations show that setting the maximum number of iterative steps, $K$, to 10 gives the largest effective time steps, i.e., $\Delta t/K$ is maximized.

\textbf{Note 2:} Other restrictions besides the convergence of the iterative scheme may be placed on accepting the solution at the next iteration or next time step, e.g. sufficiently small truncation error, thus for generality, we will refer to the collection of such conditions as the {\it solution criteria}.

If the solution criteria are not met, then the time-step is reduced and the pseudo-Newton iterative scheme is performed with the new reduced time step. Furthermore, if the new reduced time step is below some minimum, the numerical scheme halts and ends in a failed state. To improve the efficiency of the numerical scheme, the time step is increased if a minimum number of successful time steps have been consecutively computed.   To remove possible numerical errors, the time step is bounded from above.   Figure~\ref{fig:FlowChartTime} shows a flow chart of the adaptive time stepping procedure, coupled with the pseudo-Newton iterative method.

\subsection{Flux Discretization} \label{sec:FluxDiscretization}

To summarize the results so far, a second-order accurate scheme has been developed in terms of a non-linear spatial differential operator $\mathbf{D}$, representing the fluxes across cell boundaries. In this section, to complete the scheme, we specify the form of $\mathbf{D}$.  Recalling (\ref{eq:GovEqnCell2}), we may split $\mathbf{D}$ into two parts: the fluxes across the $x$ cell boundaries $\mathbf{D}_x$, and the fluxes across the $y$ cell boundaries $\mathbf{D}_y$.  The values of  the components of $\mathbf{D}$ are specified in terms of fluxes, i.e., 
\begin{equation} \label{eq:FluxDiffOper}
  \mathbf{D}_s = \frac{\mathbf{F}_{s,+} - \mathbf{F}_{s,-} } {\Delta s} \; ,
\end{equation}
where $\mathbf{D}_s$ is either $\mathbf{D}_x$ or $\mathbf{D}_y$, and $\mathbf{F}_{s,+}$ and $\mathbf{F}_{s,-}$ are non-linear matrix differential operators for the fluxes at the two boundaries.   Similarly to $\S$~\ref{sec:ConLaw}, the fluxes are generalized to a linear combination of functions of the form
\begin{equation}
	\mathbf{F} = f(u) \mathbf{L} \; ,
\end{equation}
where $f(u)$ is a function of $u$, and $\mathbf{L}$ is a linear differential operator. We may express $\mathbf{D}$ as 
\begin{equation}
	\mathbf{D} = \frac{\mathbf{F}_{x,+} - \mathbf{F}_{x,-} + \mathbf{F}_{y,+} - \mathbf{F}_{y,-} }{\Delta s} \; .
\end{equation}
To derive an expression for $\mathbf{J}_x$ and $\mathbf{J}_y$ in (\ref{eq:JacobianSystem1}) and (\ref{eq:JacobianSystem2}), we recall that mixed derivative terms have been ignored,
and therefore,
\begin{equation}
	\mathbf{J}_x = \pad{}{\mathbf{u}} \left[ \frac{\mathbf{F}_{x,+} - \mathbf{F}_{x,-} }{\Delta s} \right] \; 
	\quad \textrm{and} \quad
	\mathbf{J}_y = \pad{}{\mathbf{u}} \left[ \frac{\mathbf{F}_{y,+} - \mathbf{F}_{y,-} }{\Delta s} \right] \; ,
\end{equation}
where $\partial/\partial \mathbf{u}$ denotes the Jacobian,
and the non-linear matrix differential operator, $\mathbf{F}$, only contains derivatives with respect to its subscript. 

\begin{figure}[!h]
\centering
   \includegraphics[width=0.3\textwidth]{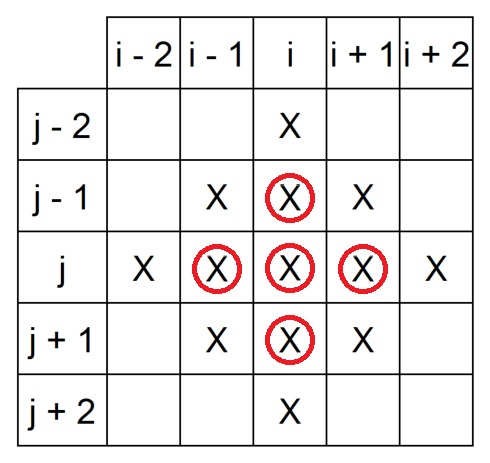}  
\caption{   X and {\color{red}$\bigcirc$} denote the solution cell-center values $u_{(i,j)}$ and the function cell-center values, $f_{i,j}$, respectively.   These values are required to numerically evaluate the flux differences (\ref{eq:FluxDiffOper}) at the cell-center $(i,j)$. } 
\label{fig:StencilF_T}
\end{figure}

To complete the derivation, expressions for $\mathbf{F}_{x,+}$, $\mathbf{F}_{x,-}$, $\mathbf{F}_{y,+}$, and $\mathbf{F}_{y,-}$ are required.
For brevity, we specify these expressions by the following descriptions: 
\begin{enumerate}
\item at the cell-centered point $(i,j)$, 
\begin{itemize}
  \item $\mathbf{F}_{x,+}$ is evaluated at $(x_{i+1},y_{j+\myfrac{1}{2}})$\; ,
  \item $\mathbf{F}_{x,-}$ is evaluated at $(x_{i},y_{j+\myfrac{1}{2}})$ \; ,
  \item $\mathbf{F}_{y,+}$ is evaluated at $(x_{i+\myfrac{1}{2}},y_{j+1})$\; , and
  \item $\mathbf{F}_{y,-}$ is evaluated at $(x_{i+\myfrac{1}{2}},y_{j})$\; ; 
\end{itemize}
\item
the linear matrix differential operator $\mathbf{L}$ is computed using cell-centered values; 
\item
 $f(u)$ at a cell boundary is evaluated using interpolation, e.g.,
\begin{equation}
  f(u_{(i+\myfrac{1}{2},j)}) = \frac{f(u_{(i+\myfrac{1}{2},j+\myfrac{1}{2})}) + f(u_{(i+\myfrac{1}{2},j-\myfrac{1}{2})}) }{2} + O (\Delta s^2) \; ;
\end{equation}
\item 
the total combined stencil of all terms is shown in Figure~\ref{fig:StencilF_T}.
\end{enumerate}

\subsection{Boundary Conditions} \label{sec:BoundaryCondition}

To complete the description of the numerical methods, we now discuss the boundary conditions.  Typically, boundary conditions of the form 
\begin{equation} \label{eq:BoundaryConditions}
\begin{split}
     \quad u_x(x_0,y) =  u_{xxx}(x_0,y) = 0 \; , \quad
     \quad u_x(x_I,y) =  u_{xxx}(x_I,y) = 0 \; , \\
     \quad u_y(x,y_0) =  u_{yyy}(x,y_0) = 0 \; , \quad
     \quad u_y(x,y_J) =  u_{yyy}(x,y_J) = 0 \; , 
\end{split}
\end{equation}
are used, and may be interpreted as symmetry conditions.  Furthermore, these boundary conditions correspond to zero flux at the boundary (recall (\ref{eq:ConservationLaw})).

To implement the boundary conditions, the stencil shown in Figure~\ref{fig:StencilF_T} includes ghost points at the boundaries.  While ghost points may be used for explicit terms ($\nstn{u}{n+1}{k}$ and $u^n$), for the implicit term, $\nstn{u}{n+1}{k+1}$, the finite difference stencil must be modified. However, since implicit terms are derivatives of a single variable, the stencil modification is relatively simple.

Using the boundary conditions to define the ghost points, the left-hand sides of the numerical scheme (\ref{eq:JacobianSystem1}) and (\ref{eq:JacobianSystem2}) are computable everywhere except at the corners of the domain, e.g., at the cell-center point $i=j=0$.  There are two choices to compute the ghost points in the corner (${x_i,y_j}$ for $i,j=-2,-1$): First, the $y$ boundary conditions are applied, computing the solution at the ghost points $(x_i,y_j)$ for $i=0,1$ and $j=-2,-1$, then the $x$ boundary conditions are applied at the new ghost points, computing the solution at the corner ghost points. Alternatively, the order of operations may be 
reversed.    It is relatively easy to show that both procedures lead to identical results.

\section{Numerical Performance} \label{sec:Num_Perf}

We now switch focus to the numerical performance and accuracy of our implementation.    The details related to implementation in the GPU 
computing environment are discussed in \ref{sec:GPUImplement}.
Here, we discuss first convergence, followed by confirmation of conservative properties, and the comparison with LSA.   Then, we discuss 
the performance by comparing  the \gls{GPU} and serial \gls{CPU} codes. 

For the purpose of the numerical tests discussed in this section, we choose an initial condition of the form
\begin{equation} \label{eq:IC_LinearWave}
u(x,y,t=0) = H_0 \left( 1 +  \left[\ep_x \cos \left( \frac{\pi x}{\lambda}\right) + \ep_y  \cos \left(\frac{\pi y}{\lambda}\right) \right] \right) \quad (x,y)\in [0,P\lambda]\times[0,P\lambda], \quad  \lambda=2\pi/q \;,
\end{equation}
where $\ep_x=\ep_y=0.1$ and $q$ is the wave number. The initial film thickness, $H_0$, is fixed at $1$ for the linear model (for simplicity), $0.5$ for the NLC model, and $3.9$ for the polymer model. For the last two models, the values are motivated by the experiments~\cite{Jacobs2008,Vandenbrouck1999}.

\subsection{Validation} \label{sec:Validation}

To test the convergence properties,  in $\S$ \ref{sec:convergence} and $\S$ \ref{sec:mass}, the adaptive time stepping is removed and we set the maximum number of Newton iterative steps to 5000 (allowing for a more robust selection of spatial step size).  
 In addition, we choose $q=q_m$ and $P=6$ in (\ref{eq:IC_LinearWave}) where $q_m$ is given by (\ref{eq:unstableLSAmax}).   {The spatial step size, $\Delta s = P \lambda_m/I$, where $\lambda_m=2\pi/q_m$ is the most unstable wavelength, $I$ is the number of grid points, and simulations are carried out on a square computational domain, $I=J$; therefore, choosing $I$ (discussed in the next sections) determines $\Delta s$. 
 Note that the choice $P=6$ is made so as to maximize the range of grid sizes that fit into the GPU's memory and maintain a sufficiently large spatial step size to maximize the (common) stable time step.  }  In $\S$~\ref{sec:LSA}, simulation results are compared with the predictions of LSA, where adaptive time stepping is utilized, the maximum number of Newton iterative steps is reverted back to 10, we set $P=1$, and we vary $q$ according to the dispersion curve from \gls{LSA}.

\subsubsection{Convergence} \label{sec:Convergence}
\label{sec:convergence}

To study the convergence properties, we first define $I_0$, the initial spatial grid size, i.e., $I=J=I_0$, and $\Delta t_0$, the initial time step size. In addition, since obtaining an analytical solution for the non-linear models is not possible, we compute the error relative to the numerical solution obtained for the most refined partition. 

\begin{figure}[!h]
	\centering
	\subfigure[]{
	\includegraphics[width=0.45\textwidth]{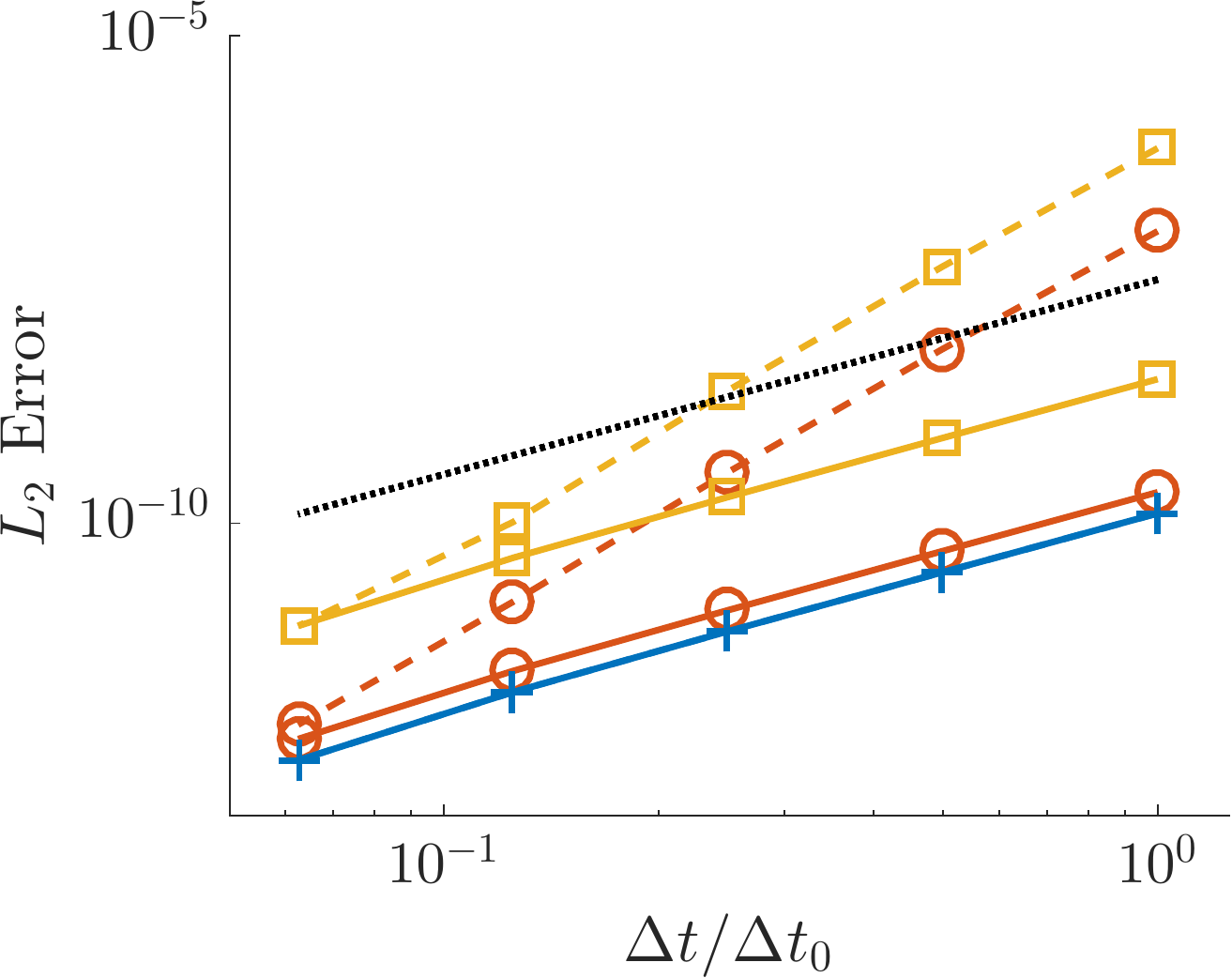} \label{fig:TimeConvergence} }
	\subfigure[]{
	\includegraphics[width=0.45\textwidth]{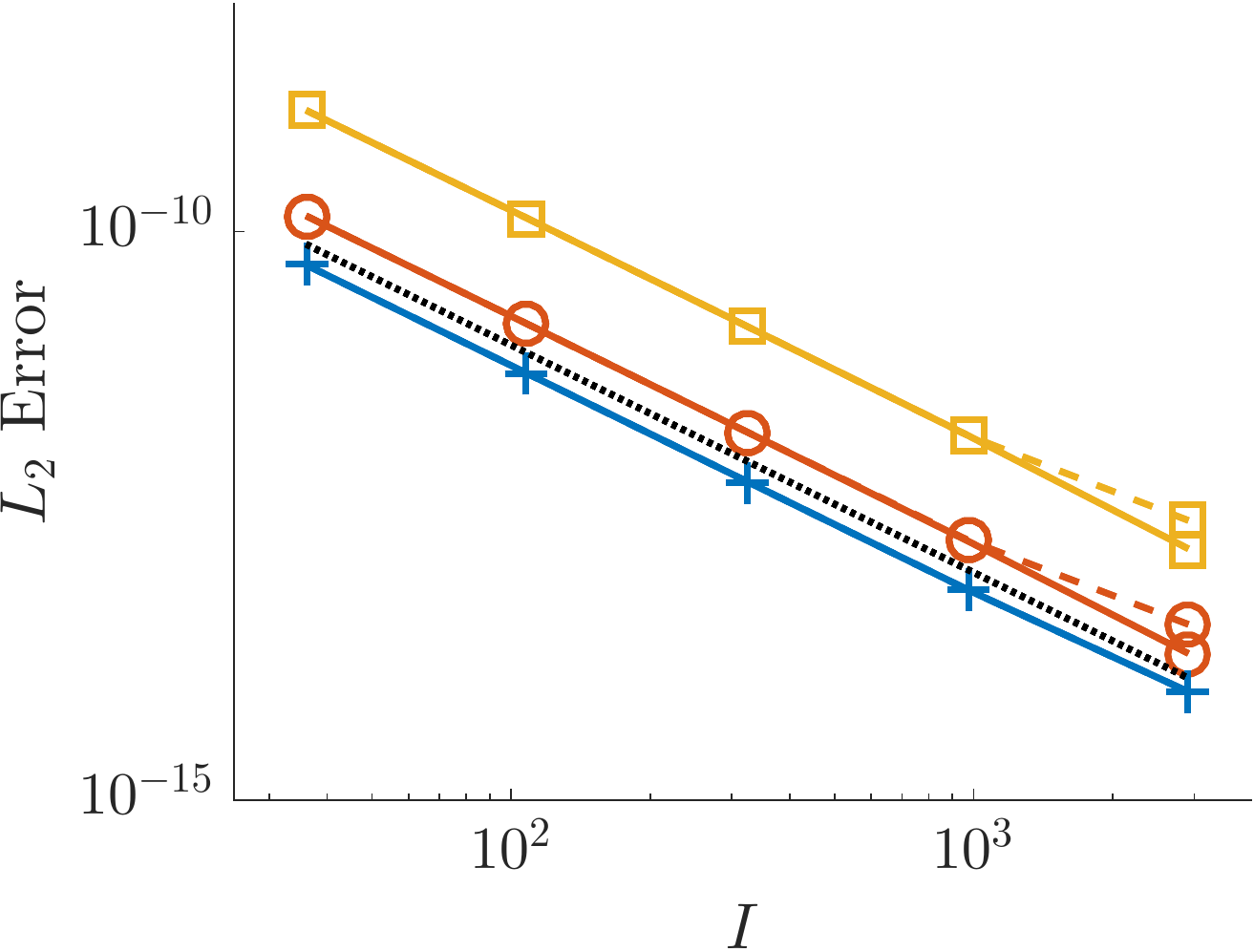} \label{fig:SpaceConvergence} }
	\caption{$L_2$ norm of the error as a function of (a) time step size and (b) number of grid points in the $x$ direction for: the linear model (blue curves with `$+$' symbols), the \gls{NLC} model (red curves with `$\circ$' symbols), and the polymer model (yellow curves with `$\Box$' symbols). Solid curves denote simulations where Newton iterative convergence error tolerance is close to machine precision, $\ep_\textrm{Tol}=10^{-14}$; and dashed curves denote simulations with error tolerances set to (a) $\ep_\textrm{Tol}=\max(\Delta s^2,10^{-14})$ and (b) $\ep_\textrm{Tol}=\max(\Delta t^2,10^{-14})$. The dotted black line denotes second order convergence. Note that in both figures for the linear model the two curves (solid and dashed) lie on top of each other, as expected. }
	\label{fig:Convergence}
\end{figure}

For the purpose of checking temporal convergence, we fix $I_0=256$ for all simulations in this section, and set $\Delta t_0=10^{-3}\omega_m^{-1}$, i.e., we scale the time step with the growth rate of the most unstable mode. The initial simulation is carried out for one time step, $\Delta t_0$, and for each subsequent simulation, the time step is halved, i.e., $\Delta t_i = 2^{-i} \Delta t_0$, where $i$ is the refinement level. Note that at each refinement level the total number of time steps doubles.  Figure~\ref{fig:TimeConvergence} shows the $L_2$ norm of the error for all three considered problems for two implementations: i) the Newton iterative convergence error tolerance, $\ep_\textrm{Tol}$ in (\ref{eq:NewtonError}), is set close to machine precision (solid curves), $\ep_\textrm{Tol}=10^{-14}$; and ii) the error tolerance is set to scale with the temporal error (dashed curves), $\ep_\textrm{Tol}=\max(\Delta t^2,10^{-14})$.  Note that we limit $\ep_\textrm{Tol} \ge 10^{-14}$ since if the relative error in the Newton iterative scheme is smaller than machine error, the iterative method will always fail.  We observe that in the first case, second-order temporal accuracy is achieved (compare to the dotted black line). Note that for the linear model, the two curves lie on top of each other, as expected.   Furthermore, the results show that it is sufficient to set the Newton error tolerance to be proportional to the temporal accuracy to maintain second-order accuracy, and even higher order convergence is achieved for the nonlinear models, although the error is larger.

{To confirm the spatial convergence, we fix $\Delta t_0=10^{-8}\omega_m^{-1}$ for all simulations and set the coarsest grid size $I_0=36$. Recalling $\S$~\ref{sec:ConLaw}, the numerical solution is evaluated at the cell-centers of the spatial grid; therefore, if the spatial step size is halved, the cell-centered points on the refined grid are not contained in the coarse grid. To avoid errors associated with interpolating the numerical solutions on different grid sizes to a common grid size, we instead triple the grid size (e.g., $I_k=3^k I_0$, where $k$ is the level of refinement) so that cell-centered points on the coarser grid are contained in the refined grid. }
 Figure~\ref{fig:SpaceConvergence} confirms second-order spatial accuracy.
 Furthermore, we observe that there is little difference between setting the Newton error tolerance to $\ep_\textrm{Tol}=\max(\Delta s^2,10^{-14})$ or close to machine precision, $\ep_\textrm{Tol}=10^{-14}$. 

To summarize, the results show that the implemented method is second-order accurate in time and space, and furthermore, that it is sufficient to set the error tolerance to scale with the order of the spatial  and temporal accuracy.   Such implementation significantly reduces computational time.

\subsubsection{Conservation of Mass}
\label{sec:mass}

\begin{figure}[!h]
	\centering
	\subfigure[]{
	\includegraphics[width=0.31\textwidth]{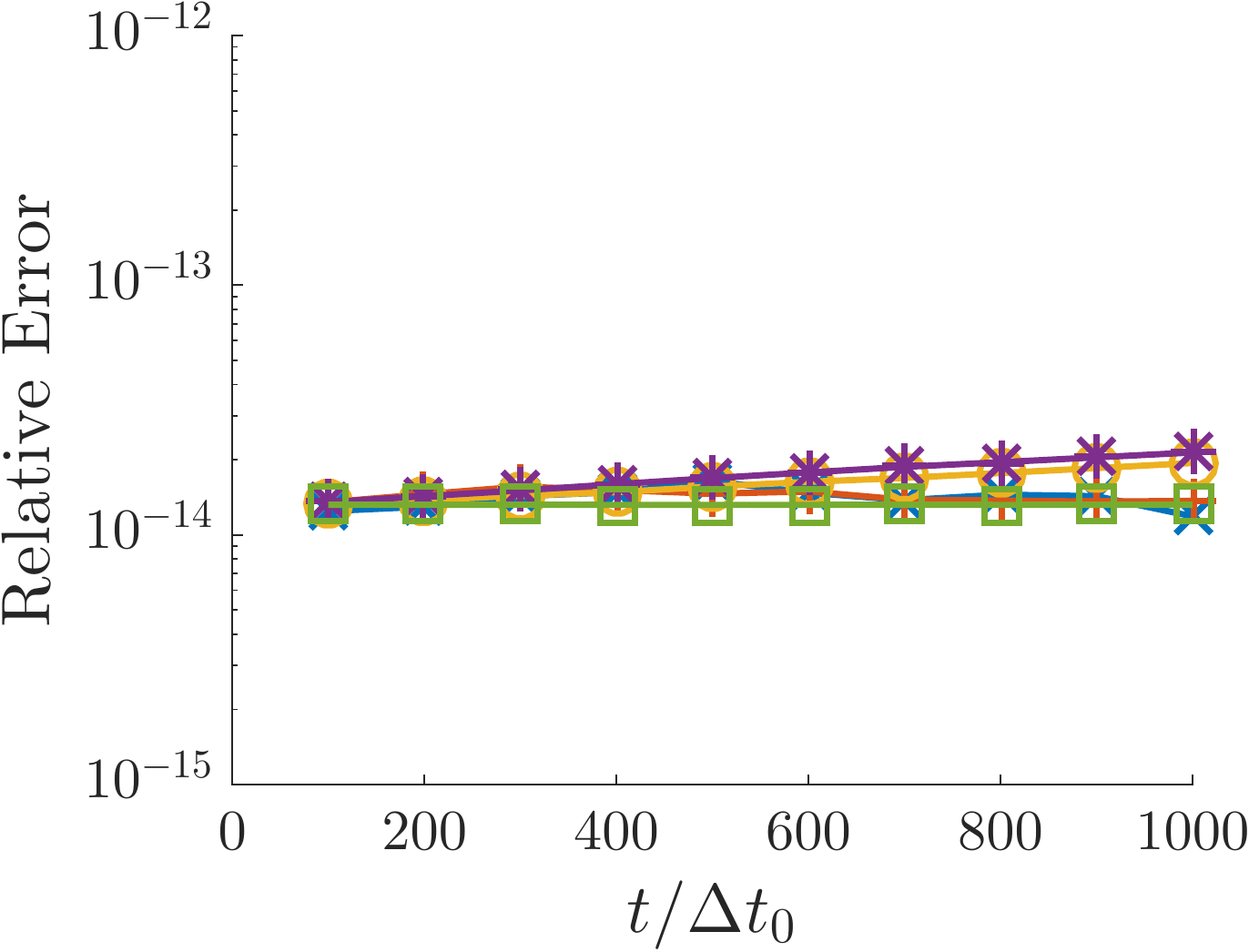} \label{fig:com_linear_space} }
	\subfigure[]{
	\includegraphics[width=0.31\textwidth]{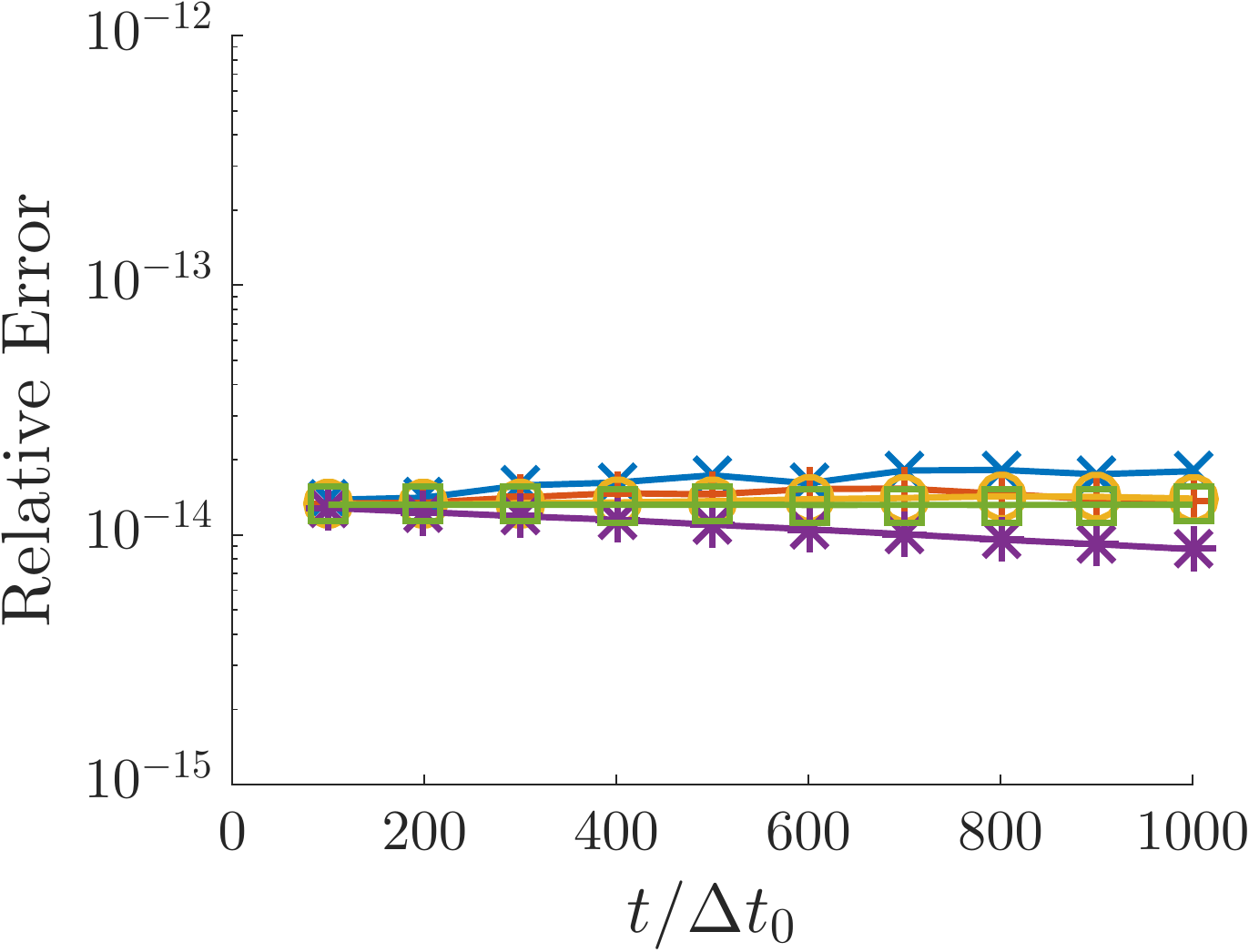} \label{fig:com_nlc_space} }
	\subfigure[]{
	\includegraphics[width=0.31\textwidth]{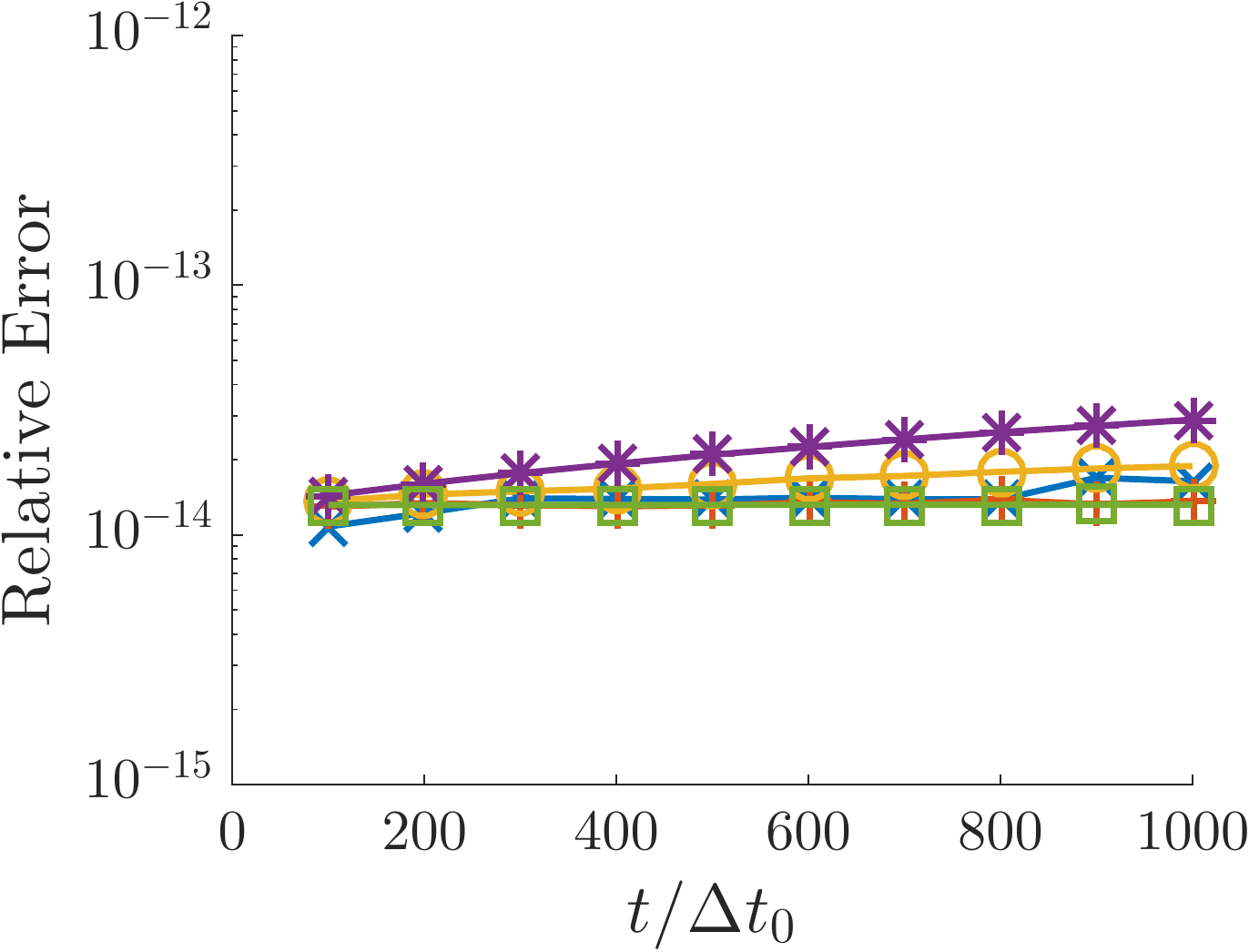} \label{fig:com_polymer_space} }
	\subfigure[]{
	\includegraphics[width=0.31\textwidth]{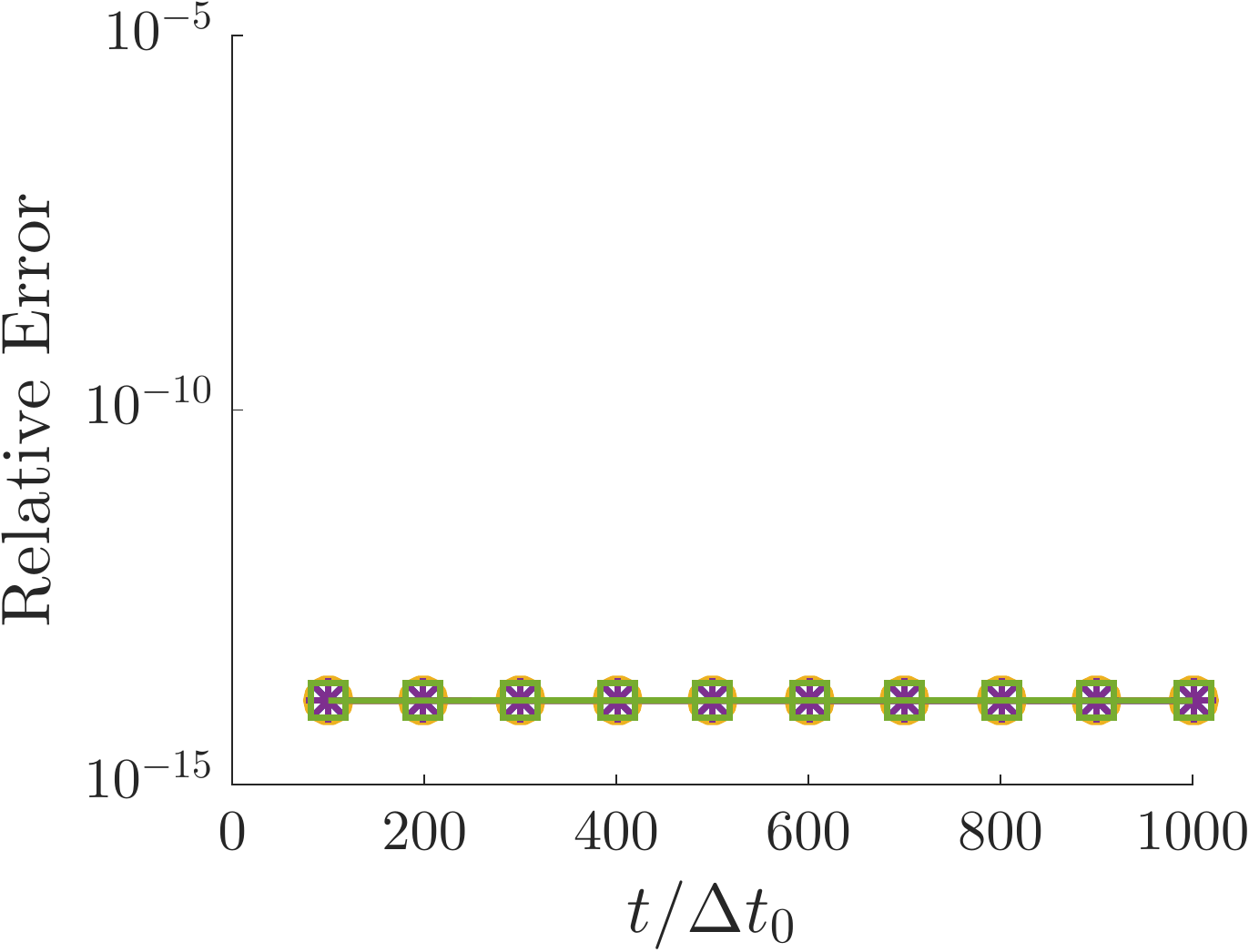} \label{fig:com_linear_time} }
	\subfigure[]{
	\includegraphics[width=0.31\textwidth]{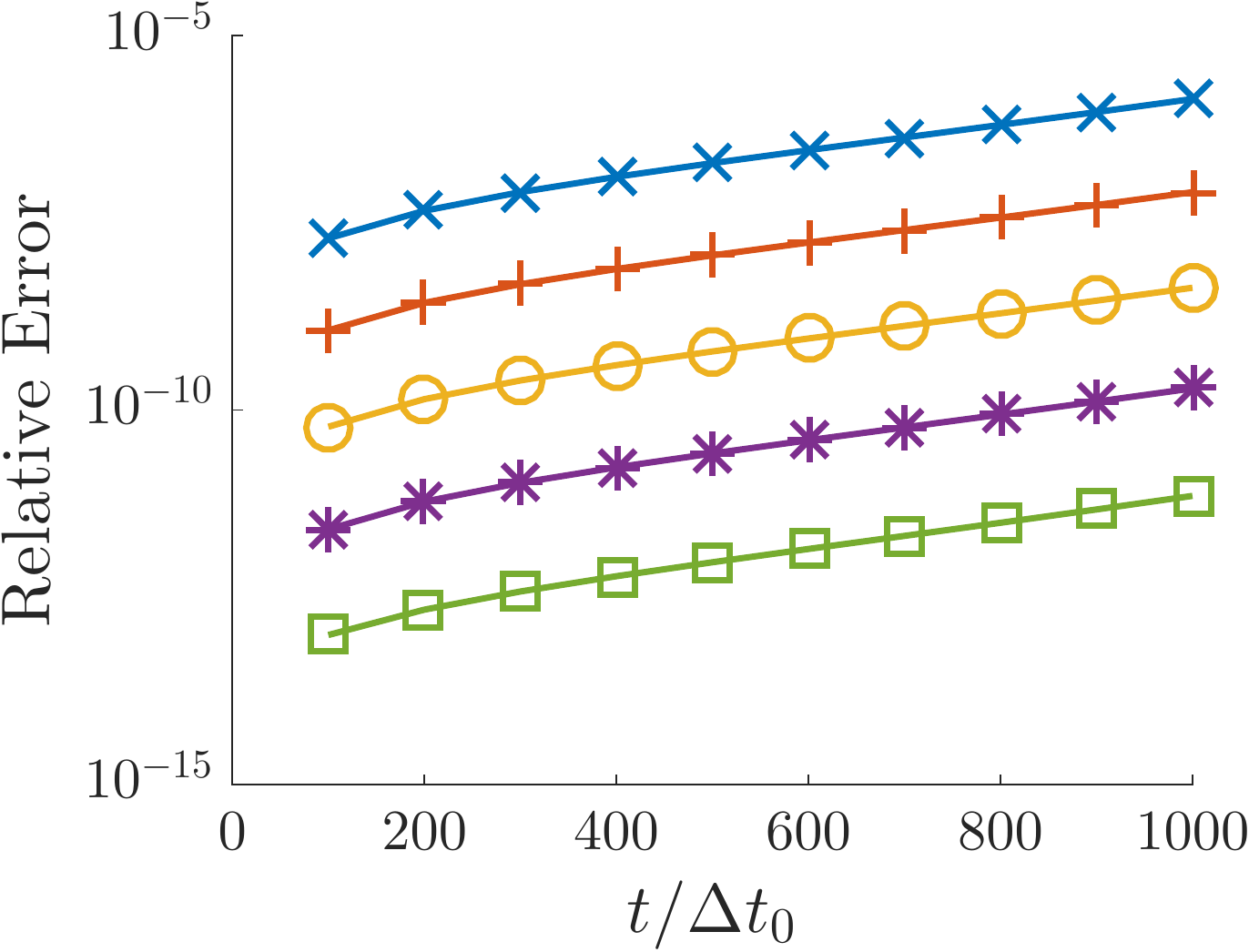} \label{fig:com_nlc_time} }
	\subfigure[]{
	\includegraphics[width=0.31\textwidth]{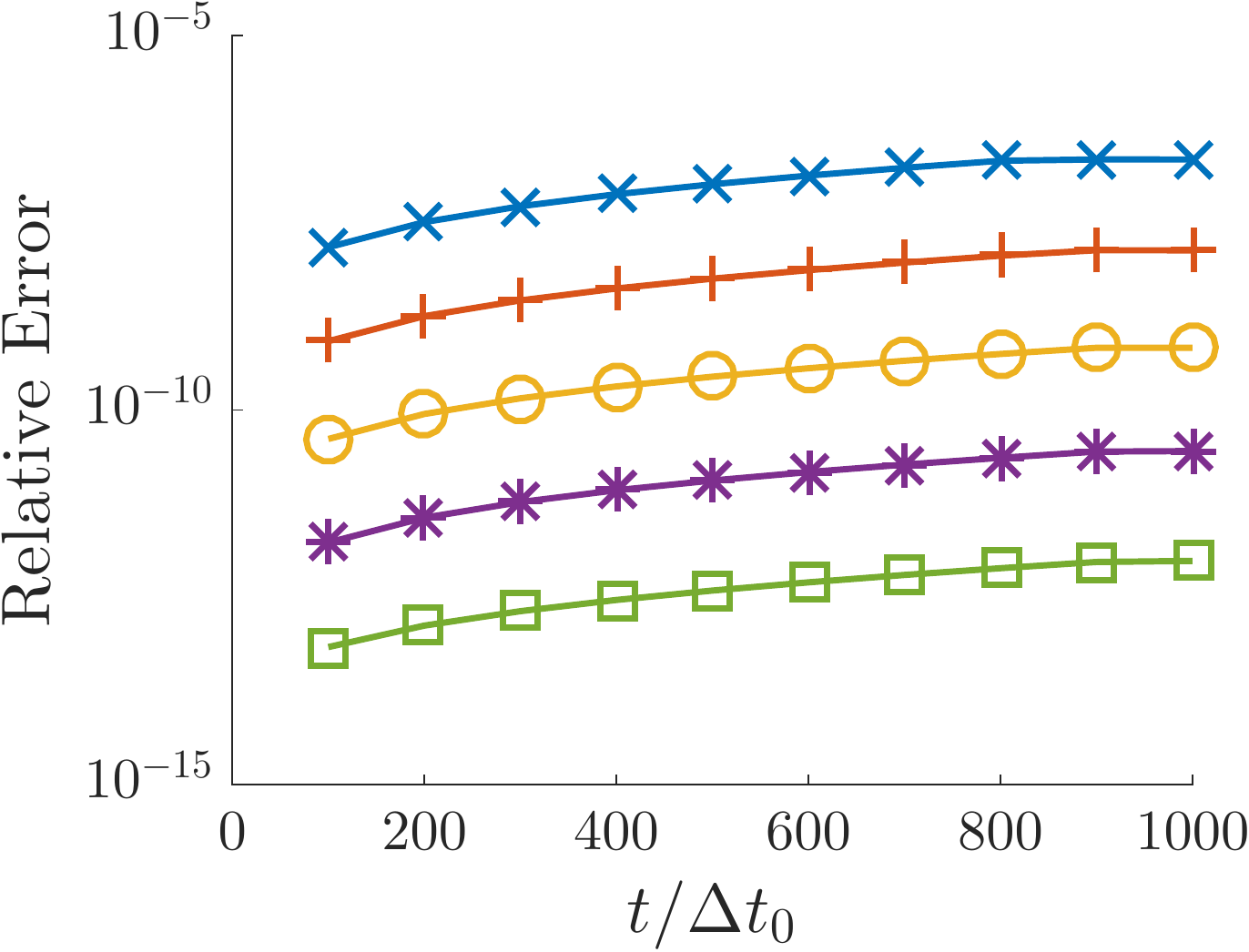} \label{fig:com_polymer_time} }
	\caption{The relative mass error as a function of time for the linear model (left column), the \gls{NLC} model (central column), and the polymer model (right column) for decreasing spatial step size (top row) and decreasing time step (bottom row). Spatial step sizes and time step sizes are the same as Figure~\ref{fig:Convergence} and `$\times$', `$+$', `$\circ$', `$*$', and `$\Box$' symbols denote progressively decreasing spatial and temporal 
step size.  Note that for the top row of figures $\Delta t= \Delta t_0 =10^{-8}\omega_m^{-1}$ and $I=I_0 3^i$, where $I_0=36$ and $i=0,1,2,3,4$; and for the bottom row of figures $I=1024$ and $\Delta t= 2^{-i} \Delta t_0$ where $\Delta t_0=10^{-3}\omega_m^{-1}$ and $i=0,1,2,3,4$ is the refinement depth (decreasing step size).  }
	\label{fig:CoM}
\end{figure}

To verify conservation of mass, we carry out the simulations until the final time, $1000\Delta t_0$, and collect data in intervals of $100\Delta t_0$. Furthermore, motivated by the results of $\S$~\ref{sec:Convergence}, we fix $\ep_\textrm{Tol}=\max(\Delta s^2,10^{-14})$ for spatial convergence simulations, and $\ep_\textrm{Tol} = \max( \Delta t^2,10^{-14})$ for temporal convergence simulations. To compute the average mass of the solution, recall that the cell average is given by the cell-centered value (with second order accuracy), therefore, the average mass on the entire domain is given by
\begin{equation}
	\bar{U} = \frac{1}{I J} \sum_{i=0}^{I-1} \sum_{j=0}^{J-1} u_{(i+\myfrac{1}{2},j+\myfrac{1}{2})} \;.
\end{equation}
Figure~\ref{fig:CoM} 
shows the change (relative to the initial condition) in total mass for the linear model (left column), \gls{NLC} model (central column) and polymer model (right column), for fixed time step and varied spatial step (top row), and fixed spatial step and varied time step (bottom row). The results obtained when the spatial step size is varied (top row) show that there is no discernible trend in the mass error (numerical noise), with the exception of the smallest step size for the \gls{NLC} model and the polymer model, indicated by the arrows in panels~\subref{fig:com_nlc_space} and ~\subref{fig:com_polymer_space}, respectively.  It is interesting to note that the conservation law formulation of the governing equation in terms of fluxes (recall equation (\ref{eq:GovEqnCell2}) in $\S$~\ref{sec:ConLaw}) indicates that conservation of mass should be second order accurate in space. 
However, while the approximations of the fluxes are second-order accurate, cancellation of these errors across a cell boundary leads to higher order accuracy.

When time step is varied, with the spatial step fixed, the bottom row of Figure~\ref{fig:CoM} shows that conservation of mass is at least second order 
accurate.  For the linear model, panel~\subref{fig:com_linear_time}, the error is close to machine precision, so no trend is observed.

\subsubsection{Comparison to LSA}
\label{sec:LSA}

\begin{figure}[!h]
	\centering
	\subfigure[$q_m=1/\sqrt{2}, \; \omega_m=1/4$]{
	\includegraphics[width=0.31\textwidth]{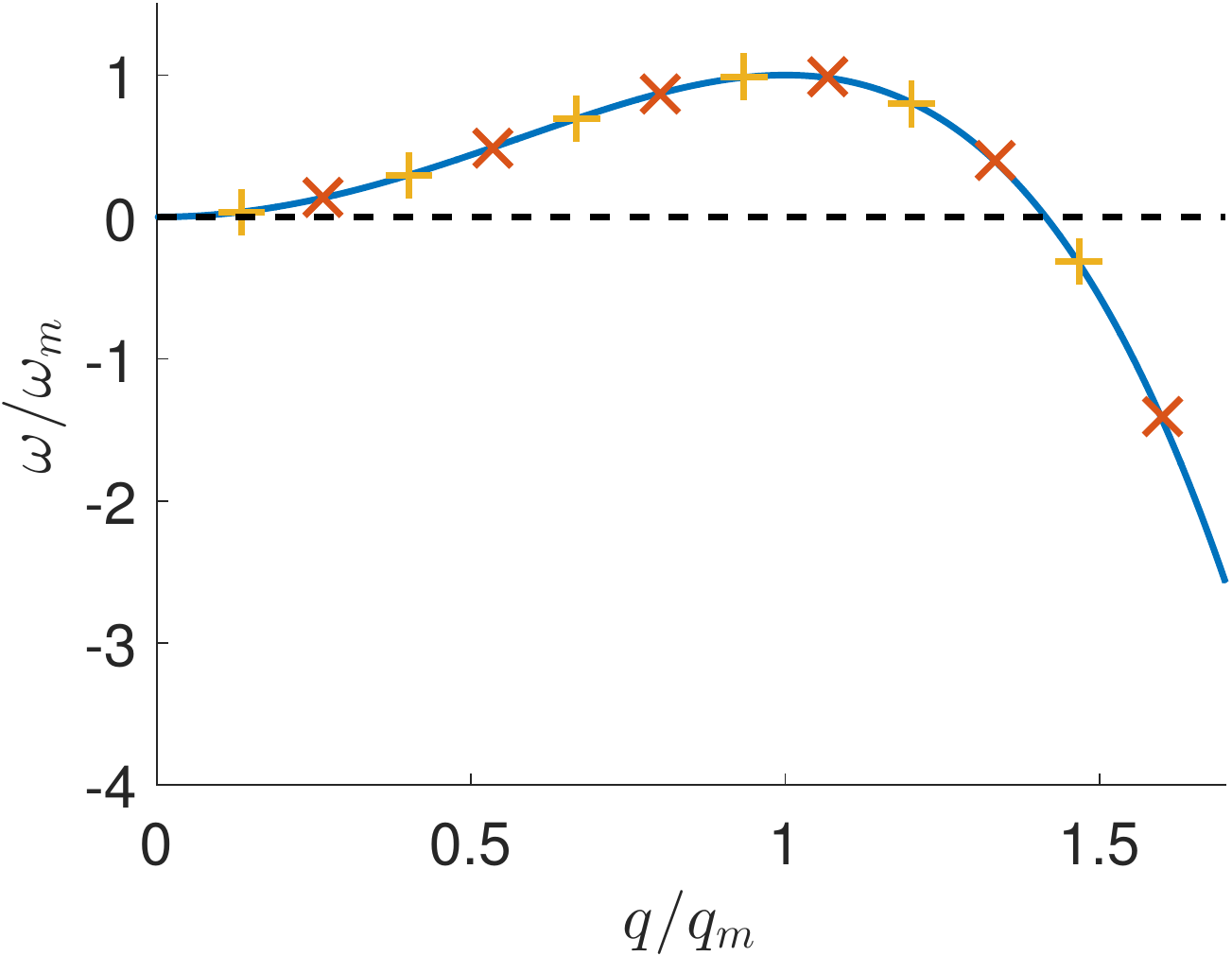} \label{fig:dispersion_curve_linear} }
	\subfigure[$q_m \approx 1.482, \; \omega_m \approx 0.05169$]{
	\includegraphics[width=0.31\textwidth]{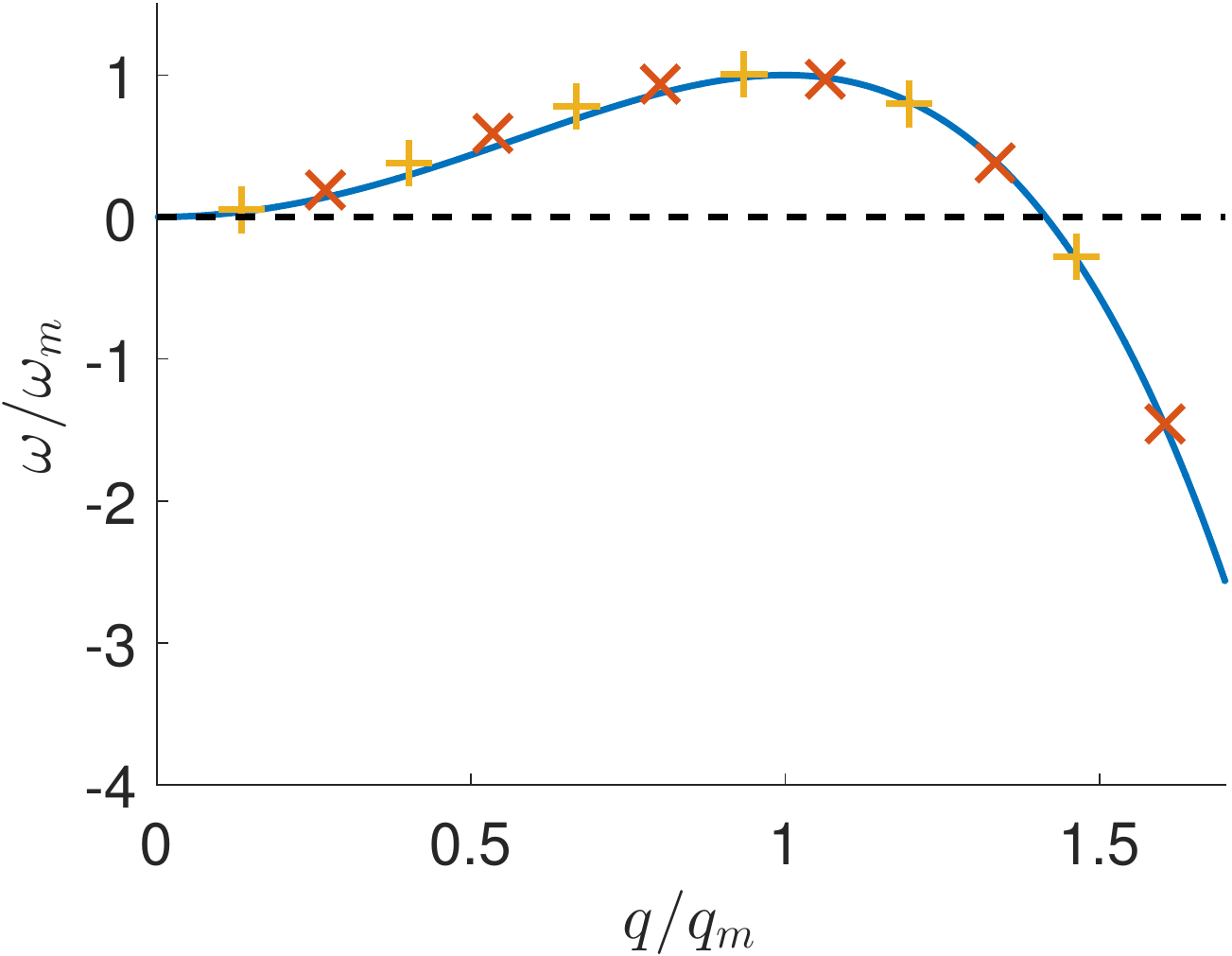} \label{fig:dispersion_curve_NLC} }
	\subfigure[$q_m \approx 1.600, \; \omega_m \approx 2.040$]{
	\includegraphics[width=0.30\textwidth]{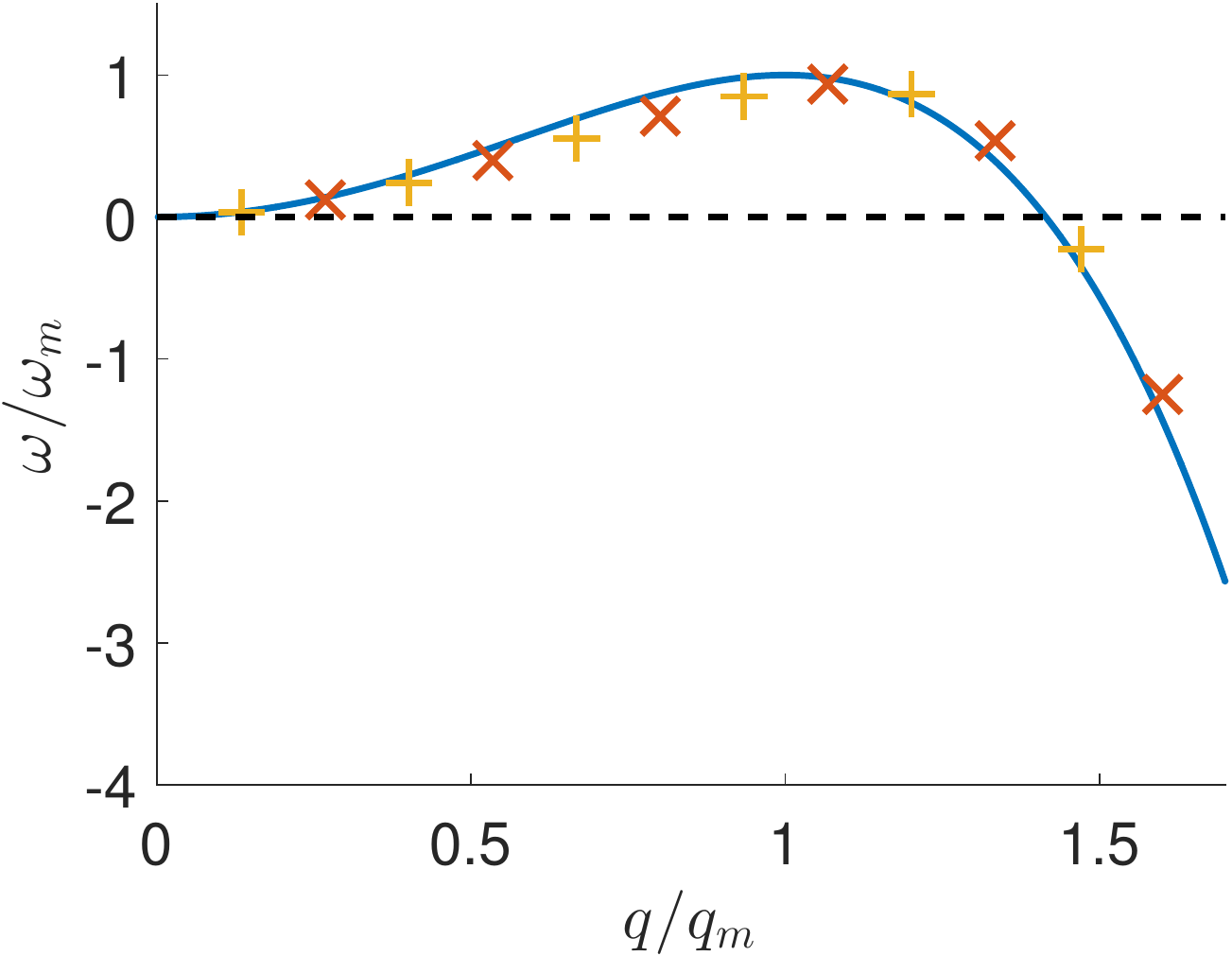} \label{fig:dispersion_curve_polymer} }
	\caption{Comparison between dispersion relations (blue curves) and growth rates extracted from numerical simulations (symbols) for a) the linear model, b) the \gls{NLC} model, and c) the polymer model. `$\times$' symbols (red) denote initial condition perturbed in the $x$ direction and `$+$' symbols (gold) denote a perturbation in the $y$ direction.} 
	\label{fig:dispersion_curves}
\end{figure}

Here, we validate our model by showing agreement with \gls{LSA}.   This is done by comparing the growth rates extracted from simulations to the dispersion relation (\ref{eq:DispersionRelation}). For each model, simulations are carried out for several $q$ values for perturbations in either the $x$ or $y$ directions. In addition, the linear domain is fixed to $\lambda_m$, i.e. $P=1$ in (\ref{eq:IC_LinearWave}) with 32 points ($I=J=32$). The final solution time is fixed at $T=\textrm{log}(1.3)|\omega(q)]^{-1}|$, where $\omega(q)$ is defined in (\ref{eq:DispersionRelation}).  Figure~\ref{fig:dispersion_curves} confirms that the numerical results agree with the LSA
predictions, validating the code.

\subsection{Performance Comparison}

Having validated the accuracy, convergence, and mass conservation of our numerical method, here we discuss the computational 
performance of our GPU implementation (details are given in \ref{sec:GPUImplement})
in comparison to our serial CPU code.  We note that CPU computations 
were performed on an Intel\textsuperscript{\copyright}\, Core\texttrademark \, i7-6700K and GPU computations were performed on a Nvidia Geforce\textsuperscript{\copyright} GTX Titan X Maxwell. For brevity, the performance results are presented only for the \gls{NLC} model. 
For simplicity we compare two major components: inverting the penta-diagonal matrices in (\ref{eq:JacobianSystem21}) and (\ref{eq:JacobianSystem22}); and computing the terms in (\ref{eq:JacobianSystem24}). These particular components were chosen since they are responsible for the dominant part of the computational cost.   

\begin{figure}[!h]
	\centering
	\includegraphics[width=0.7\textwidth]{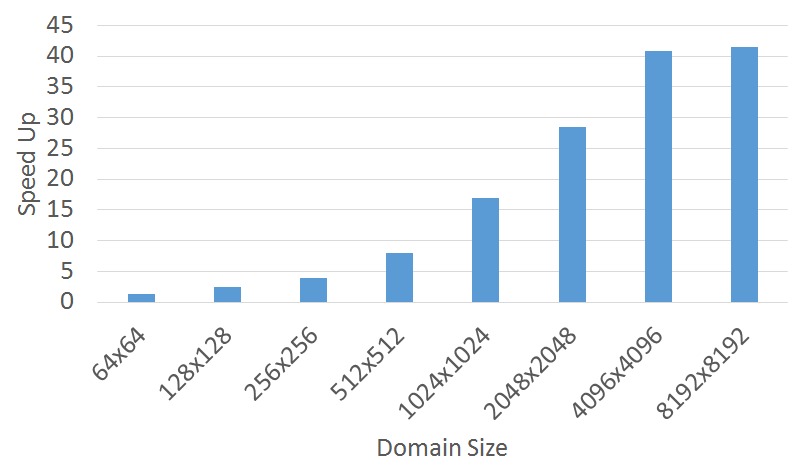}  
	\caption{ Plot of relative speedup to solve penta-diagonal system in the GPU code over the original in-house CPU code for various square domain sizes.} 
	\label{fig:GPUvsOldPenta}
\end{figure}

We begin by comparing the GPU penta-diagonal solver to the in-house serial CPU penta-diagonal solver. The serial CPU implementation is based on a method found in Numerical Recipes in FORTRAN~\cite{William1992}.   Figure~\ref{fig:GPUvsOldPenta} shows that for the smallest domain size, the GPU and CPU codes show comparable performance; however, for larger domains, the speedup with the GPU reaches factor of 40. The saturation of the speed up occurs when the total number of independent linear systems is greater than the total core count (3072 for the GPU used in the reported computations).

\begin{figure}[!h]
	\centering
	\includegraphics[width=0.7\textwidth]{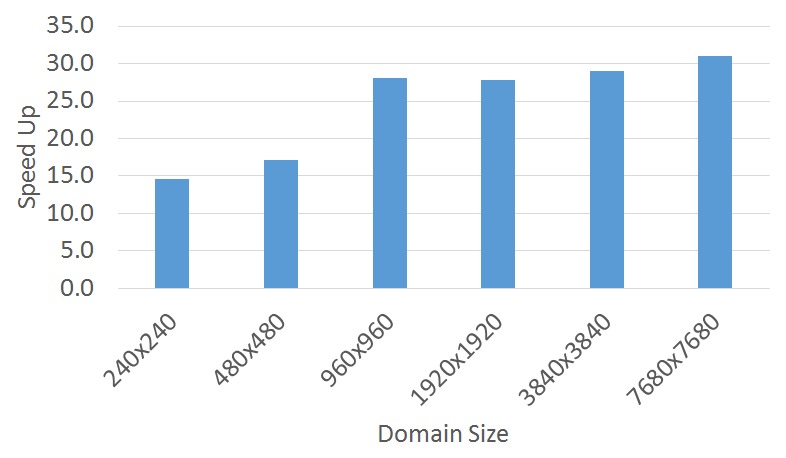}  
	\caption{ Plot of relative speedup in forming the linear system in (\ref{eq:JacobianSystem1}) in the GPU code over equivalent CPU code for various square domain sizes.  } 
	\label{fig:GPUvsNewJyF}
\end{figure}

Next, we compare the second major component of the solvers, computing the entries of the matrices in (\ref{eq:JacobianSystem1}) and (\ref{eq:JacobianSystem2}).  Since no equivalent in-house serial CPU implementation exists for computing these terms, for simplicity we only implement an equivalent version of the kernel used to compute the matrices in (\ref{eq:JacobianSystem1}).   We assume the time required for computing the matrix in (\ref{eq:JacobianSystem2}) scales (with respect to the CPU) similarly to the computation time required to evaluate the matrices in (\ref{eq:JacobianSystem1}).   Figure~\ref{fig:GPUvsNewJyF} shows that on a large enough domain, the GPU implementation is at least 25 times faster than the serial CPU code. We note that in the GPU implementation, for the smallest two domain sizes in Figure~\ref{fig:GPUvsNewJyF}, the domain is sufficiently large and all cores in the GPU should be utilized, therefore, the reduced performance (compared to the 4 largest domains) may be due to bottlenecks in the memory. 

We have therefore showed that on large enough domain sizes our GPU implementation can perform up to $25$ times faster than the CPU code. While we have not compared the GPU implementation to an equivalent serial CPU implementation in full, we have shown significant performance increase in major components of the numerical scheme, thus it is reasonable to assume similar performance gains would be observed in a full comparison.

\section{Nematic Liquid Crystals} \label{sec:NLC}

\begin{figure}[!h]
	\centering
	\includegraphics[width=0.7\textwidth]{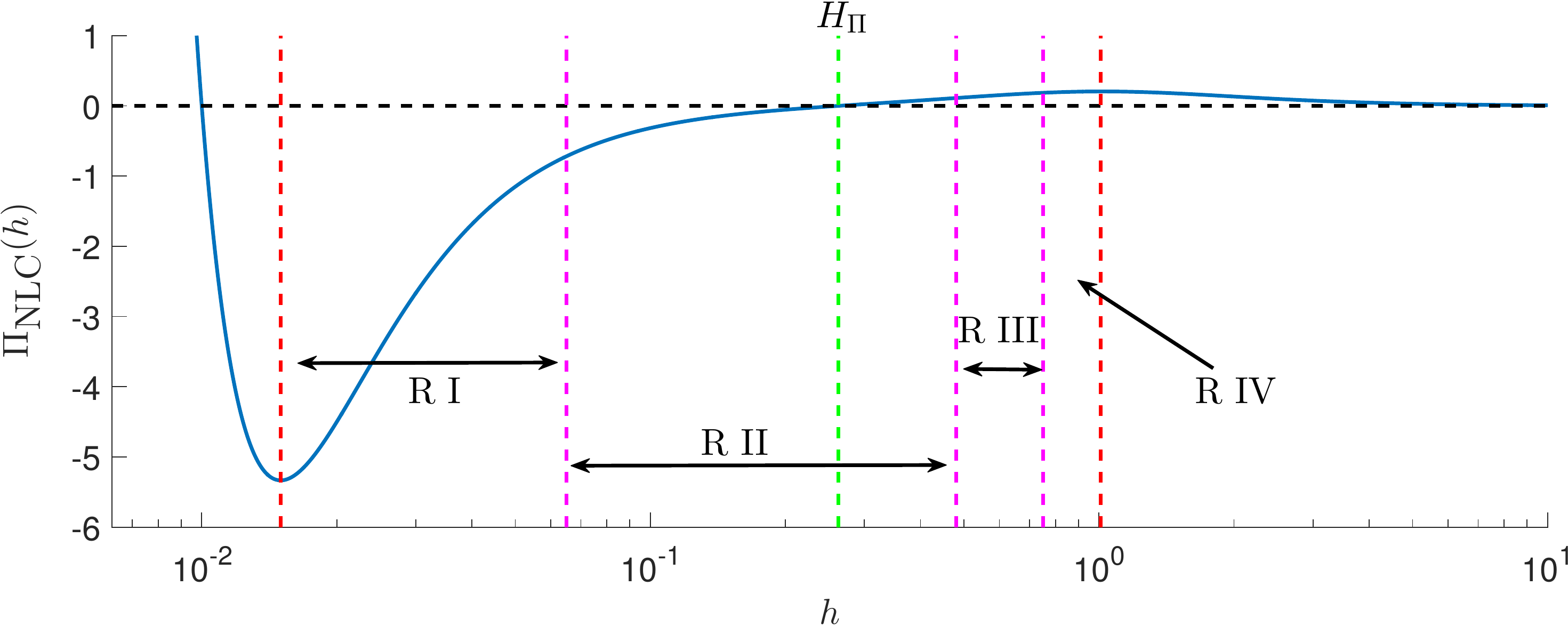}  
	\caption{ 
	The effective disjoining pressure (solid blue curve) for a NLC, (\ref{eq:DisjPressNLC}), as a function of the film thickness.  The region between the outer dashed red vertical lines denotes linearly unstable film thicknesses, and the central dashed green line denotes the zero of the disjoining pressure, $H_\Pi$ (within the linearly unstable regime). 	Stability regimes R I, R II, R III, and R IV, determined numerically for 2D films in~\cite{Lam2018}, are discussed in $\S$~\ref{sec:NucDomRegime}.
	\label{fig:StructDisjoiningPressure}}
\end{figure}

In this section, we focus on the thin film model derived for nematic liquid crystals, i.e., the governing equation given by (\ref{eq:GoverningEquation}) and (\ref{eq:DisjPressNLC}).   
The details of the model itself are discussed in detail in~\cite{Lam2018}, where a weak anchoring model was presented.  In~\cite{Lam2018} it was shown that the nematic contribution to the thin film model results in an effective disjoining pressure, given by~(\ref{eq:DisjPressNLC}), and plotted in Figure~\ref{fig:StructDisjoiningPressure}.    Similar functional forms of the disjoining pressure, characterized
by a local maximum for nonzero film thicknesses, are found in other contexts, such as polymer films~\cite{Jacobs2008}, or even simpler setups, such as 
water on quartz~\cite{homsy06}.   

In the present paper, we use large-scale simulations to discuss some of the main features of the results from~\cite{Lam2018} in the three dimensional context.   These include formation of satellite (secondary) 
drops, nucleation versus spinodal type instability, and metastability of films.  The fact that we are able to simulate accurately large domains of linear dimensions measured in tens of wavelengths
of maximum growth, allows us to obtain results that are only very weakly influenced by the presence of the domain boundaries.     This section is organized
as follows: in $\S$~\ref{sec:ThinFilmModels} we present the model in terms of the gradient dynamics formulation, $\S$~\ref{sec:model} gives a model outline, and the main part, $\S$~\ref{sec:Simulations} presents the computational results. 

\subsection{Gradient Dynamics Formulation} \label{sec:ThinFilmModels}

It is convenient to express the governing equation (\ref{eq:GoverningEquation}) in terms of the gradient dynamics formulation; in particular, it is advantageous to express analytical results in terms of a disjoining pressure, $\hat\Pi(\hat h)$.  The benefit of this formulation is that previous analytical results for 2D films~\cite{Lam2018} show that the form of the disjoining pressure primarily determines the transition between different stability types of a thin film as a function of its thickness. 

To motivate this assertion, we perform \gls{LSA} on the gradient dynamics formulation of the governing equation, which is given by
\begin{equation} \label{eq:DimGoverningEquationQD}
 \mu\hat{h}_t + \hat\nabla \cdot \left[ \hat{Q} (\hat{h}) \nabla \frac{\delta \hat F }{\delta \hat h} \right] =0 
 \;, \quad \textrm{where} \quad
      \hat F( \hat h) = \gamma \left[ 1 + \frac{ \hat\nabla \hat h \cdot \hat \nabla \hat h}{2} \right] + \hat \psi( \hat h) \; , \quad
	  \hat\Pi(\hat h) = - \frac{\partial \hat\psi}{\partial \hat h} \; ,
\end{equation}
$\hat Q( \hat h)$ is the mobility function, $\hat F( \hat h)$ is total interfacial energy (Gibbs energy), $\gamma$ is the surface tension, and $\hat\Pi(\hat h)$ is disjoining pressure.   
Relating (\ref{eq:DimGoverningEquationQD}) to the nondimensional governing equation (\ref{eq:NonDimGoverningEquation}), 
\begin{equation}
 \tilde f_0( h) = \tilde Q( h)  \;, \quad 
 \hat{F}_0 =  \gamma \hat M  \;, \quad 
 \tilde f_1( h) = \tilde Q( h) \tilde \Pi'( h) \quad \textrm{and} \quad   
 \hat{F}_1 = \frac{\hat M \hat \Psi }{ H^2} \; ,
\end{equation}
where, similarly to before, $\hat M$ and $\hat \Psi$ are dimensional constant prefactors derived from expressing $\hat Q(\hat h)$ and $\hat \Psi(\hat h)$ as dimensionless functions of the dimensionless variable $h$, e.g., $\hat Q (\hat{h}) = \hat M \tilde Q (h)$.   The nondimensional linear stability analysis results (\ref{eq:unstableLSAmax}) simplify to
\begin{equation} \label{eq:unstableLSAmaxQD}
q_m=\sqrt{\frac{\Pi'(H_0)}{2 \Gamma} } \quad \textrm{and} \quad 
\omega_m=  \frac{Q(H_0)[\Pi'(H_0)]^2}{4 \Gamma} \;, \quad \textrm{where} \quad 
\Gamma = \frac{\hat \Psi}{  \gamma \delta^2} 
\end{equation}
is a dimensionless constant and may be interpreted as the ratio of disjoining pressure to surface tension forces.   We note that the disjoining pressure determines the transition between linear stability ($\Pi'(H_0)<0$) and instability ($\Pi'(H_0)>0$).  The instability 
wavelength is independent of the mobility function, which only affects the growth rate of instabilities.

\subsection{Model Description}\label{sec:model}

\begin{figure}[!h]
	\centering
	\subfigure[]{
		\includegraphics[width=0.3\textwidth]{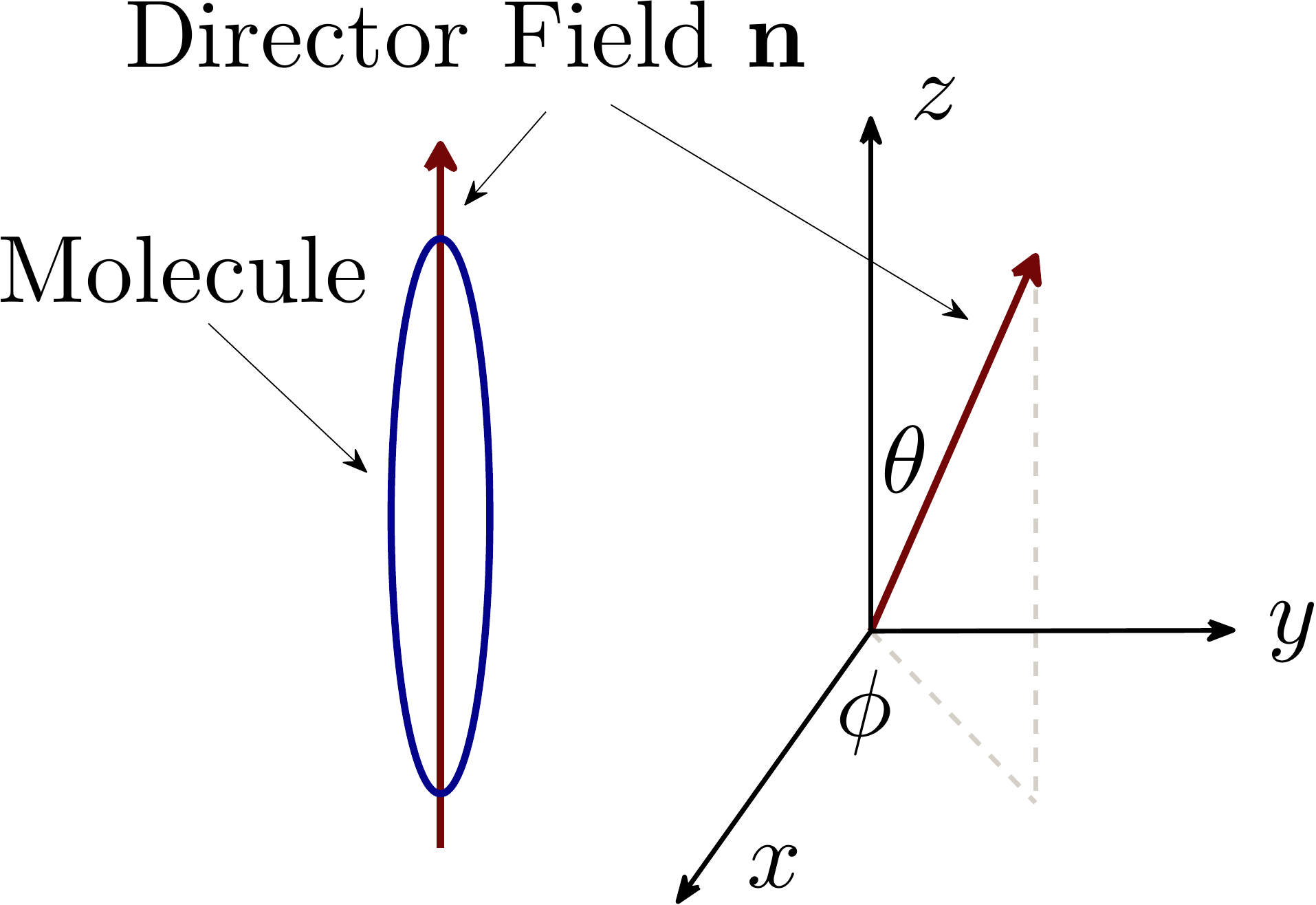} 
		\label{fig:DirectorFieldDiagram} 
	}
	\subfigure[]{
		\includegraphics[width=0.25\textwidth]{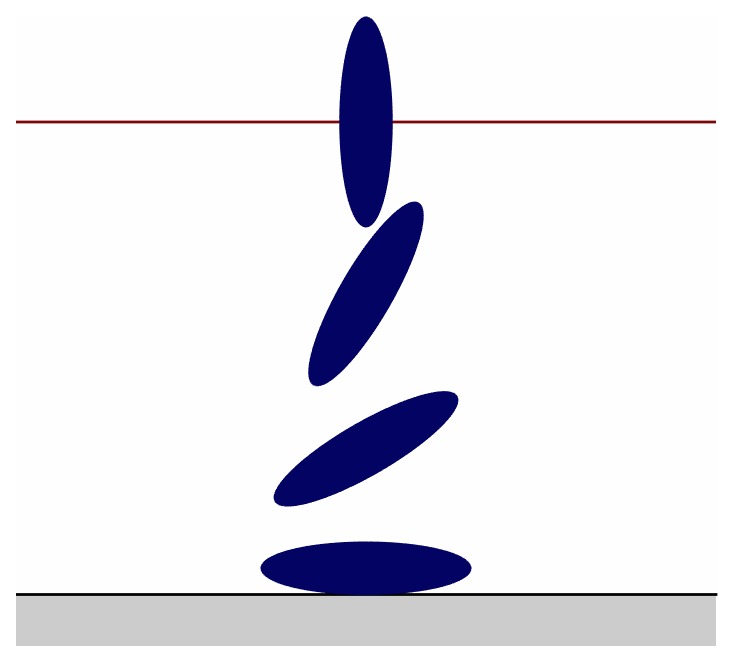}
		\label{fig:StrongAnchoring} 
	}
	\subfigure[]{
		\includegraphics[width=0.25\textwidth]{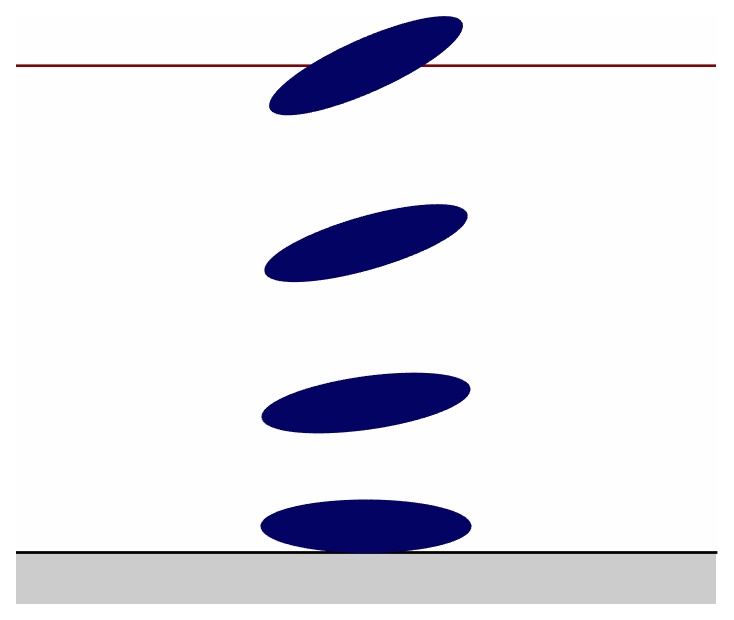} 
		\label{fig:WeakAnchoring} 
	}
	\caption{a) Schematic of director field relative to liquid crystal molecule. b) Strong homeotropic free surface anchoring model example. c) Weak homeotropic free surface anchoring model. } 
	\label{fig:DirectorField}
\end{figure}

To motivate the form of the disjoining pressure shown in Figure~\ref{fig:StructDisjoiningPressure} and defined by (\ref{eq:DisjPressNLC}), here we 
provide a brief outline of the main features of \gls{NLC}, and we refer the reader to our previous work~\cite{Lam2018} for a more detailed derivation. Typically, \gls{NLC} molecules are rod-like structures with an (electrical) dipole moment, which gives rise to a state of matter intermediate between a crystal and a liquid.   Specifically, similarly to a Newtonian fluid, there is no positional ordering to the \gls{NLC} molecules; however, similarly to a crystal, molecules have short-range orientational order due to interactions between dipoles (elastic response).  To model \gls{NLC}, in addition to modeling the velocity field of molecules (as with most fluids), short-range ordering is often modeled with a director field.  The director field is a unit vector, aligned with the long axis of the liquid crystal, see Figure~\ref{fig:DirectorFieldDiagram}, and it is often convenient to consider the polar angle, $\theta$, and azimuthal angle, $\phi$, of the director field orientation as functions of Cartesian space variables $(x,y,z)$.  

The evolution of \gls{NLC} may be modeled using the Leslie-Ericksen equations, an extension of the Navier-Stokes equations, with an additional equation modeling conservation of energy. Under the long wave approximation, and  assuming that the time scale of fluid flow is much faster than that of the elastic reorientation, the conservation of energy equation decouples from the remaining equations and only depends on the director field. Using energy minimization~\cite{Cummings2004,Lam2018,Lin2013}, the polar and azimuthal angles of the director field are determined to be of the form
\begin{equation} \label{eq:GeneralDirectorField}
\theta(x,y,z) = a(x,y)z + b(x,y) \quad \textrm{and} \quad 
\phi(x,y) = c(x,y) \; ,
\end{equation}
where $a(x,y)$, $b(x,y)$, and $c(x,y)$ are constant with respect to $z$ and are chosen to satisfy appropriate so-called anchoring boundary conditions at both the substrate and the free surface, thus the director field is a function of the instantaneous fluid height.  

At the interface between the \gls{NLC} film and another material, liquid crystal molecules molecules satisfy certain anchoring conditions. Typically, at the free surface (air/\gls{NLC} interface), molecules are normal to the free surface (a condition known as homeotropic anchoring; within the long-wave approximation, $\theta(x,y,z=h)=\pi/2$), while 
at the \gls{NLC}/substrate interface, planar anchoring is appropriate, so  $\theta(x,y,z=0)=0$, see Figure~\ref{fig:StrongAnchoring}. However, for very thin films,
or close to a contact line, this configuration induces a large energy penalty in the bulk due to rapid spatial variations in the director field; therefore, we implement a novel weak free surface anchoring model. In practice, the substrate anchoring is stronger than the free surface anchoring, thus we relax the free surface anchoring to that of the substrate for very thin films (weak anchoring model) to alleviate the large energy penalty in the bulk (compare Figures~\ref{fig:StrongAnchoring} and ~\ref{fig:WeakAnchoring}). Specifically, 
\begin{equation}
\theta(x,y,z) = \frac{\pi}{2}\left( 1 - \frac{m(h)}{h}z \right)
\end{equation}
where $m(h)\in[0,1]$ and is defined in (\ref{eq:mh_model}). The azimuthal anchoring $\phi$ in (\ref{eq:GeneralDirectorField}) is independent of $z$; therefore, assuming the  substrate anchoring dominates, the azimuthal anchoring is determined by the substrate. The governing equation (\ref{eq:GoverningEquation}) is a simplification of the long wave approximation of the Leslie-Ericksen equations that ignores substrate anchoring. In the full (long wave) model, the mobility function may be expressed as 
\begin{equation}
Q(h)  = \left[\lambda \mathbf{I} + \nu 
	\left( 
		\begin{array}{cc}
			\cos 2\phi  & \sin 2\phi \\
			\sin 2\phi  & -\cos 2\phi \\
		\end{array}
	\right) \right] h^3 \;,	
\end{equation}
where $\mathbf{I}$ is the identity matrix, and $\lambda$ and $\nu$ are anisotropic viscosities. To simplify the model, we fix $\lambda=1$ and $\nu=0$, which removes the dependence on $\phi$. Note that by definition, $\lambda>\nu$, therefore the mobility function is positive definite and does not change stability properties. We leave the investigation of the effects of substrate anchoring for future work.

\subsection{Simulations} \label{sec:Simulations}

For the remainder of this section, we focus on considering some aspects of  the previous analysis carried out for 2D films~\cite{Lam2018} in the 3D geometry. 
We note that there has been a significant amount of work discussing various aspects of instability development, see, 
e.g.~\cite{becker_nat03,neto_jphys03,pototsky_jcp05,sharma_epje03,Thie2003epje,thiele_prl01} and the relevant part of the 
review~\cite{cm_rmp09}.  The simulation results that we present here are distinguished by significantly larger computational domains that allow for more elaborate 
analysis of the results.  Direct comparison with some of the available experimental results is presented in $\S$~\ref{sec:ExperimentalResults}.   

This section is divided into three parts; in the first two parts we focus on the linearly unstable regime, investigating coarsening and satellite drop formation in 
$\S$~\ref{eq:infinitesimalPertub}; and the nucleation dominated regime in $\S$~\ref{sec:NucDomRegime}. In $\S$~\ref{sec:Metastable}, the metastable 
regime is examined.   Animations of the simulation results are available, and are provided as Supplementary Material~\cite{SM}.  
 Before presenting these results, we discuss the domain size, the spatial step size of simulations, and also the tools used to quantify 
 simulation results: (i) the Fourier transform of the film profile, and (ii) the Betti numbers,  topological measures of the film profile.  

In $\S$~\ref{sec:Validation} the spatial step size was determined by $I$, $\lambda_m$, and  $P$, in this section, we instead fix $\Delta s$, $\lambda_m$, and $P$ and then set the grid size $I = P \lambda_m/\Delta s$, with $I$ ranging between $1000$ and $4000$, giving the total number of grid points of the order of $10^6~-~10^7$.  
For all simulations, we fix the spatial step size to $\Delta s=0.05$ except for the film thicknesses $H_0=0.05,~0.01$, where $\Delta s= 0.01,~0.02$, 
respectively.  The spatial step size was changed for these film thicknesses so as to maintain a sufficient number of points per period of the most unstable wavelength, $\lambda_m$.  Note that for thicker films, while a larger $\Delta s$ may be chosen while maintaining a sufficient number of points per period (say 50), to adequately resolve the contact line, $\Delta s$ is chosen to be close to the precursor thickness, $b=0.01$. 
In addition, simulations in the linearly unstable regime are scaled in two ways: (i) the linear domain size is set to $P\lambda_m$ where 
$P$ is an integer, i.e., domains are scaled by $P$ periods of the most unstable wavelength from LSA; and (ii) when comparing results for 
different film thicknesses as a function of time, it is convenient to define a new 
timescale, $\tau = t \omega_m$, i.e., we scale time by the most unstable growth rate obtained using \gls{LSA}. 
We fix $P=40$ for all (linearly unstable) thicknesses.

One aspect of the analysis of the results focuses on the extraction of the dominant length scales.  This goal is achieved by using 
a 2D Fourier transform. Assuming that the instability pattern is isotropic,  the magnitude of the 
2D Fourier transform may be mapped to a one-dimensional (radial) function of the magnitude of the wave vector $q_{r}=\sqrt{q_{x}^2+q_{y}^2}$, 
where $q_x$ and $q_y$ are the $x$ and $y$ components of the wave vector. In addition, several smoothing techniques are required to reliably 
extract local maxima.  The details of the procedure are not trivial if one wishes to extract accurate results,  and are described in~\ref{sec:RadialFFT}.

Next, we discuss the use of Betti numbers to further quantify the simulation results.   The Betti numbers, $b_n$, are $n$ topological invariants that characterize
the connectivity of $n$-dimensional objects.   For our purposes, $n=2$ since the film thickness is a function of two spatial variables. 
To illustrate the meaning of $b_0$ and $b_1$, we use landscape analogy: consider a landscape 
sliced by a horizontal plane at the height $H^*$.  Then, $b_0$ counts the number of mountain peaks (components) above $H^*$, and $b_1$ counts the
number of valleys (loops) surrounding the peaks.  In any given landscape, the values of $b_0$ and $b_1$ change with the threshold level $H^*$. 
In the present context, we will think of the film thickness, $h$, as the variable describing the landscape; by computing Betti numbers for all relevant 
thresholds (ranging from precursor film thickness to the values larger than the initial film thickness), we will be in the position to describe in precise 
terms the topology of the evolving films, and to analyze the appearance of topological objects across scales.
To minimize the influence of boundaries, it is important to carry out simulations in large domains such that boundary-related effects are minor; 
 the computational domains that we choose are sufficiently large for the purpose.  
The calculations of Betti numbers are carried out using the Computational Homology Project (CHomP) software package.\footnote{http://chomp.rutgers.edu/}

To examine the Betti numbers for various film thicknesses, we find it convenient to normalize them by the area corresponding to the most unstable 
wavelength, $\lambda_m$. Therefore, we define normalized Betti numbers, $\grave{b}_0$ and $\grave{b}_1$ by  $\grave{b}_{0,1}=b_{0,1}/P^2$ 
where $P \lambda_m$ is the linear domain size, see (\ref{eq:3D_Random_IC}). 
The normalized Betti numbers give a measure of the number of features (components, loops) per $\lambda_m^2$, and will be 
shown as functions of the scaled threshold value $H^*/H_0$, where $H_0$ is the average
film thickness, and the rescaled time, $\tau$.

\subsubsection{Evolution of films exposed to infinitesimal perturbations of global character} \label{eq:infinitesimalPertub}

Here we analyze the evolution of randomly perturbed films of thicknesses $H_0$ that are in the linearly unstable regime. 
The initial condition is set to a flat film that has been randomly perturbed, and to excite all modes in the 2D Fourier transform independently (combinations of $q_x$ and $q_y$), pseudo-Perlin noise is used.   Specifically, the initial condition is of the form
\begin{equation} \label{eq:3D_Random_IC}
 u(x,y,t=0) = H_0( 1 + \ep \left| \zeta(x,y) \right| ) \;, \quad (x,y)\in[0,P\lambda_m]
\end{equation}
where $\lambda_m=2\pi/q_m$, $q_m$ is given in (\ref{eq:unstableLSAmax}), and $\zeta(x,y)$ is the inverse Fourier transform of 
\begin{equation}  \label{eq:PerlinNoise}
  \zeta(q_x,q_y)  =  \left| \left[ q_x^2 + q_y^2 \right]^{-\alpha/2} \exp \left(  2\pi \mathrm{i}  a(q_x,q_y)  \right) \right|  \;.
\end{equation}
Here, $\ep=0.01$, $\alpha$ is some positive constant, and $a(q_x,q_y)$ is a random variable, uniformly distributed on $[-1,1]$, for each $(q_x,q_y)$.   In addition, $\zeta(x,y)$ is scaled so that $|\zeta(x,y)|\leq1$ and we fix $\alpha=200/I$, where $I$  is the number of discretization points in the $x$ and $y$ directions,  so that the spatial scale of the noise is proportional to $\lambda_m$.

\begin{figure}[!h]
		\centering
		\includegraphics[width=0.9\textwidth]{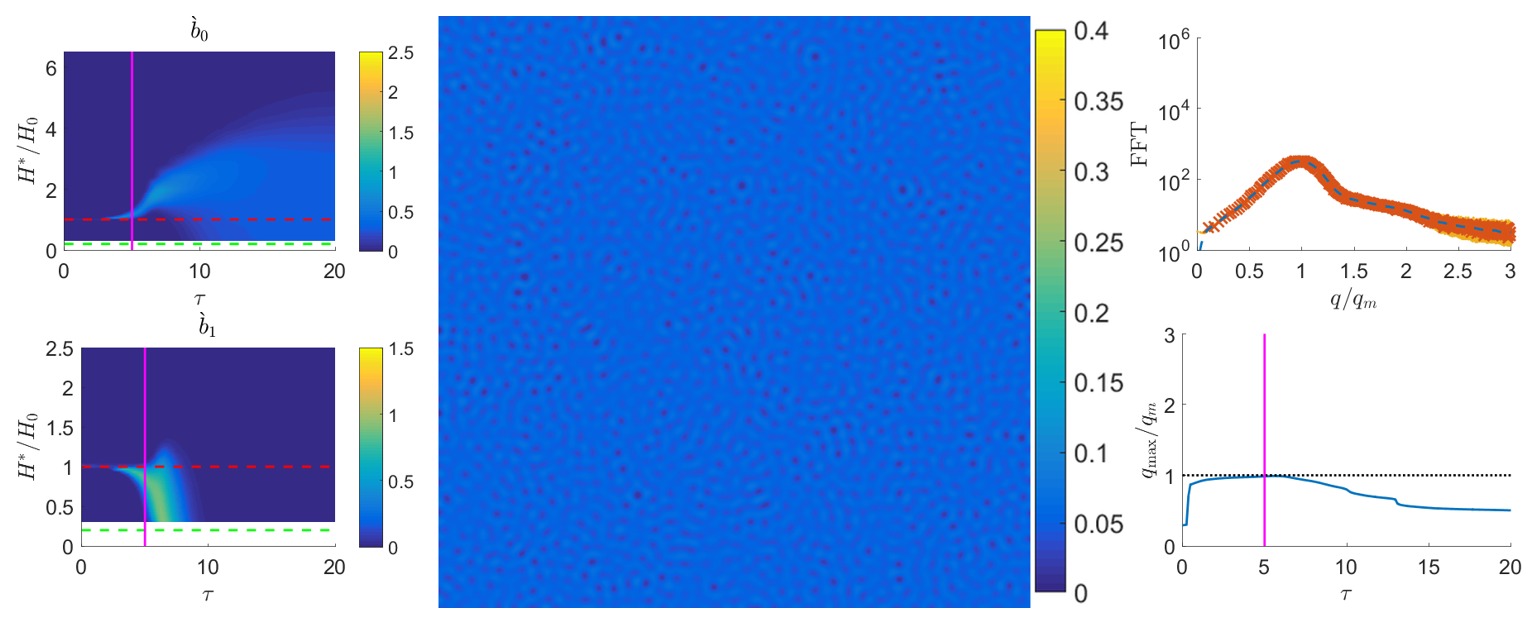}  
		\caption{Evolution of film of thickness $H_0 = 0.05$ perturbed by global perturbations as described in the text.  The central part shows a contour plot 
of the free surface height at $\tau=5$.  The domain size is $40 \lambda_m \times 40 \lambda_m$.  The 
right-hand side includes the corresponding radial Fourier transform of the film height (top), and 
local maxima of radial Fourier transform, $q_{\rm {max}}$ (bottom) as a function of $\tau$.   The left-hand 
side shows the contour plots of the average Betti numbers $\grave{b}_0$ (top) and $\grave{b}_1$ (bottom) 
as functions of $\tau$ and scaled threshold value $H^*/H_0$.  The vertical magenta line corresponds to 
the solution time shown in the central panel. The grid size is $1637\times1637$.  For animations, see Supplementary Material~\cite{SM}, movie1.  
		}
		\label{fig:H0_0_all_data_tau_5}
\end{figure}

\begin{figure}[!h]
		\centering
		\includegraphics[width=0.9\textwidth]{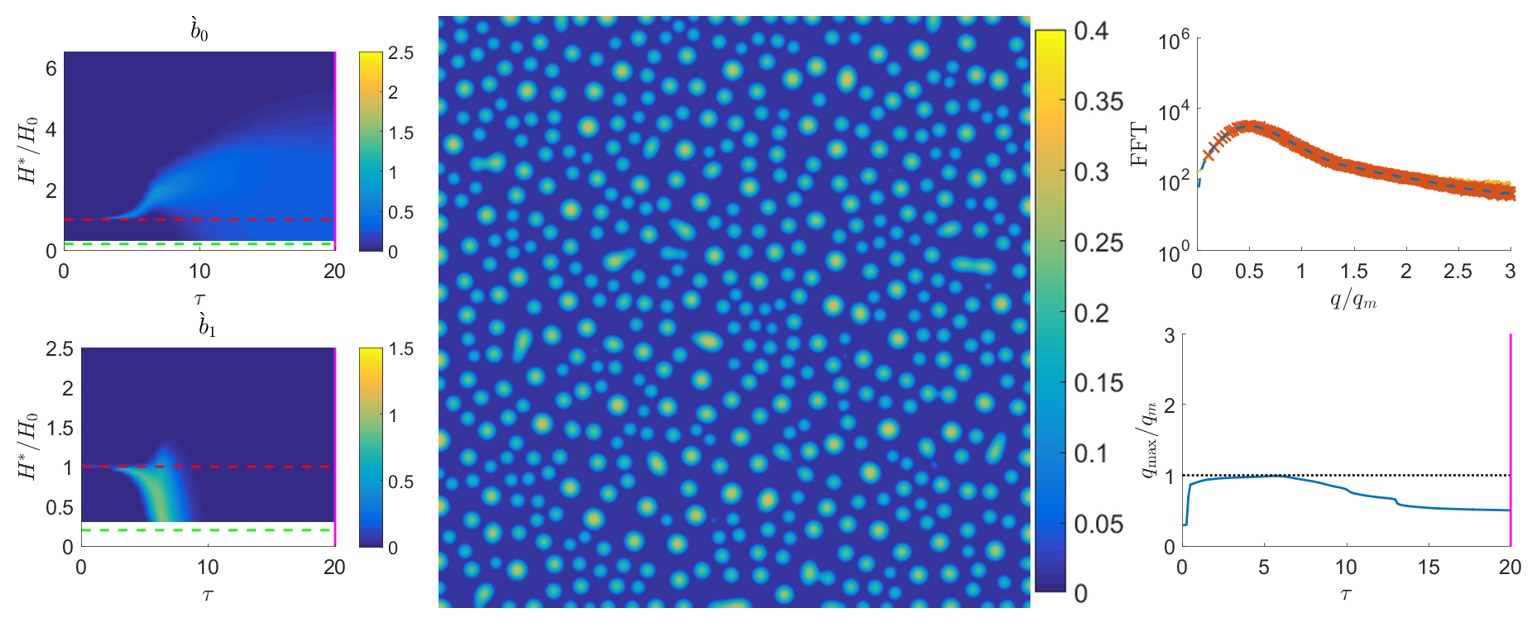}  
		\caption{Film of thickness $H_0 = 0.05$ perturbed by global perturbations. The central part shows a contour plot of the
		free surface height at $\tau = 20$.   The rest of the figure is as described in the caption of 
		Figure~\ref{fig:H0_0_all_data_tau_5}. The grid size is $1637\times1637$.  For animations, see~\cite{SM}, movie1. 
		}
		\label{fig:H0_0_all_data_tau_20}
\end{figure}

 Figures~\ref{fig:H0_0_all_data_tau_5} and~\ref{fig:H0_0_all_data_tau_20} show the results for the thinnest 
film that we consider, $H_0 = 0.05$, and for times
$\tau = 5$ and $20$, respectively.  The central parts of the figures show the free surface thickness.
The corresponding radial Fourier Transform is shown at the top right.  We see that the most unstable wavelength dominates the morphology 
of the film at $\tau=5$. On the bottom right, the local maxima of the radial Fourier transform are plotted as functions of time, 
showing consistently that for $1<\tau<6$, $q_m$ dominates the wave pattern. At later times drops form and coarsening is observed in the radial Fourier transform.
Similar coarsening effects are also observed in Volume of Fluid based simulations of nanoscale metal films~\cite{Mahady2016}.   
Additional simulations (figures not shown for brevity) uncover consistent results
for all linearly unstable film thicknesses considered, in the range $[0.05:0.8]$.   

Figures~\ref{fig:H0_0_all_data_tau_5} and \ref{fig:H0_0_all_data_tau_20} also plot the normalized Betti numbers $\grave{b}_0$ (top left) and $\grave{b}_1$ (bottom left)
as contour plots with independent variables being $H^*/H_0$ (vertical axis) and the scaled time, $\tau$ (horizontal axis).  (Note that the only difference between the Betti 
plots shown in these two figures is the position of vertical (magenta) lines, that show the current value of $\tau$ depicted in the central part of the figures.)
The basic interpretation of these Betti numbers figures is straightforward for both early and late times.  For the early times, $\tau \lessapprox 3$, the perturbations
are small and are not captured by the level of thresholding implemented;  therefore both $\grave{b}_0$ and $\grave{b}_1$  are small (the exact values
depend on the treatment of the boundaries, but this is not of particular interest here).  For long times, $\tau \gtrapprox 10$, the drops form and there are no 
separate valleys/loops (they are all connected by the precursor); therefore $\grave{b}_1$ is small.    For $H^*$ in 
the range $1.5b$ (the smallest value considered) to $\approx 4 H_0$, for long times we see essentially constant 
values of $\grave{b}_0$, suggesting that the most of the drops are of approximately similar size;
small values of $\grave{b}_0$ for $H^*/H_0\gtrapprox 4$ show that there are essentially no drops exceeding height $4H_0$.  

For intermediate times, we observe a peninsula-like light blue structure in $\grave{b}_0$ (top row), the tip of which starts at 
$\tau\approx 3$ and $H^*/H_0 \approx 1$.  This corresponds to a local increase of film thickness (therefore formation of isolated hill tops, that are not yet
drops).  These
hill tops are associated with the formation of surrounding valleys (loops), indicated by an increase of $\grave{b}_1$ around the same time.
For slightly longer times, following the contours of $\grave{b}_0$ and $\grave{b}_1$, we observe an increase in the height of the 
drops, measured by $\grave{b}_0$; and the formation 
of deeper and deeper valleys, measured by $\grave{b}_1$.   The evolution is essentially complete by $\tau \approx 10$, at which time we reach the long time 
regime already discussed above.  We note that one important utility of Betti numbers is that they provide an insight into the topology of emerging patterns across
a range of thresholds, allowing one to follow the details of instability development.

\begin{figure}[!h]
		\centering
		\includegraphics[width=0.9\textwidth]{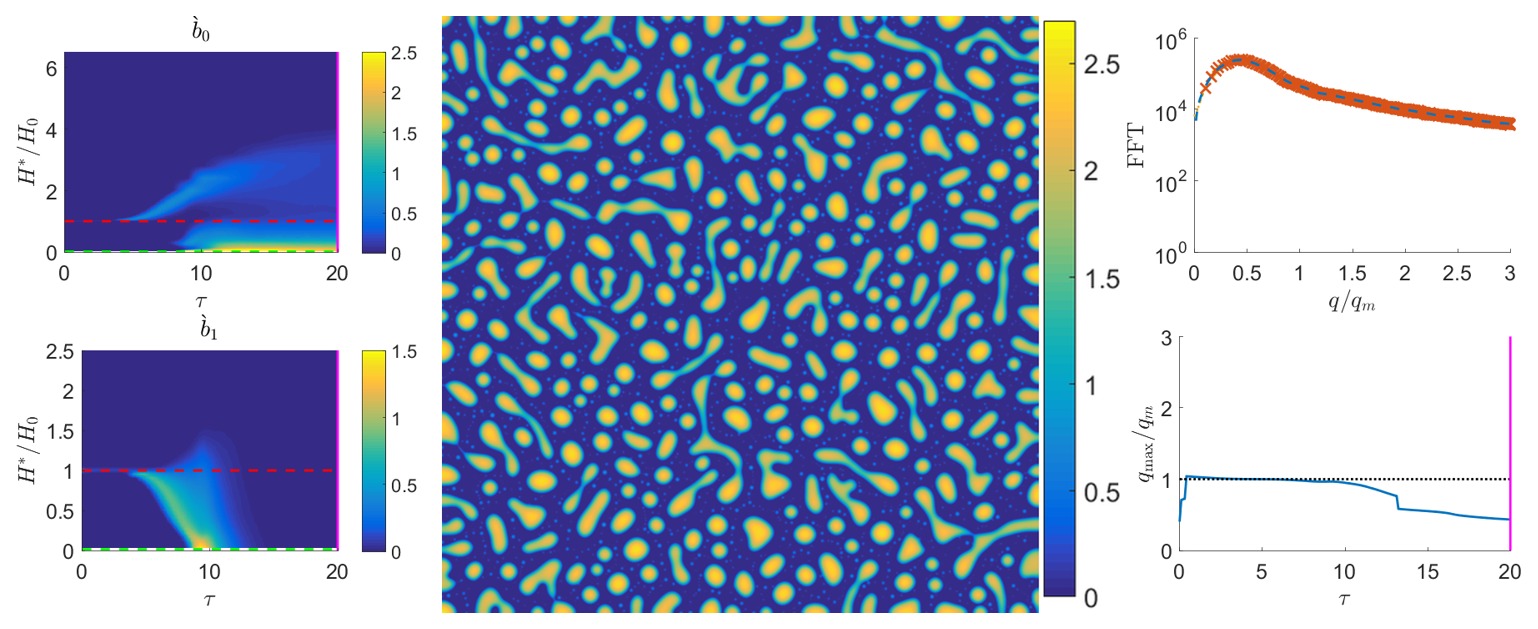}  
		\caption{Film of thickness $H_0 = 0.6$ perturbed by global perturbations. The central part shows a contour plot of the
		free surface height at $\tau = 20$. 		
				The rest of the figure is as described in the caption of 
		Figure~\ref{fig:H0_0_all_data_tau_5}.  The grid size is $3882\times3882$. For animations, see~\cite{SM}, movie2.	  }
		\label{fig:H0_6_all_data_tau_20}
\end{figure}

Our previous analysis of 2D films~\cite{Lam2018} showed the presence of satellite drops persisting on a timescale longer than that of dewetting ($\tau\gg 1$) 
for those linearly unstable film thicknesses corresponding to positive values of the disjoining pressure~\cite{Lam2018}.  For the present choice of parameters, 
$\Pi(H_0) = 0$ for $H_0 = 0.2624$, see Figure~\ref{fig:StructDisjoiningPressure}~\cite{Lam2018}.  We are now in a position to discuss whether 
the satellite drops observed in 2D simulations persist in the 3D geometry.  
Figure~\ref{fig:H0_6_all_data_tau_20} shows the results for $H_0=0.6$, for which $\Pi(H_0)>0$.    
We note that the Fourier transform (in particular, the dominant wave number) shows results similar to those for thinner films presented in 
Figures~\ref{fig:H0_0_all_data_tau_5} and~\ref{fig:H0_0_all_data_tau_20}.  However, there is a significant difference in the normalized Betti 
numbers for threshold values near the precursor thickness; specifically we note the presence of the thin rectangular bright yellow region in $\grave{b}_0$ 
bounded by $10<\tau<20$. This region corresponds to the presence of features characterized by heights that are much smaller than $H_0$: 
these features are readily identified as small satellite drops.
Therefore, our earlier finding that positive values of the disjoining pressure lead to the existence of satellite drops for 2D films, is found to extend to 3D films.
We note in passing that the configuration shown in the central part of Figure~\ref{fig:H0_6_all_data_tau_20} at $\tau=20$ is still evolving; the evolution has not yet concluded.  For longer times, the main drops form and the satellite drops persist for the entire simulated time interval ($\tau = 40$).  

\begin{figure}[!h]
		\centering
		\includegraphics[width=0.9\textwidth]{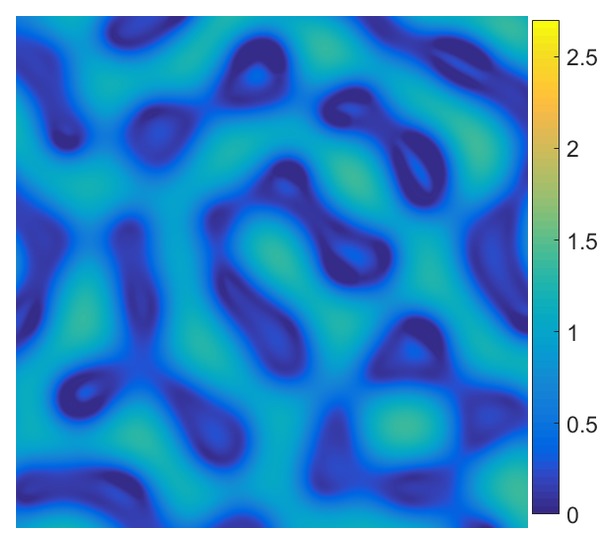}  
		\caption{Zoom-in for the film shown in Figure~\ref{fig:H0_6_all_data_tau_20} at $\tau = 9$. Linear domain size is $5\lambda_m$ and the sub-grid size is $485\times485$. Note formation of satellite 
		drops (light blue islands) in the centers of growing holes. 		  }
		\label{fig:zoom}
\end{figure}

The results for $\grave{b}_1$ provide additional insight about satellite drops and the nature of film breakup.  Note the bright yellow (approximately) circular region 
in the plot of $\grave{b}_1$ in Fig.~\ref{fig:H0_6_all_data_tau_20}, bounded by $9<\tau<10$, which indicates the formation of holes for $H^*/H_0 \approx 0.1$.  It turns out that this region is due to 
formation of satellite drops in the centers of holes, which formed due to the film instability.  To illustrate this process further, Figure~\ref{fig:zoom} shows a small part of the 
domain where one can clearly see these satellite drops (see also 
Supplementary Material~\cite{SM}). These drops are surrounded by a thinner film layer, 
leading therefore to a temporary increase of $\grave{b}_1$.  For later times, the holes disintegrate, and $\grave{b}_1$ decreases to small values again.

\subsubsection{Evolution of films exposed to both localized and global perturbations} \label{sec:NucDomRegime}

We now switch focus to linearly unstable film thicknesses exposed to both localized and global (random) perturbations.  
Previous results for 2D films~\cite{Lam2018} show that a localized perturbation propagates into the flat regime, dewetting the film and 
successively forming drops.  Based on the average distance between the drop centers, four regimes were identified 
numerically: regions R I and R III, where the mean distance between the drop centers is characterized by $\lambda_m$, the 
most unstable wavelength (spinodal dewetting); and regions R II and R IV, where the mean distance between drops 
is not characterized by $\lambda_m$ (nucleation dominated regimes). These stability regions, for 2D films, can be 
related to the disjoining pressure as shown in Figure~\ref{fig:StructDisjoiningPressure}.  In this section we will use these 2D
results as a guide to analyze 3D film instability in the presence of both local and global perturbations.  

To begin, we consider the evolution of a flat film of thickness $H_0$, perturbed by global random perturbations, with
superimposed larger localized perturbations, with a specified average mean distance apart. Specifically, the initial condition is given by   
\begin{equation} \label{eq:3DLocalAndRandom}
 u(x,y,t=0) = H_0( 1 + \ep  |\zeta(x,y)| + 10\ep \eta(x,y) ) \;,
\end{equation}
where $\zeta(x,y)$ is the pseudo-Perlin noise (\ref{eq:PerlinNoise}), and localized perturbations are defined by $\eta(x,y)$.  
We will not attempt here a careful study of the effect of the choice of functional form $\eta(x,y)$ on results, but consider instead
just a single realization that specifies the functional form of each perturbation in a manner similar to~\cite{Lam2018}, with 
perturbations distributed at an average distance $\bar d$ that is large compared to $\lambda_m$.  As an illustrative example, we choose $\bar d = 8$ in the present study, a 
choice motivated by our desire to place the local perturbation centers at a reasonable distance from each other, while simultaneously allowing a relatively large number of such perturbations.  The form chosen for $\eta(x,y)$ is
\begin{equation} \label{eq:3DRandomLocalTerm}
 \eta(x,y) = \sum_{i=1}^4 \sum_{j=1}^4  \textrm{exp} \left( - \frac{ (x-\chi_i)^2 - (y-\xi_j)^2 }{0.04\lambda_m^2} \right) \;,
\end{equation}
where $\chi_i = (8 i-4 + 4 a_i)\lambda_m $, $\xi_j = (8 j-4 + 4 b_j)\lambda_m$, and $a_i$ and $b_j$ are random variables uniformly distributed on $[-1,1]$ 
(i.e., we generate a two-dimensional array of localized perturbations with mean distance $8\lambda_m$ apart).  
We note that the motivation for considering random global perturbations alongside local ones is that random perturbations are always inevitably present in experiments, hence
carrying out simulations where both local and global perturbations are present brings us closer to understanding experimental results.   

\begin{figure}[!h]
		\centering
		\includegraphics[width=0.9\textwidth]{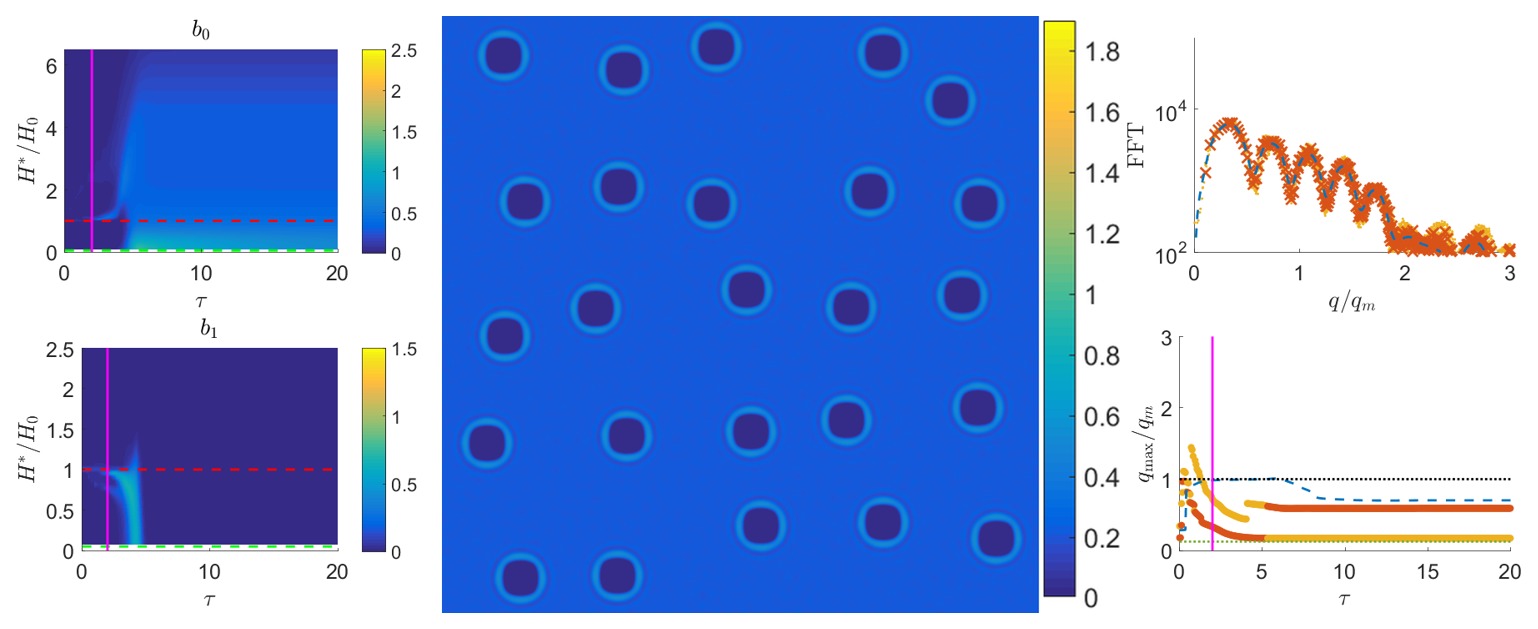}  
		\caption{Film of thickness $H_0 = 0.2$ perturbed by local and global random perturbations as described in the text. 
		 The central part shows a contour plot of the free surface height at $\tau = 2$. The corresponding radial Fourier transform (top right) of the free surface height in the center of the domain, and the large local maximum (red dots) and second largest local maximum (yellow dots) of the radial Fourier transform (bottom right) are plotted as functions of $\tau$. Here the dashed blue curve corresponds to the local maximum of the radial Fourier transform for a simulation carried out with the same film thickness, but with random perturbations only, and the horizontal green line is the wave number associated with the average distance between localized perturbations.  Left panels plot the Betti numbers $\grave{b}_0$ (top) and $\grave{b}_1$ (bottom). The vertical magenta line corresponds to the solution time in the respective panels. The grid size is $1970\times1970$.
		For animations, see~\cite{SM}, movie3.	
		}
		\label{fig:H0_2_mixed_all_data_tau_2}
\end{figure}

 Figure~\ref{fig:H0_2_mixed_all_data_tau_2} shows the simulation results  for $H_0=0.2$ (nucleation dominated regime) at $\tau=2$. 
 The figure is presented after the fashion of Figures 
 \ref{fig:H0_0_all_data_tau_5}--\ref{fig:H0_6_all_data_tau_20}, 
 except that Fourier spectra figures (right hand side) also show the results obtained in the corresponding simulation with only random perturbations applied (dashed blue lines).  
 From the central panel of this figure,  it may be seen that the localized perturbations grow into large 
 holes and dominate the radial Fourier transform for $1<\tau<5$.   To further clarify the influence of random perturbations, in the Fourier
 transform plots (right hand side) we plot not only the main maximum (red dots), but also the second largest maximum, if well defined (yellow dots).  
 For the early times shown in the central panel of this figure, the dominant mode corresponds to the long wavelengths that are due to 
 the localized perturbations.  In  Figure~\ref{fig:H0_2_mixed_all_data_tau_2} (and in Figure~\ref{fig:H0_0_mixed_all_data_tau_20}
 below) the second maximum is plotted only if its amplitude is at least $5$\% of the largest maximum.   
 
\begin{figure}[!h]
		\centering
		\includegraphics[width=0.9\textwidth]{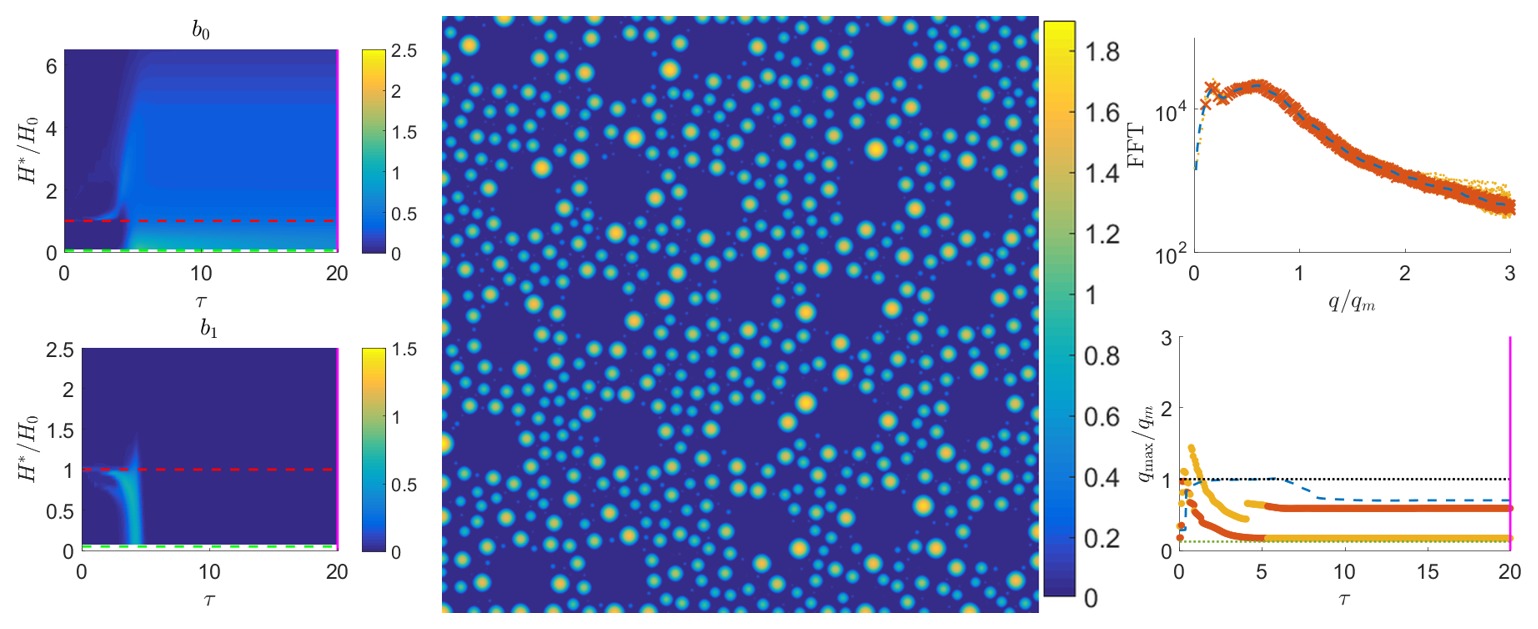}  
		\caption{
		Film of thickness $H_0 = 0.2$ perturbed by local and global perturbations as described in the text. 
		The central part shows a contour plot of the free surface height at $\tau = 20$. The rest of the figure is as described in the caption of 
		Figure~\ref{fig:H0_2_mixed_all_data_tau_2}. The grid size is $1970\times1970$. For animations, see~\cite{SM}, movie3.  		}
		\label{fig:H0_2_mixed_all_data_tau_20}
\end{figure}

Figure~\ref{fig:H0_2_mixed_all_data_tau_20} shows the results of the same simulation at a later time.  We note that, while drops form, 
the holes are still visually distinguishable (see central panel). Furthermore, by considering Fourier spectra again, we observe that 
 for $\tau>5$, there is a switch in the dominant wave number from the long wavelength one (corresponding to localized perturbations)
 to the shorter wavelength one corresponding to random perturbations; the latter is clearly very close to that found if only 
 random perturbations are imposed (see green dashed line in the bottom right panel of Figure~\ref{fig:H0_2_mixed_all_data_tau_20}).  
 These results indicate that it is possible to identify the nucleation-dominated regime by considering Fourier transforms, as long as the 
 considered domains are sufficiently large to be able to extract the second, smaller, wavenumber from the radial Fourier transform. 
 We note that a more elaborate approach, based on the use of Minkowski functionals, was considered for this purpose previously~\cite{becker_nat03}.  
We should also note that, as with our previous work for 2D films~\cite{Lam2018}, the nucleation-dominated regime is confirmed only numerically: 
we do not yet have an analytical framework for identifying this regime.  For all considered film thicknesses, the coarsened wavenumber 
appears much earlier than in the simulations with global random perturbations only (compare yellow/red dotted lines with the blue dashed
line in Fig.~\ref{fig:H0_2_mixed_all_data_tau_2}); this may be due to the localized perturbations being much larger than 
global random perturbations, initializing dewetting at an earlier time.

Figure~\ref{fig:H0_0_mixed_all_data_tau_20} plots, for comparison, the results for $H_0=0.05$ (spinodal dominated regime) at $\tau=20$. The results show that 
for long times, the localized perturbations have negligible effect on the film morphology, and only a single wave number (which is the same as in the simulation with 
random perturbations only) dominates the wave pattern. Simulations were also carried out for $H_0=0.3$ (nucleation-dominated) and $H_0=0.6$ 
(spinodal-dominated), confirming the results (figures not shown for brevity). Therefore, our simulations are able to distinguish 
between spinodal-dominated and nucleation-dominated regimes. This achievement is difficult if not impossible to reach without resorting to the kind of
large-scale simulations presented here.  

\begin{figure}[!h]
		\centering
		\includegraphics[width=0.9\textwidth]{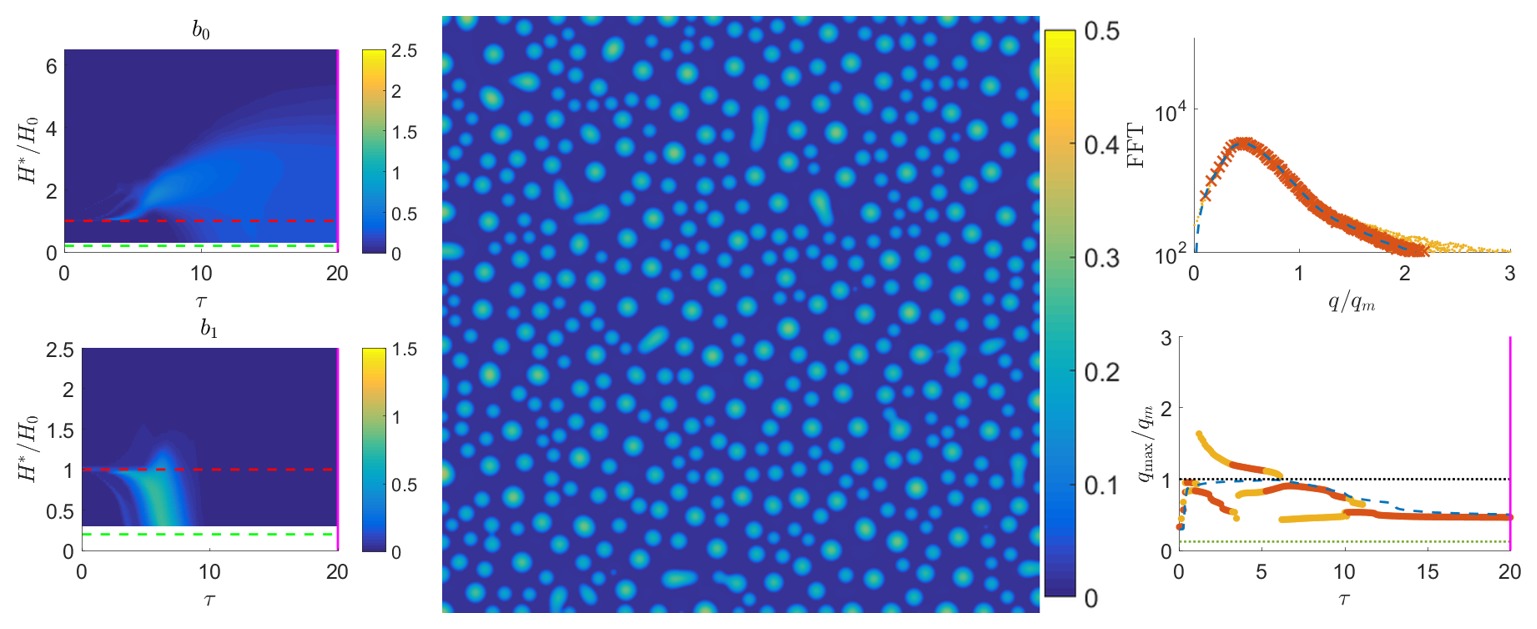}  
		\caption{Film of thickness $H_0 = 0.05$ perturbed by local and global perturbations.
		The central part shows a contour plot of the free surface height at $\tau = 20$.	The rest of the figure is as described in the caption of 
		Figure~\ref{fig:H0_2_mixed_all_data_tau_2}. The grid size is $1637\times1637$. For animations, see~\cite{SM}, movie4.	 }
		\label{fig:H0_0_mixed_all_data_tau_20}
\end{figure}

Next we discuss briefly the Betti numbers for the simulations where both local and global perturbations are present, with 
the focus on a few revealing features of the results.   The comparison of results for the Betti number $\grave{b}_0$ between Figures~\ref{fig:H0_2_mixed_all_data_tau_2}
and \ref{fig:H0_0_mixed_all_data_tau_20} shows two main differences. First, there are some much larger (taller) drops in the nucleation
regime: in Figure \ref{fig:H0_2_mixed_all_data_tau_2}, top left panel, we see a relatively large number of components (drops), even for thresholds
as large as $5H_0$, while in the spinodal regime shown in Figure \ref{fig:H0_0_mixed_all_data_tau_20}, such tall drops are not present.  The second difference, 
which is perhaps even more interesting, is the gradient (with respect to the threshold value) of the number of components (drops) 
for films in the nucleation regime, see Figure \ref{fig:H0_2_mixed_all_data_tau_2}, 
top left panel.   We see a larger number of components (drops) for smaller values of the threshold, all the way down to the precursor film thickness. 
This finding suggests (i) larger differences in the drop sizes in the nucleation regime; and (ii) a significant presence of small satellite drops, similar to the
RII regime in the case of global perturbations only, see Figure~\ref{fig:H0_6_all_data_tau_20}.  Further inspection (see also Supplementary Materials)
shows that the variety of drop sizes, including rather small ones, is due to breakup of the ridges between growing holes.   
Furthermore, there are no satellite drops forming inside of the growing holes, in contrast to the results shown in Figure~\ref{fig:H0_6_all_data_tau_20}.

\begin{figure}[!h]
		\centering
		\includegraphics[width=0.9\textwidth]{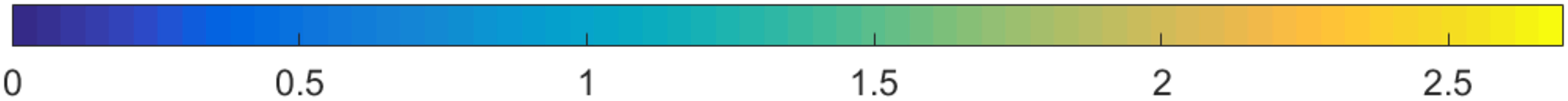}  	
		\subfigure[]{
			\includegraphics[width=0.45\textwidth]{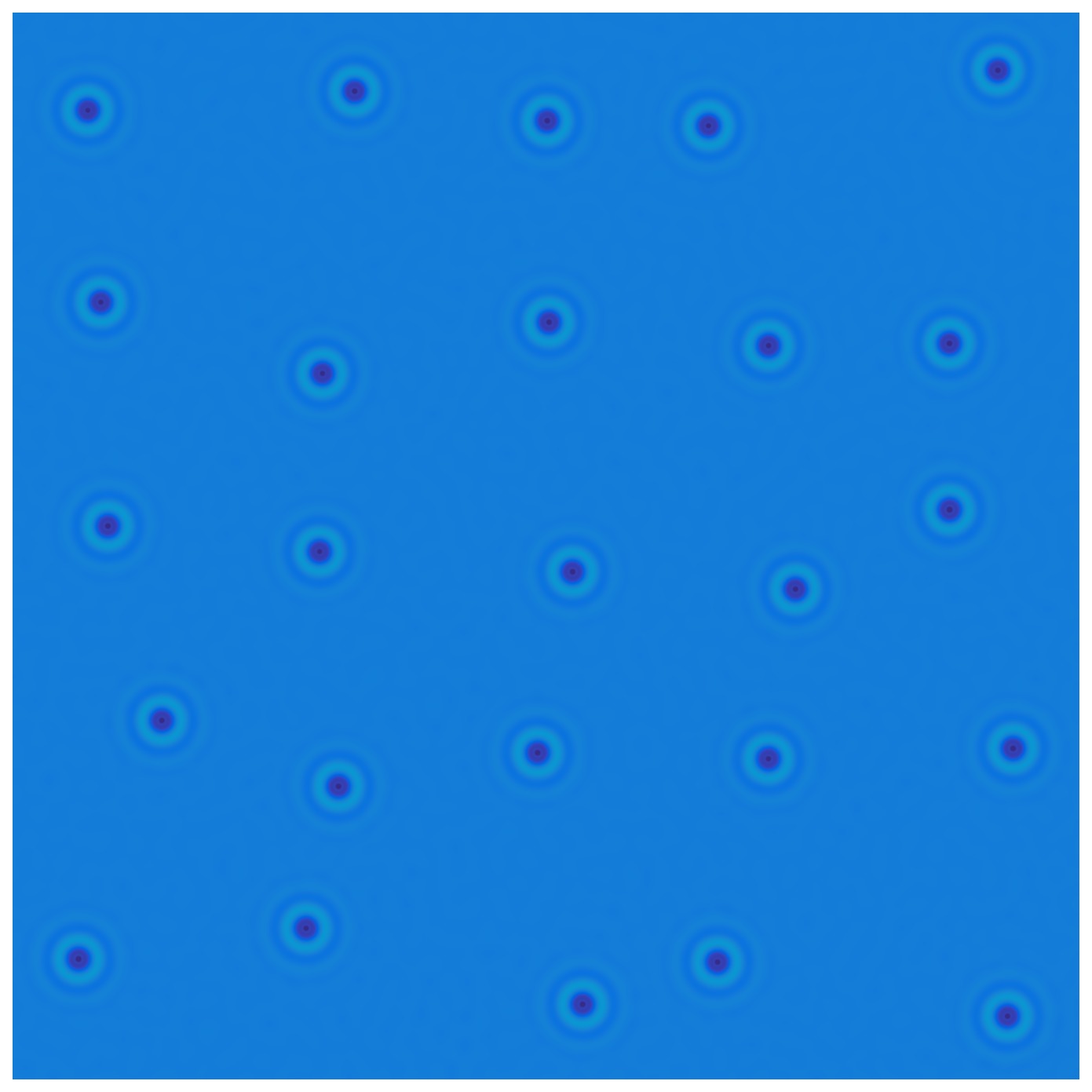} 
			\label{fig:H0_6_mixed_all_data_1} }  	
		\subfigure[]{
			\includegraphics[width=0.45\textwidth]{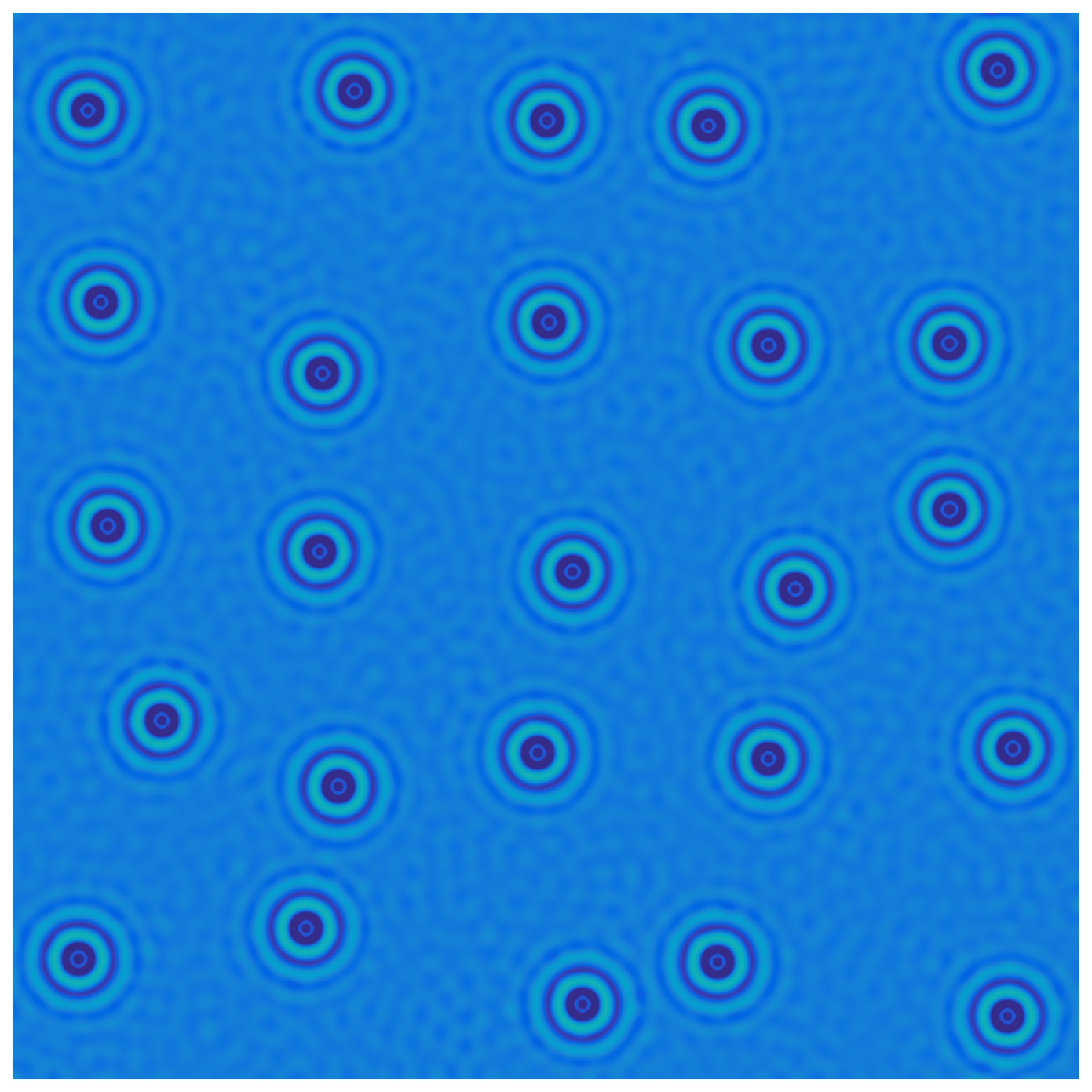} 
			\label{fig:H0_6_mixed_all_data_2} }  	
		\subfigure[]{
			\includegraphics[width=0.45\textwidth]{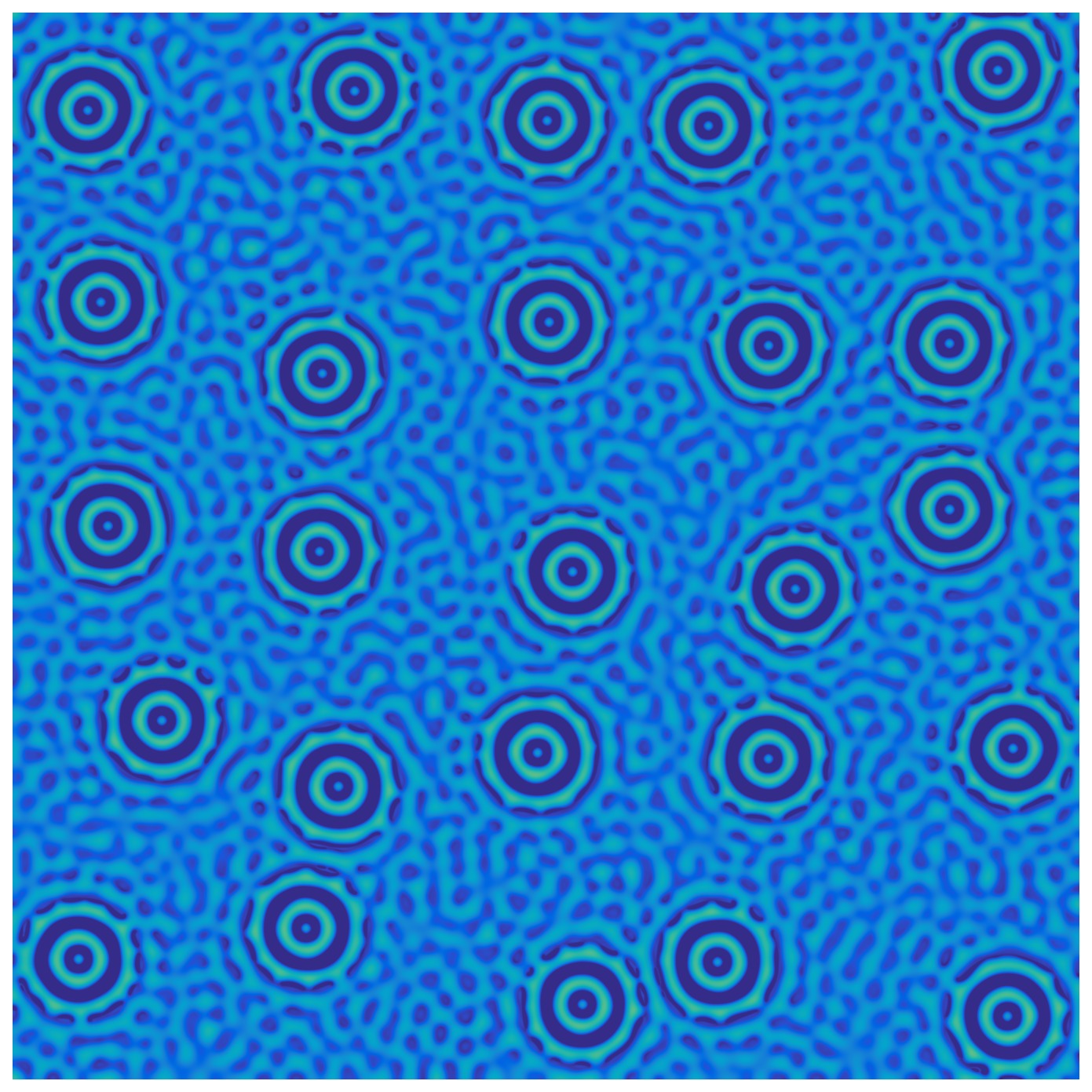} 
			\label{fig:H0_6_mixed_all_data_3} }  	
		\subfigure[]{
			\includegraphics[width=0.45\textwidth]{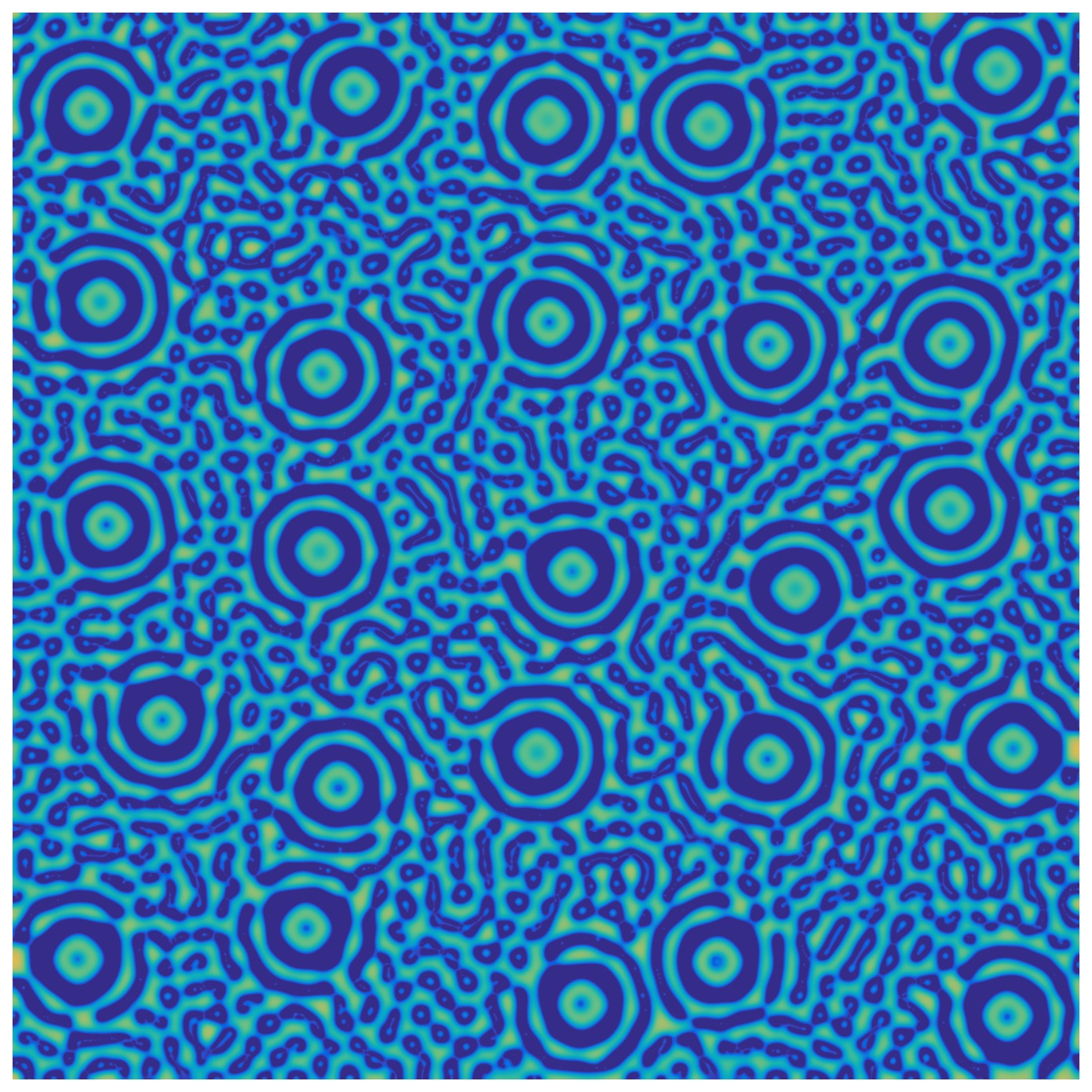} 
			\label{fig:H0_6_mixed_all_data_4} }
		\caption{Film of thickness $H_0 = 0.6$ perturbed by local and global perturbations shown at $\tau = 2.5, ~5.0,~7.5,~10$.  Note the expanding/collapsing
		ridges close to the perturbation centers, resulting in formation of large drops there at the last time shown. 
		The grid size is $3882\times3882$.	For animations, see~\cite{SM}, movie5.	 }
		\label{fig:H0_6_mixed_all_data}
\end{figure}

Figure~\ref{fig:H0_6_mixed_all_data} shows another example of interesting and non-trivial dynamics.  Here, the expanding fluid film 
breaks up into a series of ridges.  The innermost of these ridges stops its expansion and reverses its motion, collapsing back into itself. 
This process resembles the collapse of liquid rings, discussed recently by Gonzalez~{\it et al.}~\cite{gdk_jfm13}.   The outcome is the formation of a large drop, instead of a 
small satellite at a perturbation center.  This type of unusual dynamics is referred to below as an `expanding/collapsing ridge'.  

In our previous work considering the 2D geometry~\cite{Lam2018} we have shown that the evolution of unstable films may be very complex; some aspects of that complexity
could be analyzed in detail in 2D since we were able to run simulations in large domains for long times.  Doing such detailed simulations is not feasible in 3D, even with our GPU code, so we limit ourselves here to illustration of the main features of the results. Still, the simulations that we have carried out for various 
film thicknesses and perturbation types produce consistent results, in particular with respect to the sign of the disjoining pressure, the instability type, and 
the type of perturbation applied. Table~\ref{table} illustrates some of these common features.  While new simulations may uncover additional features 
of the results and patterns that form, we believe that the generic features of the results described so far already provide significant insight. 

We also comment briefly on the speed at which the localized perturbation propagates into the linearly unstable flat film regime investigated
in previous work for 2D geometry~\cite{Lam2018}. Our previous work applied the Marginal Stability Criterion (MSC)  to derive an analytical expression for 
the speed at which the localized perturbation propagates (the details are beyond the scope of the present paper and we refer the reader to other works~\cite{Saarloos2003,Lam2018} for a more detailed  discussion). However, we note that when we extract this speed from simulations of a single localized perturbation for $H_0=0.2$ and $0.6$ in 3D geometry, the numerically-extracted speeds agree with the analytical MSC result, demonstrating that 
the MSC result carries over from the 2D geometry. In addition, as noted in our previous work~\cite{Lam2018}, MSC can also be applied to gravity driven flow of a NLC film
down an inclined substrate~\cite{Lam2014,Lam2015} (where van der Waals forces are ignored); to flow of Newtonian films down an inverted substrate~\cite{lin_pof12,Lin2010} and down the outside of a vertical cylinder~\cite{Mayo2013}; and may also be applied to other models such as the 
Cahn-Hilliard and Kuramoto-Sivashinsky type of equations~\cite{Saarloos2003}.

  \begin{table}[!th]
\begin{tabular}{|lllllll|} \hline \hline
$H_0$             &Perturbation &Regime  & $\Pi$        &Satellites          &FFT  & Comment  \\ \hline \hline
 $0.05$     &	global  &	spinodal    &	$<0$  & no	      &	single mode &  agreement with LSA and long time \\
                 &                 &                        &               &             &                         &  coarsening  \\ \hline  
  $0.05$     &	local    &	spinodal    &	$<0$  & no	      &	single mode &  no influence of local perturbations  \\ \hline 
 $0.05$     &	mixed  & 	spinodal    &	$<0$  & no	      &	single mode &  no influence of local perturbations  \\ \hline 
 $0.2$       &	global  &	nucleation &	$<0$  & yes	&	single mode &  satellites due to ridge breakup \\ \hline  
 $0.2$       &	local    &	nucleation &	$<0$  & yes	&	bimodal        &   satellites due to ridge breakup \\ \hline  
 $0.2$       &	mixed  &	nucleation &	$<0$  & yes	&	bimodal        &   satellites due to ridge breakup \\ \hline    
 $0.3$       &     global  &      nucleation &     $>0$  & yes      &    single mode  &  satellites at perturbation centers   \\ 
                &                 &                        &               &             &                         &  and due to ridge breakup  \\ \hline 
 $0.3$       &     local    &      nucleation &     $>0$  & yes      &    bimodal         &  satellites at perturbation centers   \\ 
                 &                 &                        &               &             &                         &  and due to ridge breakup  \\ \hline  
 $0.3$       &     mixed   &      nucleation &     $>0$  & yes      &    bimodal         &  satellites at perturbation centers   \\ 
                 &                 &                        &               &             &                         &  and due to ridge breakup  \\ \hline  
 $0.6$       &	global  &	spinodal    &	$>0$  & yes      &	single mode &   satellites at perturbation centers   \\ 
                 &                 &                        &               &             &                         &  and due to ridge breakup  \\ \hline  
 $0.6$       &	local    &	spinodal    &	$>0$  & yes	 &   single mode &  satellites at perturbation centers  \\ 
                 &                 &                        &               &             &                         &  and due to ridge breakup  \\ \hline  
 $0.6$       &	mixed  & 	spinodal    &	$>0$  &  yes      &	single mode  & satellites at perturbation centers  \\ 
                 &                 &                        &               &             &                         &  and due to ridge breakup  \\ \hline  
     
  \end{tabular}
    \caption {Summary of the results for linearly unstable films. }  
    \label{table}
\end{table}

\subsubsection{Metastable Regime} \label{sec:Metastable}

Here we briefly investigate the metastable regime previously studied for 2D films~\cite{Lam2018}.   This regime 
corresponds to film thicknesses that are linearly stable; but unstable with respect to perturbations of 
finite amplitude. Previous analytical results have shown that there exists a thickness regime, $H_0\in(H_+,H_{m_3})\approx(1.01,3.04)$ 
(using the notation from~\cite{Lam2018} here) that is metastable, and was derived analytically by extending previous results for a Newtonian
flow under gravity~\cite{Diez2007}.  Furthermore, for 2D films, extensive numerical simulations 
have shown that for an initial condition of the form
 \begin{equation} \label{eq:NanoICLocal}
      h(x,y,t=0) =  H_0 \left[ 1 - d \exp \left( - \frac{x^2}{0.04 W^2}  \right)  \right] \;, \quad W=4\pi \, ,   
\end{equation}
for dewetting to occur, the minimum magnitude of the perturbation required for dewetting to occur, $d_c$, approaches $1$ as $H_0\rightarrow H_{m_3}$.  In other words, 
 for films of thickness close to $H_{m_3}$, the initial perturbation has to essentially reach the precursor thickness to produce dewetting.

\begin{figure}[!h]
	\centering
	\subfigure[]{
	\includegraphics[width=0.27\textwidth]{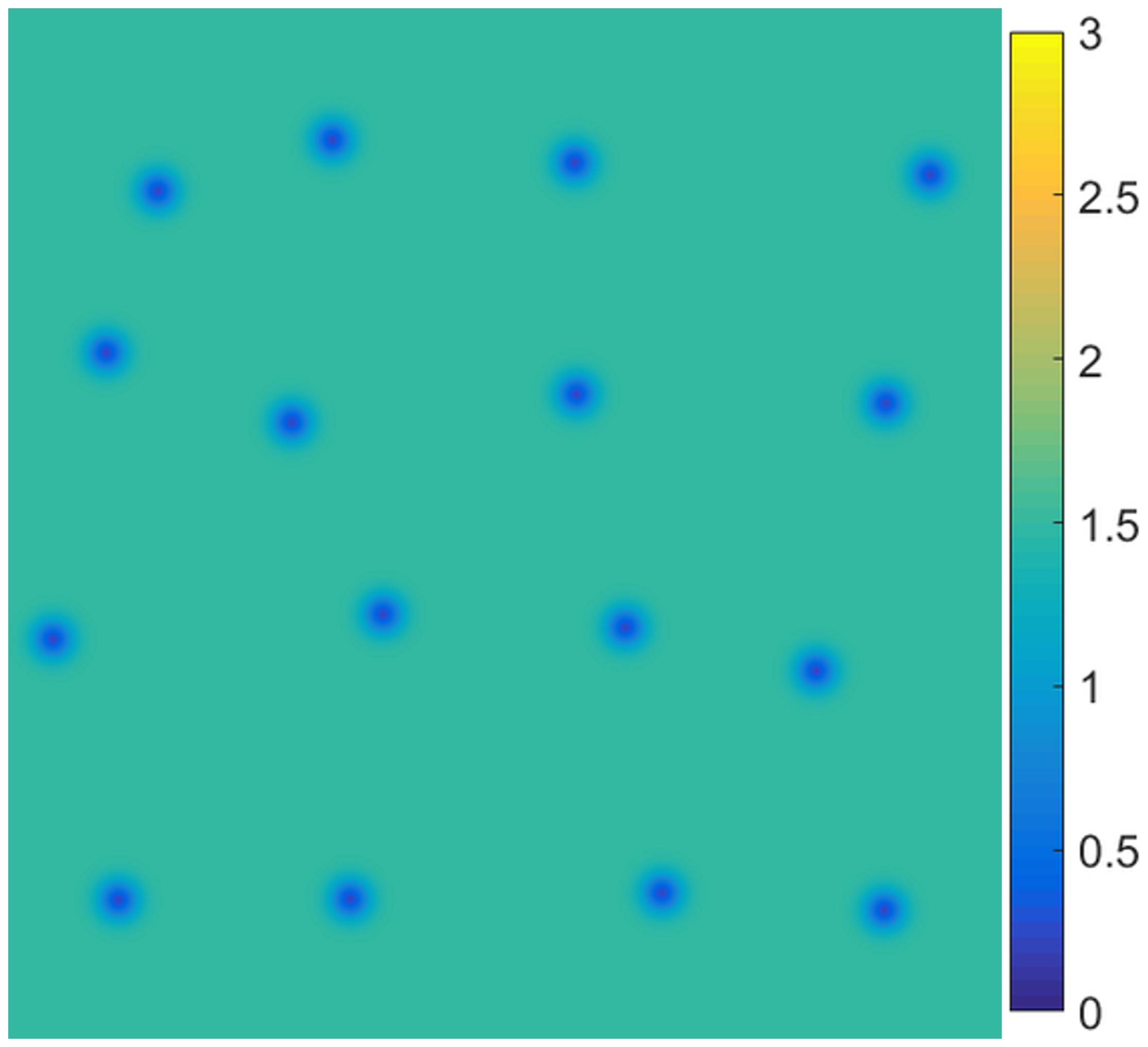}\label{fig:MetastabilityExample_1} }
	\subfigure[]{
	\includegraphics[width=0.27\textwidth]{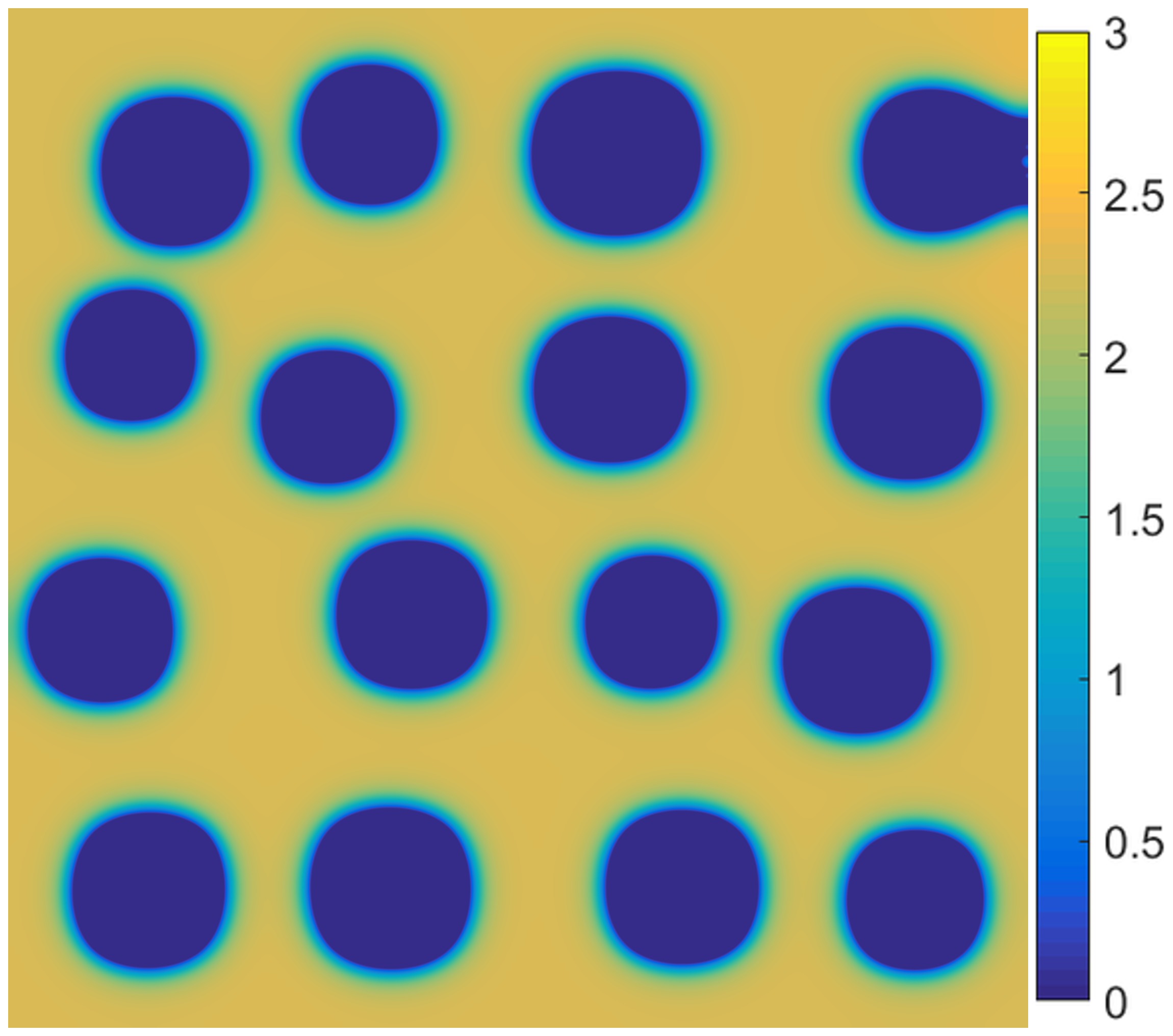}\label{fig:MetastabilityExample_2} }
	\subfigure[]{
	\includegraphics[width=0.33\textwidth]{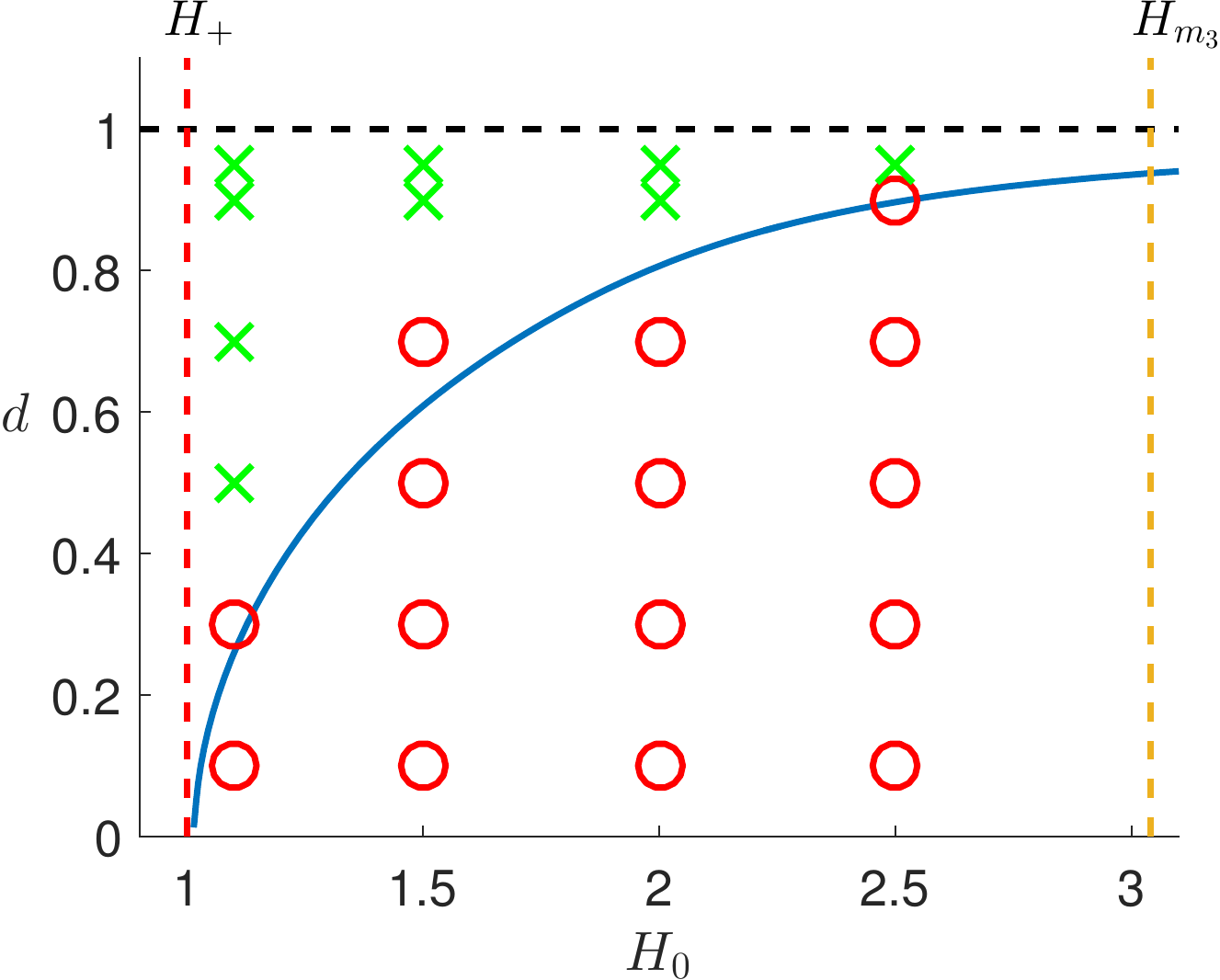} \label{fig:MetastabilityTest}}
	\caption{Contour plot of metastable film, with a) the initial condition given by (\ref{eq:NanoICLocalMulti}), and b) the  free surface at $t=1000$. 
	c) Metastability results for initial condition (\ref{eq:NanoICLocal}) as a function of the initial film thickness, $H_0$, and initial magnitude of the localized 
	perturbation, $d$.  The blue curve denotes $d_c$, the minimum value of $d$ required to induce dewetting in 2D films~\cite{Lam2018}.
	For the 3D films, $\times$ in green font denotes initial condition parameters that induce dewetting, and $\bigcirc$ in red font denotes parameters that do not induce dewetting. 
	The grid size is $2000\times2000$. For animations, see~\cite{SM}, movie6. }
	\label{fig:Metastability}
\end{figure}

For 3D films, we use an initial condition of the form
 \begin{equation} \label{eq:NanoICLocalMulti}
   h(x,y,t=0) =  H_0 \left[ 1 - d \eta(x,y)\right] \;,
\end{equation}
where the linear domain size is $100$, $H_0=1.5$, $d=0.85$, and $\eta(x,y)$ is given by (\ref{eq:3DRandomLocalTerm}) with $\lambda_m=2.5$ (for convenience, we use the value from (\ref{eq:3DRandomLocalTerm}) although $\lambda_m$ is not relevant for metastable films).  Figure~\ref{fig:MetastabilityExample_1} shows the contour plot of 
the initial condition. The imposed local perturbations grow, leading to the profile characterized by growing holes shown in 
Figure~\ref{fig:MetastabilityExample_2}  at $t = 1000$ (note that in the linear stable regime, $\omega_m$ is undefined, therefore,
we use $t$ instead of $\tau$ in this section).  We note that the profile shown in this figure is transient: the holes continue to expand for later times. 
In this work we focus only on the regime during which the perturbations evolve essentially independently of each other.  

Figure~\ref{fig:MetastabilityTest} focuses on the comparison between the minimum thicknesses, $d_c$, needed to cause dewetting, 
for 2D and 3D films.   In 3D, we carry out simulations with a single perturbation of the form 
specified by (\ref{eq:NanoICLocalMulti}), with the perturbation centered in the middle of the domain.   
In Figure~\ref{fig:MetastabilityTest}, crosses and circles denote the parameters that do/do not induce dewetting, respectively, in 
3D, and the blue curve shows the analogous (analytical) results 
 for 2D films.  It may be seen that the 2D findings essentially extend to 3D. The minor differences seen are to be expected, due to the different geometry
 of perturbations in 2D and 3D: a localized perturbation in 2D is essentially a `trench' since the perturbation is invariant with respect to the $y$ direction.   

\begin{figure}[!h]
	\centering
	\subfigure[Linearly unstable regime, $H_0 = 0.2$. ]{
	\includegraphics[width=0.45\textwidth]{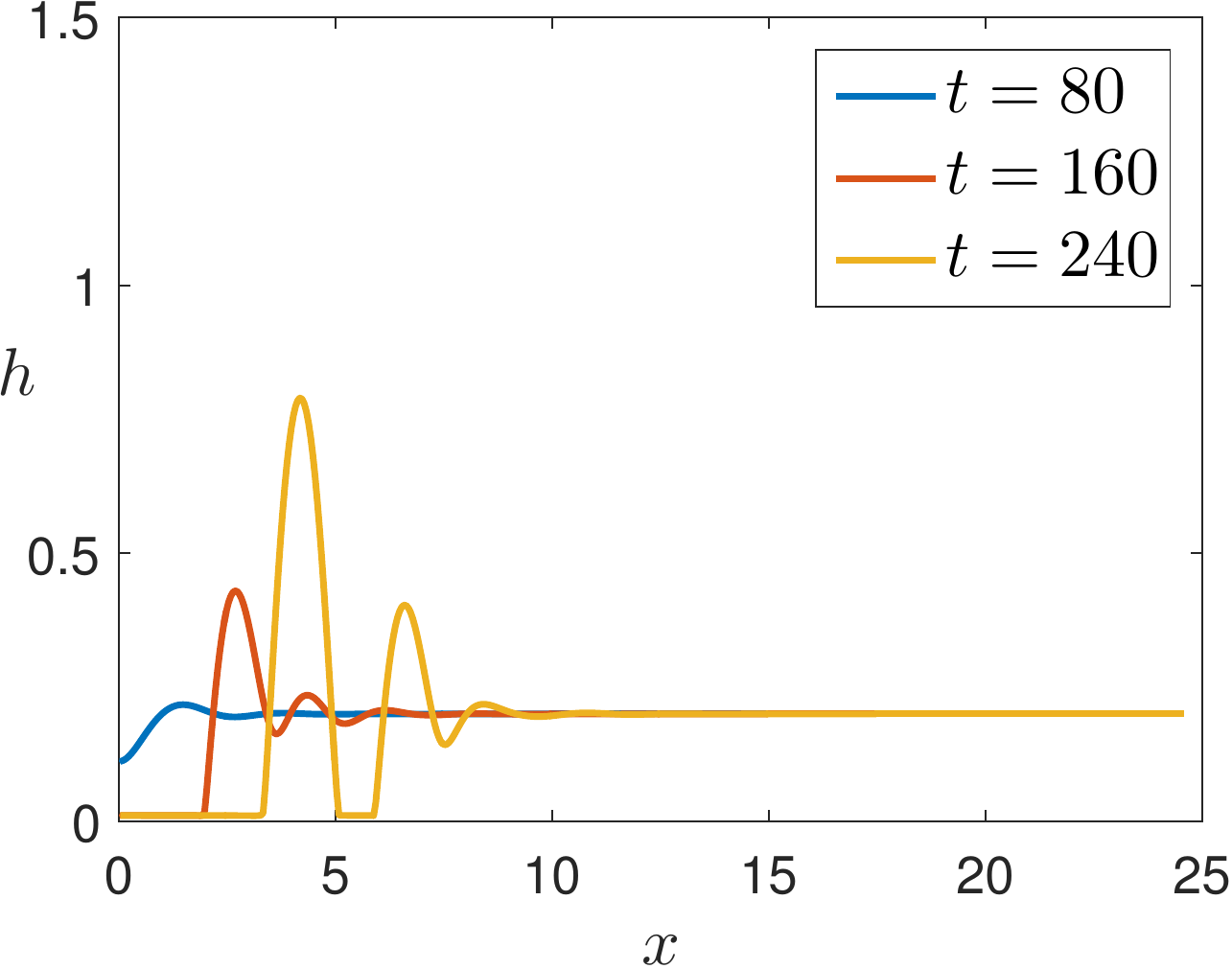}\label{fig:cross_section_RII} }
	\subfigure[Metastable regime, $H_0 = 1.5$.]{
	\includegraphics[width=0.45\textwidth]{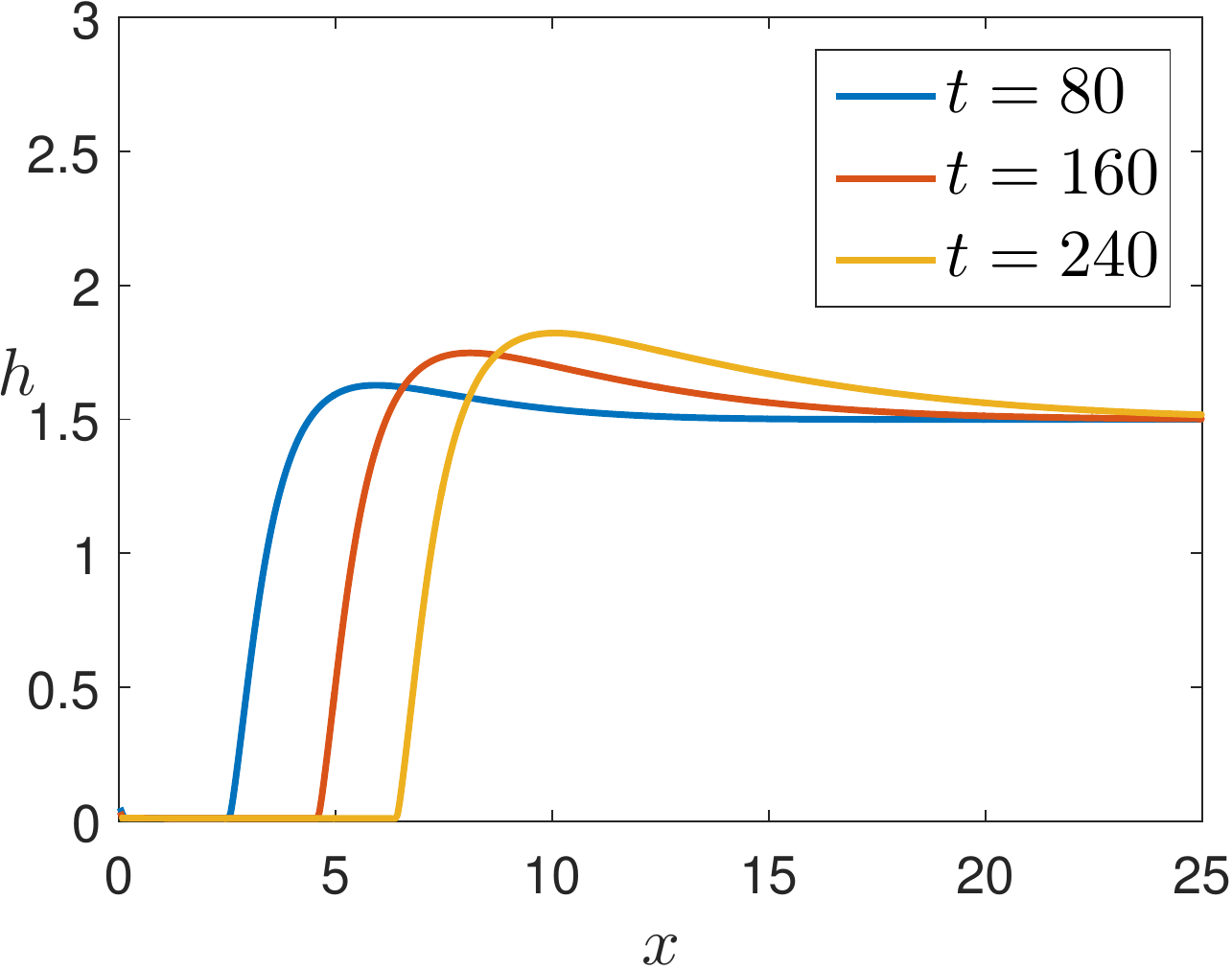}\label{fig:cross_section_meta} }
	\caption{Cross-section of the film profile as a function of the distance from localized perturbation centered at $x=0$. The grid sizes are
    a) $492\times492$ and b) $750\times750$.	}
	\label{fig:cross_section}
\end{figure}

Finally, we briefly comment on the shape of the expanding fronts away from localized perturbations for linearly unstable nucleation-dominated films  versus
metastable ones.   Figure~\ref{fig:cross_section} shows a few cross-sectional snapshots of one of perturbations from Figures~\ref{fig:H0_2_mixed_all_data_tau_2}
and~\ref{fig:Metastability}.
  The expanding front for a linearly unstable film is characterized by the presence of a well-developed capillary ridge, followed by an
  oscillatory damped profile, similar to the 2D films considered
previously in~\cite{Lam2018}.  For metastable films, however,  the ridge is much less prominent and the connection to the rest 
of the film is monotonous.   We are not aware that these different front shapes have been discussed in this context 
in the literature so far.

\section{Comparison to Experimental Results} \label{sec:ExperimentalResults}

In this section, we compare simulation results for \gls{NLC} and polymeric films to available experimental results. 
The disjoining pressures considered are given by (\ref{eq:DisjPressNLC}) and (\ref{eq:DisjPressPolymer}), respectively.   To scale the domain and initial condition for different film thicknesses, simulations are performed on a square domain, and the linear domain size is scaled with $\lambda_m$.   In addition, we fix $\Delta s = 0.05$ for both models.
The initial condition is given by (\ref{eq:3D_Random_IC}).  For all simulations in this section, we fix $P=40$.

\begin{figure}[!h]
	\centering
	\subfigure[Our simulation]{
	\includegraphics[width=0.31\textwidth]{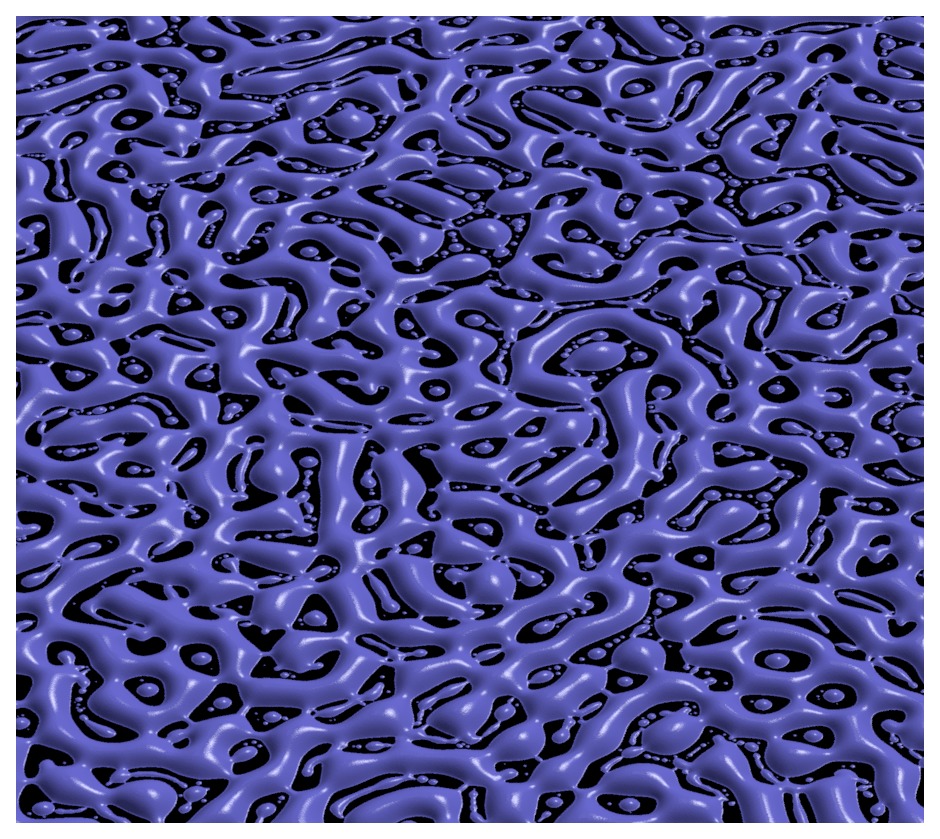} \label{fig:Sim_H0_5_Random} }
	\subfigure[Schlagowski {\it et al.} (2002)]{
	\includegraphics[width=0.31\textwidth]{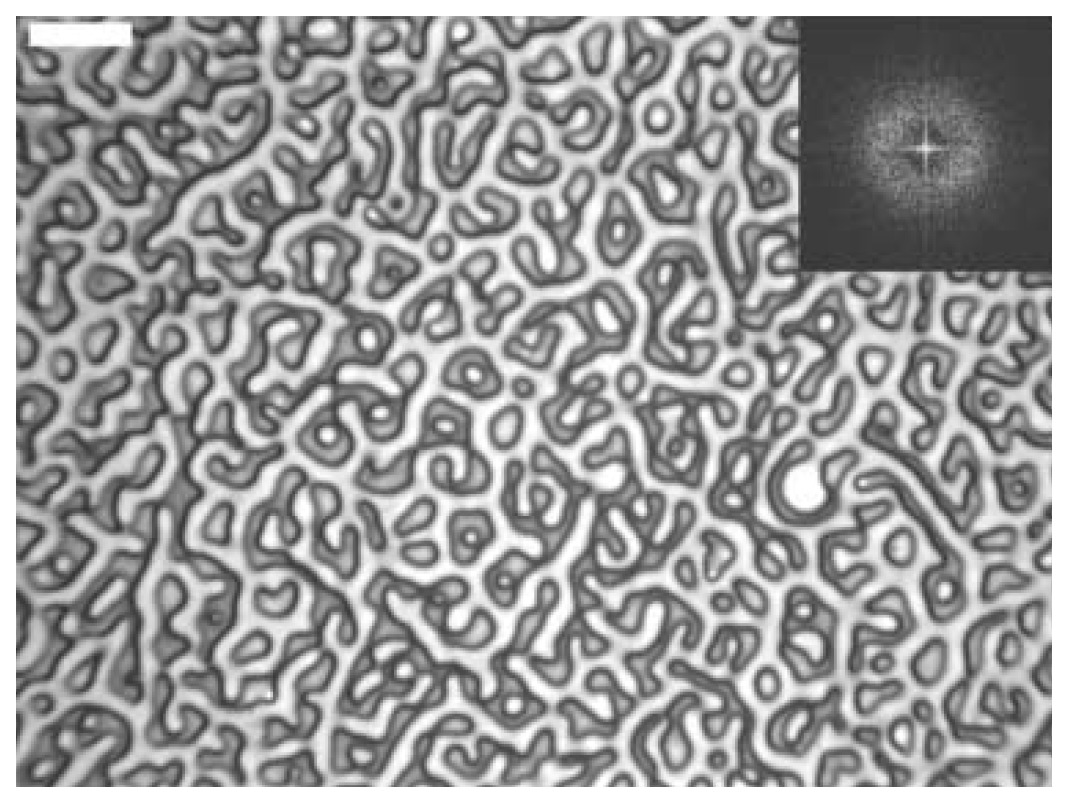} \label{fig:Schlagowski2002_1} }
	\subfigure[Vandenbrouck {\it et al.} (1999)]{
	\includegraphics[width=0.31\textwidth]{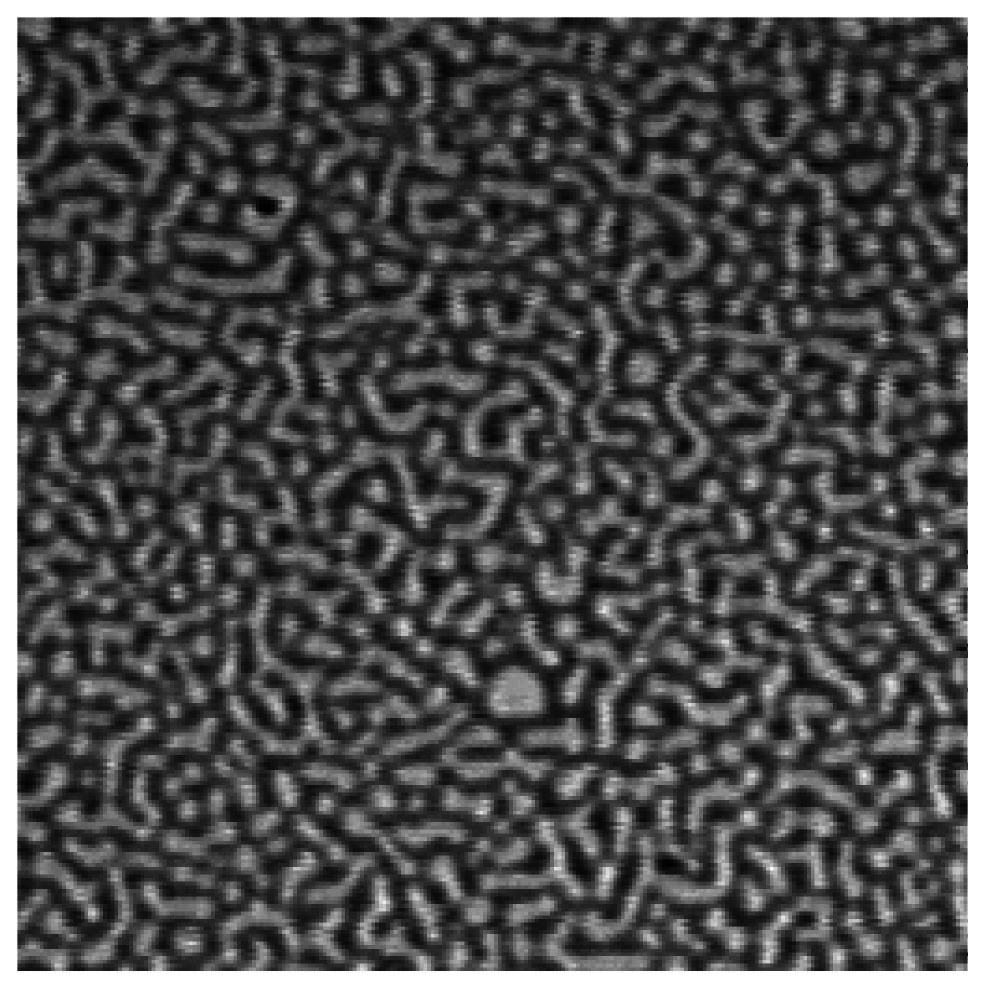} \label{fig:Vandenbrouck1999_1} }
	\caption{ a) Simulation result for a 50 nm ($H_0=0.5$) thick film of 4-Cyano-4'-pentylbiphenyl (5CB) nematic liquid crystal vs experimental results for b) a 85 nm thick 4-Octyl-4’-Cyanobiphenyl (8CB) film and c) a 43 nm thick 5CB film. The grid size in a) is $3400\times3400$.} 
	\label{fig:ExperimentsRandom}
\end{figure}

The first set of experimental results considered is dewetting experiments for NLC films by Schlagowski {\it et al.}~\cite{Schlagowski2002} and Vandenbrouck {\it et al.}~\cite{Vandenbrouck1999}.   In the experiments by Vandenbrouck {\it et al.}~\cite{Vandenbrouck1999}, a flat film is formed by increasing the temperature of the NLC such that a flat film is stable at the considered film thickness of interest (isotropic phase).   The flat film is formed by using a spin coating method, and then cooled to a temperature so that it becomes unstable (nematic phase). Schlagowski {\it et al.}~\cite{Schlagowski2002} varied the temperature of the NLC sample, alternating between the isotropic and nematic phases, observing the morphology of the film as the temperature crosses the threshold between the isotropic phase and nematic phase. Figure~\ref{fig:ExperimentsRandom} compares a 3D rendering of our numerical results to experimental results of two different types of NLC. In both experiments, the characteristic wavelength of undulations is appropriately 30 $\mu$m, which is similar to $\hat\lambda_m = 42$ $\mu$m for the parameters stated in (\ref{eq:NLC_scales}) and (\ref{eq:NLC_paras}).

\begin{figure}[!h]
	\centering
	\subfigure[Our simulation]{
	\includegraphics[width=0.32\textwidth]{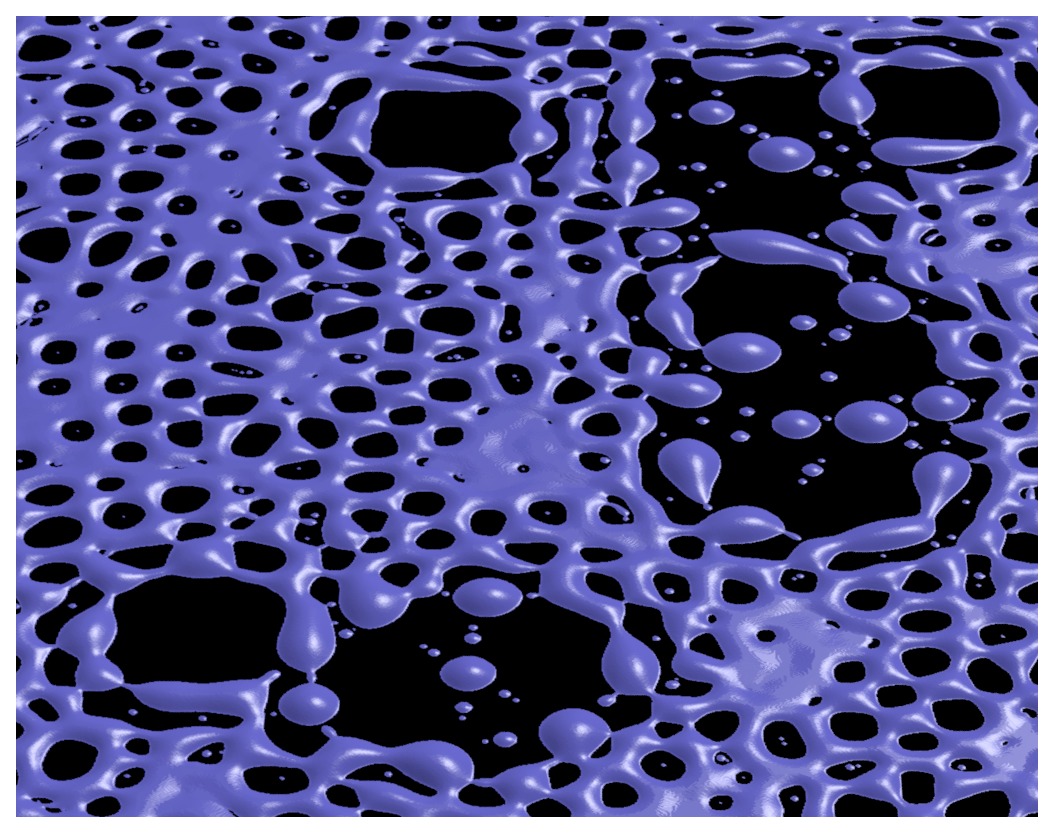} \label{fig:Sim_H0_2_Mixed} }
	\subfigure[Herminghaus {\it et al.} (1998)]{
	\includegraphics[width=0.32\textwidth]{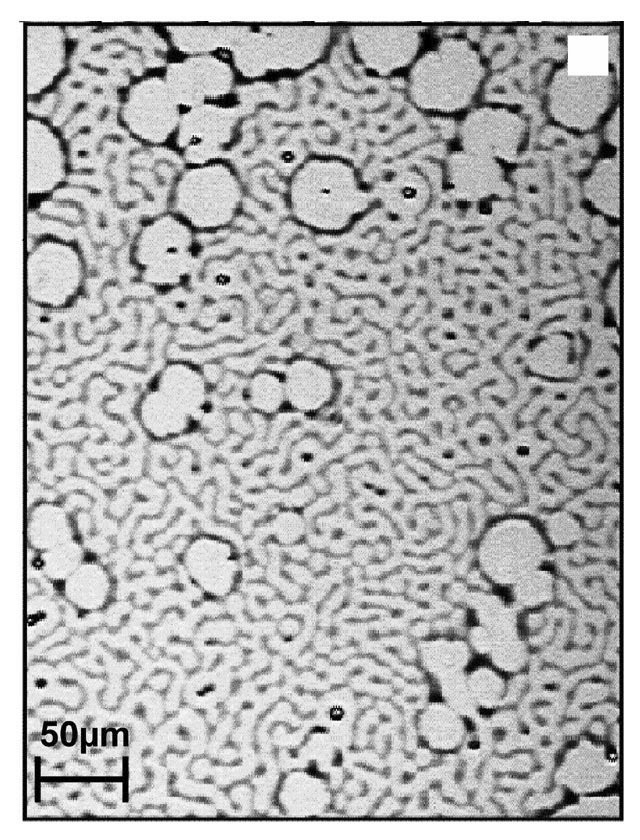} \label{fig:Herminghaus1998_2a} }
	\caption{a) Simulation result for a 20 nm ($H_0=0.2$) thick film of 4-Cyano-4'-pentylbiphenyl (5CB) nematic liquid crystal film vs b) an experimental result from Herminghaus {\it et al.}~\cite{Herminghaus1998} for a 40 nm thick tris(trimethylsiloxy) silane-ethoxycyanobiphenyl (5AB\textsubscript{4}) film. The grid size in a) is $1970\times1970$.} 
	\label{fig:ExperimentsMixed}
\end{figure}

The next experiment we consider is that by Herminghaus {\it et al.}~\cite{Herminghaus1998}, who prepared flat films of NLC in a Langmuir trough.  Unlike the spin coating method, transferring the flat film from the Langmuir trough to the solid substrate may induce defects in the film (localized perturbations).   To simulate this initial state, large localized perturbations at random locations are added to the initial condition (\ref{eq:3D_Random_IC}).   Figure~\ref{fig:ExperimentsMixed} demonstrates that the localized perturbations can induce the formation of large holes, leading to patterns that are visually similar to the experimental images.

\begin{figure}[!h]
	\centering
	\subfigure[Our simulation]{
	\includegraphics[width=0.32\textwidth]{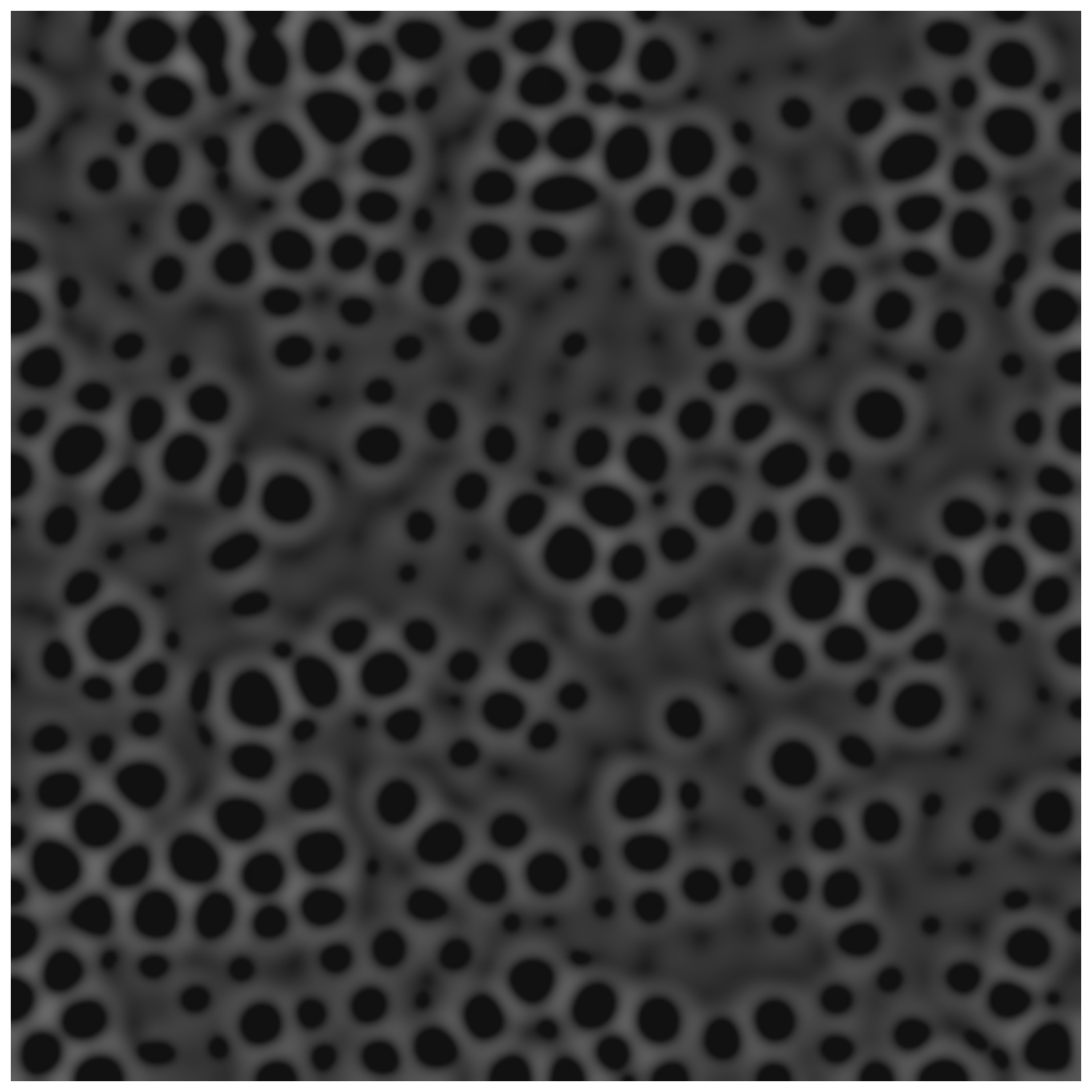} \label{fig:Polymer_Sim} }
	\subfigure[Jacobs {\it et al.} (2008)]{
	\includegraphics[width=0.32\textwidth]{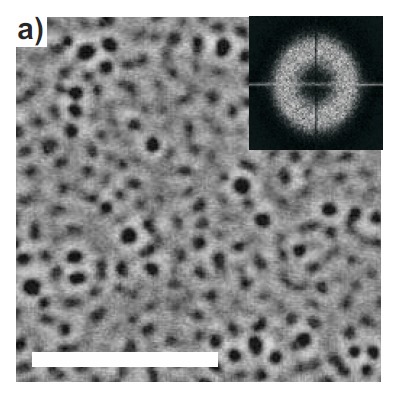} \label{fig:Jacobs2008_6a} }
	\caption{Comparison between a) our numerical simulations at $t=60s$ and b) the experiment by Jacobs {\it et al.} (2008)~\cite{Jacobs2008} for a 4.9 nm film on a silicon substrate coated with a 191 nm layer of silicon oxide. Note color scales on both panels are the same; specifically, black corresponds to 0 nm and white corresponds to 20 nm. The linear domain size in both panels is 8 $\mu$m and the scale bar for (b) corresponds to 5$\mu$m. The grid size in a) is $1600\times1600$.}
	\label{fig:ExperimentsPolymer}
\end{figure}

The last experiment we consider is a dewetting experiment with polymeric films by Jacobs {\it et al.} (2008)~\cite{Jacobs2008}. Figure~\ref{fig:ExperimentsPolymer} compares simulation results with experimental results for a 3.9 nm thick polymeric film on a silicon substrate coated with a 191 nm layer of silicon oxide. It may be seen that the dominant wavelength in our simulations matches the experimental results; however, the timescales of dewetting differ. 
This difference may be due to the choice of the initial perturbation size.   

\section{Conclusions}\label{sec:Conclusions}

In this paper, we describe a GPU implementation of an ADI numerical scheme applied to evolution of thin films in a three dimensional setting.  The 
scheme has been generalized to conservation equations with nonlinear fluxes, and validated by extensive testing of its convergence and 
conservative properties, as well as by comparison with linear stability analysis.   The computational gains obtained by carrying out simulation on a GPU 
are substantial, and allow for carrying out simulations in large domains with rather basic computational resources.   

The main set of results presented in this paper focuses on three dimensional (3D) simulations of unstable films.  We focus in 
particular on instabilities of nematic liquid crystal films, and to a lesser extent, those of spreading polymer films. Direct comparison 
of the computational results to available experimental data has been carried out, with striking visual similarity observed between the simulations and experiments. 

The large scale simulations that we present in the paper have uncovered a number of results.  Regarding linearly unstable films, we find that both evolution of instability and the final outcome may depend strongly on the source of the instability,
and on film thickness. In particular, films that are dominated by a spinodal instability mechanism lead to formation of drops whose
typical distances and sizes are consistent with the predictions of linear stability analysis, independently of the source of instability. 
In the nucleation dominated regime, however, we observe the signature of the source of instability both in the manner in which the 
instability evolves, and in the final outcome.   While we have not presented an extensive study of this aspect 
of the results, our results suggest that one may be able to deduce the instability source by observing the outcome.  
Attaining such a goal requires large computational domains, as used in this work, or comparably large experimental domains.

Another outcome of our results that is made possible by large computational domains is analysis of satellite drop formation.  
Our previous works~\cite{Lam2018} on 2D films predicted formation of satellite drops during breakup of films
characterized by positive values of (effective) disjoining pressure.  The 3D results in the present paper are consistent with the 
2D results, showing that the number of physical dimensions does not influence the basic features of instability development.  
The details of the results are influenced by the geometry however; in particular, we find two possible ways for satellite drops to 
form: they can form either at the centers of imposed perturbations (equivalently, at the centers of the expanding holes), or 
they can form during the process of ridge breakup. We find that both the instability regime, and the sign of the disjoining pressure,
influence the process of satellite drop formation, with only the thinnest of the considered films not showing satellites.  In some cases, 
typically for thicker films, we also observe expanding/collapsing ridges -- a ridge growing radially away from a perturbation center
may detach from the rest of the film and collapse back to its center due to azimuthal curvature.  

For the purpose of analysis of results, we have found that careful Fourier analysis combined with computing topological 
invariants (Betti numbers) provides insight.  Betti numbers, in particular, are computed across all relevant film thicknesses, 
giving a detailed picture of the film evolution.  

We have also analyzed metastable films, which are linearly stable, but unstable with respect to finite size perturbations.   The 
most interesting aspect of the results here includes the distinctive shape of the expanding holes due to imposed
local perturbations.  While for linearly unstable films one observes the formation of capillary ridges at the expanding fronts, for 
metastable films we find a monotonous increase of thickness from the equilibrium layer to the scale of the original film thickness.  

There are several possible avenues for extending the work presented in this paper. Regarding satellite drop formation, and the nucleation-dominated and  
metastable regimes, investigation of other thin film models would verify the generalization of our results in 
terms of a form of disjoining pressure. There are also several possible improvements that may be made to our \gls{GPU} code, for example, a multi-\gls{GPU} 
implementation; implementing adaptive mesh refinement; or improving numerical accuracy. Another important improvement would be to improve the 
parallelization of the linear solver using techniques such as the SPIKE method~\cite{Polizzi2006} or cyclical reduction~\cite{Heller1975,Meier1985}.

Regarding future applications of the presented computational methods, there are numerous possible directions.  In our work, we have
so far considered relatively simple setups where the film evolution could be described by a single partial differential equation.  An extension 
of the presented approach to problems involving multiple miscible or immiscible films, or to problems involving thermal
effects, for example, is possible.

\appendix

\section{GPU Implementation} \label{sec:GPUImplement}

In this Appendix, the specifics of the GPU implementation are discussed.   To begin, we give an overview on key concepts of GPU computing utilized in our implementation.  The GPU is a vector type processor which processes data in the \gls{SIMD} framework of parallel computing~\cite{Flynn1972}, i.e., the same finite difference scheme (single instruction) is applied at each grid point (multiple data).   A GPU consists of several multi-core processors (MPs) which share a \textit{global} memory bank (analogous to the CPU memory).   Analogous to the levels of cache memory on CPUs\footnote{Technically, the shared memory bank is not a level of cache memory, and MPs already have a L1 and a L2 cache (register memory).}, each MP has a fast (compared to global memory) \textit{shared} (between cores of a MP) memory bank, and an even faster \textit{register} memory (of the individual cores); however, unlike CPU cache memory, the shared memory bank has to be explicitly handled in a code in order to reduce redundant access to the (slow) global memory. Use of register memory is automatic when defining variables within the scope of the kernel (similar to the automatic use of cache memory on a CPU).   Typically, the shared memory bank is used to store data that will be reused by other threads within a thread block, whereas register memory is used to store values that will only be used by an individual thread.

In \gls{CUDA}, instructions on the GPU (\textit{device}) are executed from the CPU (\textit{host}) in terms of \textit{kernel} calls.   There are two input parameters unique to a kernel call, each composed of 3 numbers, specifying the size of a vector, matrix, or tensor.   The purpose of these parameters is to divide a domain into sub-lengths, sub-squares or sub-cubes; and assign each \textit{sub-domain} a vector, matrix, or tensor of threads, which we will refer to as a \textit{thread block}.  The input parameters provide an intuitive way of sub-dividing a multi-dimensional domain for parallelization (sub-domains), and a natural (geometric) way of mapping threads to points on the sub-domain.   On the hardware level, threads within a thread block are further sub-divided into groups called warps  (32 threads to each warp for the \gls{GPU} used in this paper).   Instructions to the MPs are batched in terms of warp (vectors) and are executed in the \gls{SIMD} architecture framework.

The last important concept for GPU computing (for our implementation) is \textit{coalesced} data access to the global memory, i.e., adjacent data values that are adjacent in memory may be accessed in a single \gls{SIMD} transaction (as opposed to multiple transactions), improving efficiency.   This is an 
important issue for the \gls{ADI} method, in particular, solving the system of penta-diagonal matrices in each direction, the details of which will be discussed in \ref{sec:LinerSystemSolver}.

\textbf{Note 1:} It is important to note that to fully utilize all threads on a MP, the size of the thread block is typically an integer multiple of the warp size.

\textbf{Note 2:} Since warps are executed in the \gls{SIMD} framework, warps with threads with different instructions (\textit{warp divergence}) are not computed in parallel, but each different instruction is executed in series, e.g., for $n$ different sets of instructions,
the warp is computed in $n$ serial steps (each sub-group of threads, within the warp, with the same instructions, are computed in parallel). Ghost points remove warp divergence at the boundaries; however, this is only an order $I+J$ improvement (the complexity of the complete scheme is $IJ$).

\subsection{GPU and CPU Interactions}

In addition to the three levels of GPU memory, the GPU may also access the CPU memory; however, the access time for this level of memory (with respect to the GPU memory) is prohibitively slow and may drastically affect the performance of a GPU code, e.g., for simple computations, the time required for transferring the data between the GPU and CPU can be comparable to (or even greater than) the time required to perform the computation on the GPU.   Furthermore, data have to be explicitly (in code) transferred between the GPU and CPU, and memory must be allocated on both the CPU and GPU\footnote{Newer GPU models support NVDIA's \textit{Unified Memory} model, where explicit transferring of data and allocation of memory on both the CPU and GPU is no longer necessary. Our implementation was originally developed on an older GPU that did not support the Unified Memory model.}\textsuperscript{,}\footnote{Data in memory are typically organized in blocks of fixed length, often called `pages'. For higher data bandwidth between CPU and GPU, the CPU memory is page-locked to GPU memory (called `pinned' memory in \gls{CUDA}), i.e., similar to coalesced data access, page locking provides a direct mapping between pages in the GPU's memory and pages in the CPU's memory, removing offsets between pages, allowing \gls{CUDA} to improve the data bandwidth by transferring the page in a single instruction (as opposed to multiple instructions). }.
To remove this bottleneck, unless necessary to do otherwise, all computations are performed on the GPU.   Recalling the outline of the numerical scheme,  controlling the adaptive time stepping is independent of grid size, and is therefore performed on the \gls{CPU}. Computationally expensive steps (center column of steps in Figure~\ref{fig:FlowChartTime}) may be (naively) parallelized on the GPU, with the exception of evaluating the convergence criteria (restrictions on accepting the solution at the new time).

To compute the convergence criteria, we first note that for serial computations, it is simple to determine if the convergence criteria are violated at any of the grid points, i.e.,  the serial code may iterate through the grid points until one of the convergence criteria is not satisfied; once determined, the appropriate changes may be made to the time step.  However, this procedure is slightly non-trivial on a \gls{GPU}; specifically, a thread cannot terminate the execution of warps in the queue (remaining grid points at which the convergence criteria are to be computed) and control cannot be immediately returned to the CPU.  Therefore, once one of the convergence criteria is determined to be violated at a grid point, this information (state) cannot be immediately returned to the \gls{CPU}, nor can control be immediately returned to the \gls{CPU} in this manner; thus, all warps must be executed before appropriate changes to the time step can be made.  The naive approach is to transfer the data from the \gls{GPU}'s memory to the \gls{CPU}'s memory. However, as mentioned before, this transfer is slow; therefore, a reduction method is used to reduce the amount of data to transfer. In more exact terms, each thread evaluates the convergence criteria at a grid point. The data are reduced on each thread block (sub-domain) by determining if any of the convergence criteria are not met within the thread block, and the reduced data (information for each thread block) are transferred to the \gls{CPU}.

In our implementation, the size of the thread block is $16\times16$, thus reducing the total number of data to be transferred by a factor of 256.  To reduce further the data volume to be transferred, we do not store the $8$ byte double precision value for the point that violates the convergence criteria, but instead store an integer value (an ID representing the convergence criterion that was not satisfied) in a $1$ byte char (flag variables).  In total, this method reduces the total volume of data to be transferred by a factor of 2048, and the cost of transfer is therefore negligible compared to the GPU computations.  The relevant information regarding convergence criteria may then be compressed (reduction) by the GPU, transferred quickly to the CPU, and using a simple loop, the CPU checks the flag on each thread block and adjusts the time-step accordingly.

\textbf{Note 1:} To the authors' knowledge, on a GPU, each entry in a boolean array is stored as a byte (8 bits) so is not a true bit array (1 bit per entry); thus there is no reduction in memory usage when using a boolean array instead of a char array. While a bit array could have been implemented, this was not done as: a) memory has to be carefully accessed by threads so as to not incur inefficiencies; b) the use of a char datatype allows information about the failure state on a sub-domain to be passed back to the host CPU (thus logged); and c) the total data size has been sufficiently reduced.

\textbf{Note 2:} The reduction on the GPU may be done recursively, i.e. one could further divide the flags into sub-domains and check whether each flag on the sub-domains (of flags) is equal to one of the convergence condition IDs.   However with each recursive step, there are diminishing returns in efficiency (less parallelization due to recursively smaller domain sizes), and furthermore, a sufficient reduction in the size of the data has already been achieved.

Having developed a hybrid method (CPU and GPU) to evaluate the convergence criteria, all computationally expensive operations (center column of steps in Figure~\ref{fig:FlowChartTime}) can be parallelized on the GPU, and only a small volume of data (relative to size of the solution) has to be transferred between the CPU and the GPU at every time step and iteration step.   While the remaining computationally expensive operations (center column of steps in Figure~\ref{fig:FlowChartTime}) may have a naive parallelization, for fast computation, effective uses of the various GPU memory types is required.  This applies in particular to solving (\ref{eq:JacobianSystem1})--(\ref{eq:JacobianSystem3}), on which we focus in the next section.

\subsection{GPU Kernels} \label{sec:GPU_Kernels}

\begin{figure}[!h]
	\centering
	  \subfigure[Generic]{
	   \includegraphics[width=0.3\textwidth]{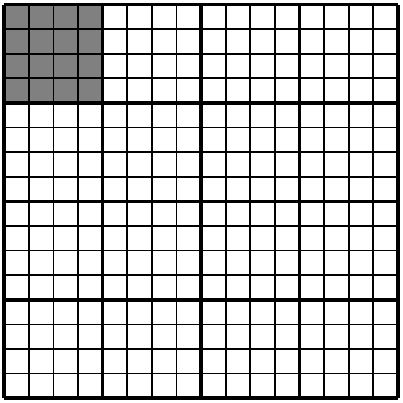} \label{fig:GPU_Box} }
	  \subfigure[Linear System Solver]{
	   \includegraphics[width=0.3\textwidth]{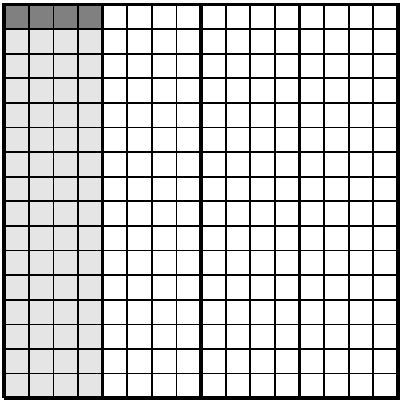} \label{fig:GPU_LinearSolver} }
	  \subfigure[Form Linear System]{
	   \includegraphics[width=0.3\textwidth]{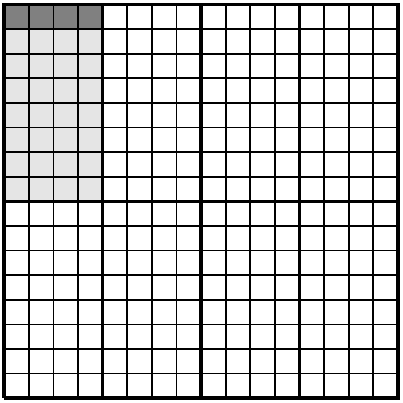} \label{fig:GPU_JyF} }
	\caption{Graph of the various sub-domain block and thread block combinations used in this GPU implementation.  Thin lines denote borders between cell centers and thicker lines denotes sub-domain blocks. Light gray shaded areas mark an example subdomain, with darker gray shared areas marking thread block. Note in b) thread block and sub-domain block are the same.} 
	\label{fig:GPU_Kernel_Domain}
\end{figure}

Analogous to loop fusion (removal of loop overheads by combining similar loops),  reducing the number of kernel calls (kernel fusion) can improve efficiency, in particular by reducing repeated requests to the same data in global memory, and reducing the use of global memory to store intermediary data.   Before discussing the details of the kernels solving (\ref{eq:JacobianSystem1}) and (\ref{eq:JacobianSystem2}), we first present the remaining GPU computations, which are computed with two kernels: the \textit{preprocessing} kernel, which computes the solution at the ghost points and the nonlinear functions and their derivatives at the cell-center points ($x_{i+\myfrac{1}{2}},y_{j+\myfrac{1}{2}}$) (see $\S$~\ref{sec:FluxDiscretization} and~\ref{sec:BoundaryCondition}); and the \textit{reduction and update} kernel, which computes the solution criteria and reduces the dataset, and updates the iteration step (\ref{eq:JacobianSystem3})\footnote{At this point it is unknown if the iteration step satisfies the solution criteria; therefore, this calculation is possibly redundant if the solution is rejected, however, this inefficiency is negligible in comparison to splitting the kernel.}.
In both kernels, the domain is sub-divided into squares (16$\times$16 in our implementation) and assigned a thread block of the same size, see Figure~\ref{fig:GPU_Box}. Shared memory and register memory are naively used to remove repeated access to global memory.

To solve efficiently the penta-diagonal system defined in (\ref{eq:JacobianSystem1}) and (\ref{eq:JacobianSystem2}) requires care.   Recall that at each time step and each iteration step, there are  $I$ penta-diagonal matrices of size $J\times J$ and $J$  penta-diagonal matrices of size $I\times I$ to invert, (\ref{eq:JacobianSystem1}) and (\ref{eq:JacobianSystem2}), respectively. To show this, we arrange solution values $u_{(i,j)}$ in terms of a matrix, $\mathbf{u}$; and use the subscripts $(,j)$ and $(i,)$ to refer to the $j$th column of $\mathbf{u}$ (fixed $y$) and the $i$th row of $\mathbf{u}$ (fixed $x$), respectively. Similar notation will be used for $\mathbf{v}$ and $\mathbf{w}$.  We may then express (\ref{eq:JacobianSystem1}) to (\ref{eq:JacobianSystem3}) as

\begin{eqnarray} 
	\mathbf{A}_j \left[ \mathbf{w}_{(,j)} \right]^T & = & \left[ \mathbf{a}^{n+1}_{(k),(,j)} + \mathbf{b}_{(,j)}^{n} \right]^T  \quad 0 \le j < J \;, \label{eq:JacobianSystem21} \\
	\mathbf{B}_i \mathbf{v}_{(i,)}                  & = & \mathbf{w}_{(i,)}  \quad \quad \quad \quad \quad 0 \le i < I \;,                   \label{eq:JacobianSystem22} \\
	\mathbf{u}^{n+1}_{(k+1)}               & = & \mathbf{u}^{n+1}_{(k)} + \mathbf{v} \;,                                   \label{eq:JacobianSystem23} 
\end{eqnarray}
where
\begin{equation}\label{eq:JacobianSystem24}
 \mathbf{A}_j = \mathbf{I} + \frac{\Delta t }{2} \mathbf{J}_{y_j} \; , \; \;
 \mathbf{B}_i = \mathbf{I} + \frac{\Delta t }{2} \mathbf{J}_{x_i} \;, \; \;
 \mathbf{a}^{n+1}_{(k),{(,j)}} = -\left( \mathbf{I} - \frac{\Delta t }{2} \mathbf{D}_{(,j)} \right) \mathbf{u}^{n+1}_{(k),(,j)} \;, \; \;
 \mathbf{b}^{n}_{(,j)} = \left( \mathbf{I} - \frac{\Delta t }{2} \mathbf{D}_{(,j)} \right) \mathbf{u}^{n}_{(,j)} \;.
\end{equation}
Note that $\mathbf{A}_j$ and $\mathbf{B}_i$ are penta-diagonal matrices representing the implicit $y$ derivatives for a fixed $x_i$, and the implicit $x$ derivatives for a fixed $y_j$, respectively. 

We therefore solve (\ref{eq:JacobianSystem21}) and (\ref{eq:JacobianSystem22}) in four kernels: 1) Compute $\mathbf{A}_i$ for each $i$, $\mathbf{a}^{n+1}_{(k)}$ and $\mathbf{b}^{n}$; 2) Invert the penta-diagonal system in (\ref{eq:JacobianSystem21}) and compute the transpose (coalesce data access); 3) Compute $\mathbf{B}_j$ for each $j$; and 4) Invert penta-diagonal systems in (\ref{eq:JacobianSystem22}) and compute the transpose. 

\textbf{Note 1: }$\mathbf{b}^{n}$ is independent of the iteration step, $k$, and is therefore only computed at the initial iteration step (for each time step), and stored in memory for use in future iterations. Furthermore, when initializing the iterative method, $\nstn{\mathbf{u}}{n+1}{0}=\mathbf{u}^{n}$, we do not explicitly copy $\mathbf{u}^{n}$ into $\nstn{\mathbf{u}}{n+1}{0}$, but instead manipulate array pointers to access the correct terms.

\textbf{Note 2: } While there are pre-existing subroutine libraries to perform a matrix transpose, to remove redundant global memory access, the inverse solver is fused with an efficient transpose method.

\textbf{Note 3: } Kernel fusion may be used to combine steps 1) and 3); however, the time saving is negligible compared to the prohibitive total global memory cost, e.g., if $M$ bytes are needed to store the solution, $6M$ bytes are required to store one set of penta-diagonal linear systems, plus an additional $M$ bytes for the fixed implicit term in (\ref{eq:JacobianSystem24}), $\mathbf{b}^{n}$.   Our current implementation restricts the total domain size to be able to fit into global memory; therefore, not fusing the kernels allows almost 25\% more memory to be utilized (an additional $4M$ bytes are needed for storing the nonlinear functions and their derivatives)\footnote{Our implementation uses synchronous kernel calls i.e. kernels are executed in serial by the CPU. By sub-dividing the domain into partitions (which would be further sub-divided by the kernels), clever use of asynchronous kernel calls may be used to reduce the total memory requirement, e.g., if each partition contains $M_1 \ll M$ bytes of the solution, then $5M$ plus some small multiple of 10$M_1$ bytes would be required, i.e., the total memory usage may be halved (approximately).}.

\subsubsection{Linear System Solver} \label{sec:LinerSystemSolver}

We begin with the most difficult computation to parallelize, inverting the penta-diagonal linear systems, (\ref{eq:JacobianSystem21}) and (\ref{eq:JacobianSystem22}).   At best, a single penta-diagonal linear system may be solved with linear complexity; however, the scheme is highly non-parallelizable.   To parallelize a single penta-diagonal system would require a reduction type method; however, such a method would increase the total number of calculations.   To avoid this issue, we instead assign to each thread on a GPU a linear system to solve.

To invert an individual penta-system, a LU factorization\footnote{While Gaussian elimination may be used, it requires more divisions to be computed, which, on a GPU, is more expensive to evaluate when compared to a multiplication. On a CPU, typically, there is no significant difference. }
is implemented e.g., $\mathbf{A}_i = \mathbf{L}_i \mathbf{U}_i$. In general, solving the linear system $\mathbf{L}_i \mathbf{U}_i \mathbf{a}_i = \mathbf{c}_i $ may be separated into two operations,
\begin{equation} \label{eq:LUFactorization}
	\mathbf{U}_i \mathbf{a}_i = \mathbf{b}_i \;,\; \textrm{and} \; \mathbf{L}_i \mathbf{b}_i = \mathbf{c}_i \; ,
\end{equation}
which may be inverted with linear complexity. To sub-divide the $(x,y)$ domain for parallelization, the domain is divided according to the spatial direction, i.e., the domain is divided into strips.   The direction that is sub-divided depends on the current direction of the \gls{ADI} method being computed, e.g., solving the implicit $y$ derivatives for fixed $x$, the domain is divided in the $x$ direction.   The width of the strips (16 in our implementation) is the number of penta-diagonal systems in each sub-domain (strip) e.g., different $i$ in (\ref{eq:LUFactorization}).   Each sub-domain is assigned a vector of threads (equal to the width of the strips) which invert the subset of penta-diagonal systems by LU factorization.   Figure~\ref{fig:GPU_LinearSolver} gives an example of a 16$\times$16 domain with strips of length $4$. Since each penta-diagonal system is independent (for a given \gls{ADI} direction), each step of the LU factorization is essentially a three point iterative method, therefore, register memory is used to store the previous two iterative points.

As mentioned previously, to utilize memory coalescence (threads access data points that are adjacent to each other in global memory), data must be transposed in memory. To fuse the transpose with LU factorization, the transpose is computed while inverting the second step of the LU factorization, e.g., inverting $\mathbf{U}_i a_i = b_i$ in (\ref{eq:LUFactorization}).   Since coalesced memory access is not relevant for shared memory access, it may be used to temporarily store data.  For example, in the $y$ implicit direction of the \gls{ADI} method, 16 threads (each $x$ value) compute an iteration step (fixed $y$ value) of the U factorization, and then store the data row-wise (fixed $y$) in a square block of shared memory, the size of which depends on the strip width ($16\times16$ in our implementation). Repeating the U factorization for the next 15 iteration steps, the solution is computed on a $16\times16$ sub-block.  The transpose may be performed in a block-wise fashion by accessing the shared memory column-wise (fixed $x$) without any significant inefficiency, and then stored in global memory with coalesced access.

\subsubsection{Computing the Linear System}

We now switch focus to the remaining two kernels not yet defined, computing $\mathbf{A}_j$, $\mathbf{a}^{n+1}_{(k,(,j))}$, $\mathbf{b}^{n}_{(,j)}$, and $\mathbf{B}_i$ in (\ref{eq:JacobianSystem24}).   It is mostly trivial to parallelize the computation of $\mathbf{A}_j$, $\mathbf{B}_i$, $\mathbf{a}^{n+1}_{(k),(,j)}$ and $\mathbf{b}^{n}_{(,j)}$, as at a given time step, threads do not depend on the data computed by other threads; however, to compute numerical derivatives, adjacent points are required. It is therefore prudent to use shared memory to reduce redundant global memory access.

To compute $\mathbf{B}_i$ in (\ref{eq:JacobianSystem24}), the domain is divided into squares and assigned a thread block of equal size (16$\times$16 in our implementation).   While we may define a shared memory block of equal size to the sub-domain, threads near the $x$ boundaries of the sub-domain depend on data in the adjacent sub-domain; therefore analogous to the idea of ghost points, we padded that shared memory block with ghost points (corresponding to solution values in adjacent sub-domains). In our implementation, adding ghost points (2 points at each boundary) results in a 20$\times$16 block of shared memory.   Note that data must be in the transpose form; however, since shared memory is already being used and a square thread block is operating on the sub-domain, the transpose is performed block-wise when transferring the data to global memory.

While we may apply the same procedure for computing $\mathbf{A}_j$, $\mathbf{a}^{n+1}_{(k),(,j)}$, and $\mathbf{b}^{n}_{(,j)}$ in (\ref{eq:JacobianSystem24}), unlike computing $\mathbf{B}_i$, threads near either the $x$ or $y$ boundaries of the sub-domain depend on data in the adjacent sub-domain; therefore, the shared memory bank is padded with ghost points in both directions (20$\times$20 in our implementation).   For the $\mathbf{B}_j$ shared memory block, ghost points could be transferred from global memory to shared memory with little inefficiency; however, for the $\mathbf{A}_i$, $\mathbf{a}^{n+1}_{(k),(,j)}$, and $\mathbf{b}^{n}_{(,j)}$ shared memory blocks, memory transfer is inefficient with the additional ghost points.   A faster approach is to subdivide the domain into sub-rectangles and assign each sub-domain a vector of threads, e.g., in our implementation the sub-domain size is 32$\times$15, where the 32 threads are mapped to 32 $x$-values, and each thread loops through 15 $y$-values, see Figure~\ref{fig:StencilF_T}.   Furthermore, we note that the $y$ derivatives depend on at most five points ($j-2,~j-1,~j,~j+1$ and $j+2$); therefore, instead of a (32+4)$\times$(15+4) (sub-domain plus ghost points) shared memory block, we implement five $36\times1$ shared memory blocks.   Similarly, function values and their derivatives require three points, therefore six (three each) $36\times1$\footnote{While only a $34\times1$ shared memory block is required to store function values plus ghost points, a $36\times1$
domain is chosen to align shared memory indices to those of the shared memory used to store the solution, simplifying coding. }
shared memory blocks are used for each nonlinear function (there are two in our model).   While the initial iteration (first $y$-value on sub-domain) requires transferring seventeen $36\times1$ blocks of data from global memory to shared memory, in subsequent iterations, only five $36\times1$ blocks
of data are transferred from the global memory to shared memory: the solution, the two non-linear functions and their derivatives.

\textbf{Note: } Numerical tests show that the choice of 15 in the sub-block size for $\mathbf{A}_i$, $\mathbf{a}^{n+1}_{(k),(,j)}$, and $\mathbf{b}^{n}_{(,j)}$ leads to the fastest computation time (seemingly independent of the thread block size), which is related to our implementation.  To remove explicitly copying data between shared memory location, computations for each looped $y$ value are grouped into a subroutine, and at each loop step, the input parameters are alternated by pointer manipulation. For example, for a generic three point method, first define $v_{-1}=u_{j-1}$, $v_{0}=u_{j}$, and $v_{1}=u_{j+1}$, where $u$ is the global memory variable and $v$ is the shared memory variable. Performing the following operations at the next three iteration steps (three cycle):
\begin{enumerate}
 \item Compute $f(v_{-1},v_{0},v_{1})$, and then set $v_{-1}=u_{j+2}$;
 \item Compute $f(v_{0},v_{1},v_{-1})$, and then set $v_{0}=u_{j+3}$;
 \item Compute $f(v_{1},v_{-1},v_{0})$, and then set $v_{1}=u_{j+4}$;
\end{enumerate}
it may be seen that by the third step, $v_{-1}=u_{j-1+3}$, $v_{0}=u_{j+3}$, and $v_{1}=u_{j+1+3}$, i.e., the definitions of the shared memory variables are realigned but offset by 3. Similarly, a five point method results in a five cycle.   Recalling the combined stencil in Figure~\ref{fig:StencilF_T}, the method is a three-point method with respect to function values, and a five-point method with respect to solution values; therefore, combining the 3-cycle and 5-cycle, results in a 15-cycle, the most efficient choice for the $y$ domain size according to numerical tests.

\section{Radial Fourier Transform} \label{sec:RadialFFT}

Here we give an overview of the procedure for computing the radial Fourier Transform used to analyze simulations
in $\S$~\ref{sec:Simulations}. To map the magnitude of the 2D Fourier transform
to the radial Fourier Transform, a one dimensional (1D) function of the magnitude of the wave vector $q_{r}=\sqrt{q_{x}^2+q_{y}^2}$, six processing steps are performed. The first four steps smooth the magnitude of the Fourier transform and map the smoothed data from a 2D Cartesian definition of the wavenumber to a 1D function of the magnitude of the wave number. The remaining two steps further smooth the data so as to reliably extract the local maxima.

\begin{figure}[!h]
\centering
  \subfigure[]{
   \includegraphics[width=0.23\textwidth]{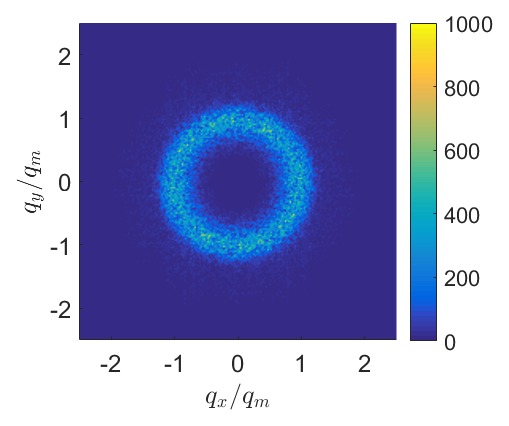} 
   \label{fig:eg_2DFFT} }
  \subfigure[]{
   \includegraphics[width=0.23\textwidth]{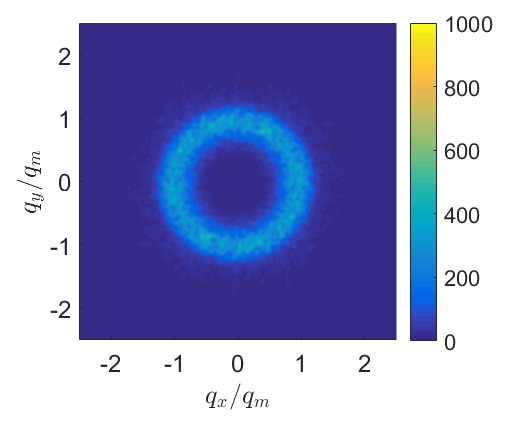} 
   \label{fig:eg_2DFFT_Smooth} }
  \subfigure[]{
   \includegraphics[width=0.23\textwidth]{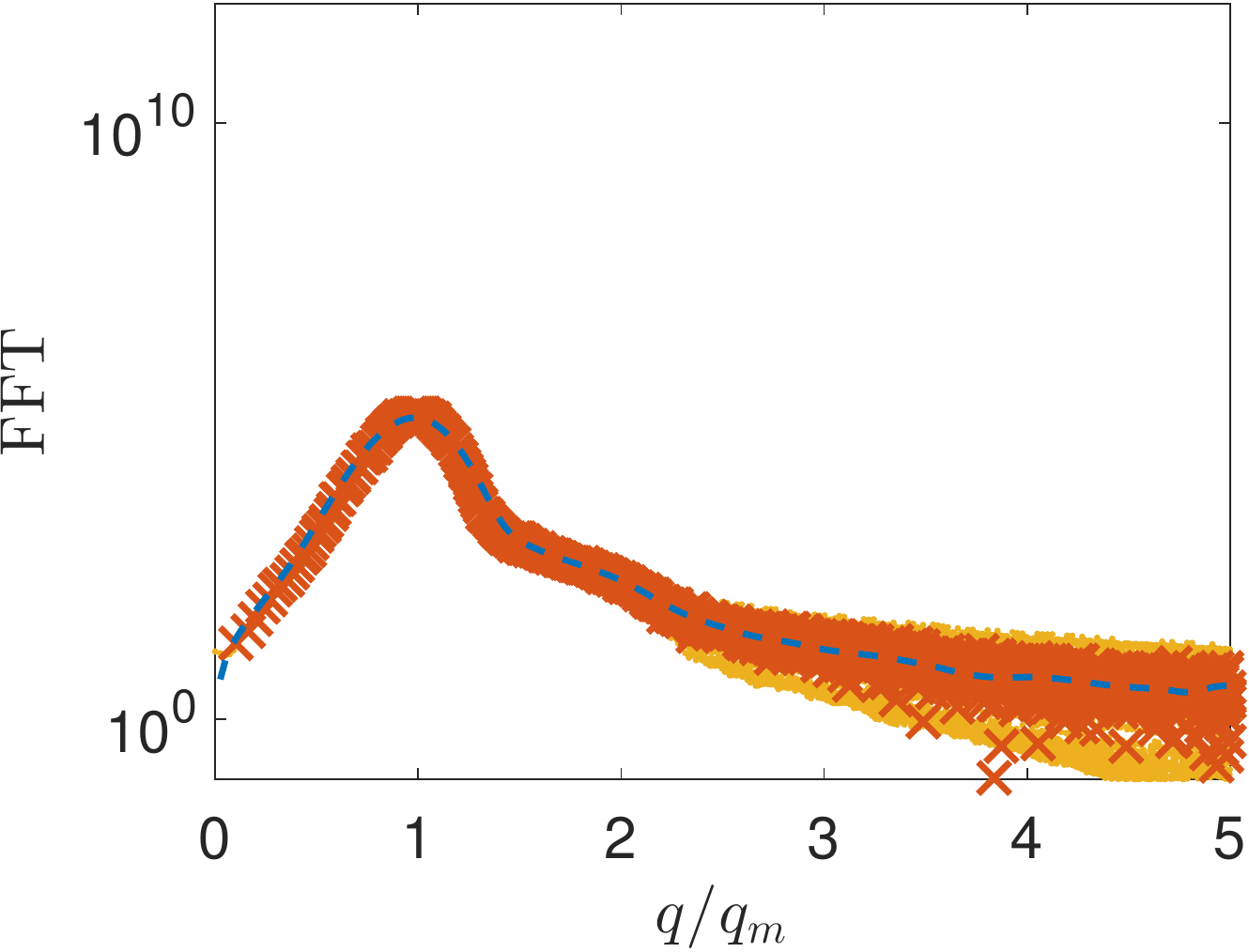} 
   \label{fig:eg_1DFFT} }
  \subfigure[]{
   \includegraphics[width=0.23\textwidth]{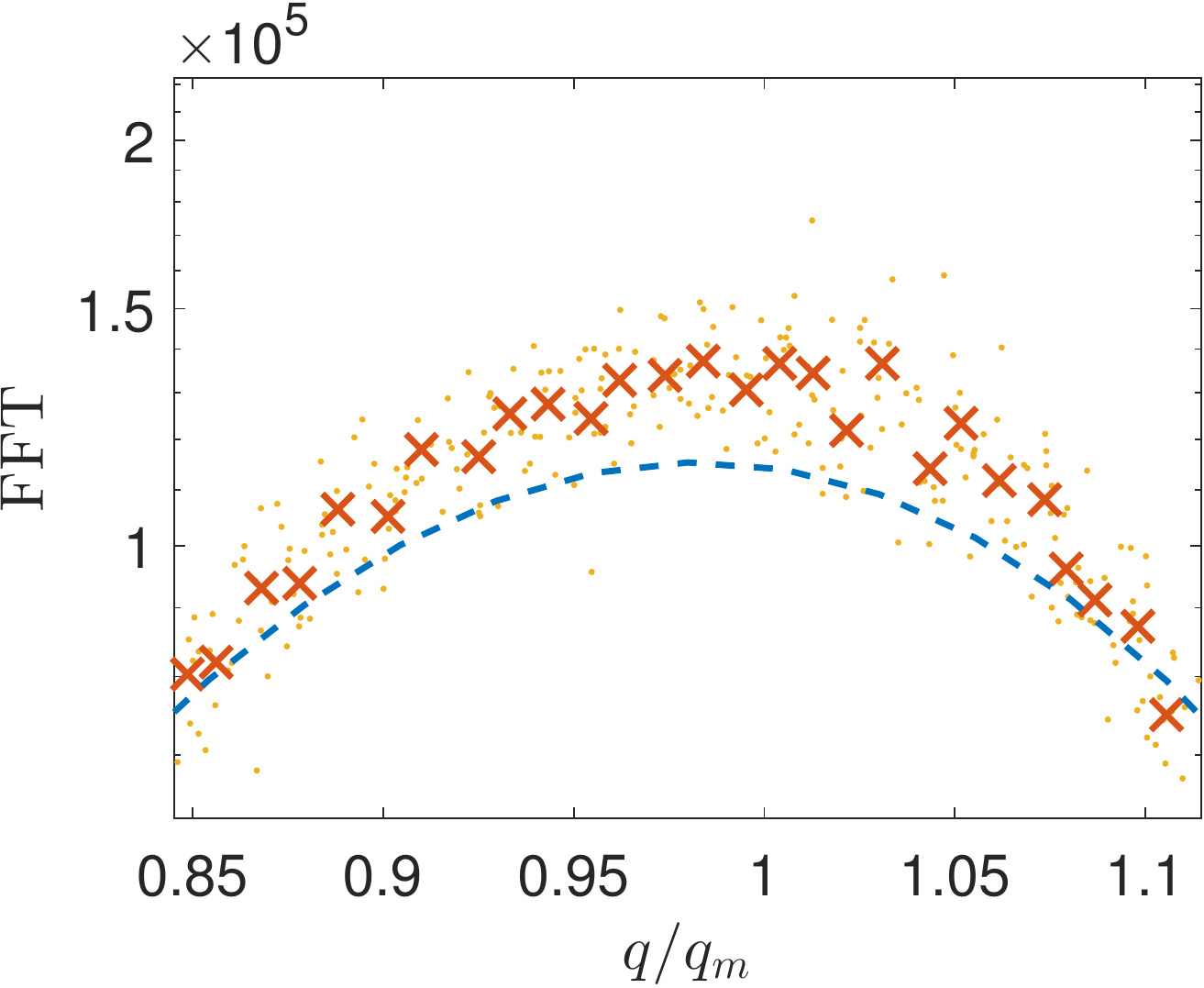} 
   \label{fig:eg_1DFFT_large} }
	\caption{a) An example of the log of the magnitude of the two-dimensional Fourier transform. b) An example of smoothing the data in (a) using a convolution with a Gaussian filter.  c) A plot of: i) the data in (b) mapped to a radial one-dimensional function (yellow scatter plot) using equation (\ref{eq:FFT_radial_averaging}); ii) a linear least-squares fit of the scatter plot data to the piece-wise linear function (red `$\times$' symbols); and iii) the piece-wise linear function smoothed with a moving average filter (dashed blue curve). d) Zoom of panel (c) demonstrating the improved smoothness using the moving average filter.} 
	\label{fig:eg_2D_to_1D_FFT}
\end{figure}

First,  the peak at the zero wave number is removed by subtracting the initial film thickness $H_0$, i.e., the magnitude of the Fourier transform of $\zeta(x,y)=h(x,y)-H_0$ is computed, an example of which is shown in Figure~\ref{fig:eg_2DFFT}.  Second, to reduce noise and smooth the data, the magnitude of the Fourier transform data is convoluted with a Gaussian function/filter. Third, the smoothed data, $\bar{\zeta}$, are mapped from the Cartesian coordinates $(q_{x_i},q_{y_j})$ to polar coordinates  $(q_{r_i},q_{\theta_j})$. Fourth, the polar coordinate data are mapped to a (radial) function, $\bar{\bar{\zeta}}(q_{r_i})$, by averaging over the $\theta$ direction for a fixed $q_{r_i}$, i.e., we compute
\begin{equation} \label{eq:FFT_radial_averaging}
  \bar{\bar{\zeta}}(q_{r_k}) = \frac{1}{N_k}\sum_{(i,j) \in P_k}  \bar{\zeta}(q_{r_i},q_{\theta_i}) \;, \quad P_k = \{ q_{r_i},q_{\theta_i} \in \mathbb{R}
  | q_{r_i} = q_{r_k} \},
\end{equation}
where $N_k$ is the number of points in the set $P_k$.  An example of (\ref{eq:FFT_radial_averaging}) is given by the yellow scatter plot in Figure~\ref{fig:eg_1DFFT} with Fourier transform data in Figure~\ref{fig:eg_2DFFT}. To extract a local maximum  (dominant wavenumber) numerically, sufficiently smooth data are required. To smooth the scattered data two final processing steps are performed: first, using a linear-least squares method, the scattered data are fitted to a piece-wise continuous function of the form
\begin{equation} \label{eq:PiecewiseFunction}
  \bar{\bar{\bar{\zeta}}}(q) = a_k + b_k \frac{s-q_k}{q_{k+1}-q_{k}} \; , \quad \textrm{where} \quad 
  s \in [0,1] \;, \quad
  q = q_k + s(q_{k+1} - q_{k}) \;, 
\end{equation}
and the domain is restricted to $q\in [0,5 q_m]$ (reducing computational cost). Note that we choose $q_k$'s such that each interval contains 10 points. Denoting the number of intervals as $K$, continuity implies $b_k=a_{k+1}-a_{k}$ for $0\le k < K -1 $, and choosing $a_0=0$ (i.e. $\zeta(q=0)=0$) leaves $K$ free parameters: $b_{K-1}$ and $a_k$ for $1\le k \le K-1$, therefore the fitting procedure is fully determined.

An example of this fitting procedure is given by the red `$\times$' scatter plot in Figure~\ref{fig:eg_1DFFT}. This step accurately captures the trend of the scattered data; however, we could not produce a sufficiently smooth function, see Figure~\ref{fig:eg_1DFFT_large} for a comparison between the piece-wise continuous function (red `$\times$' symbols), and the additional smoothing step (dashed blue curve) discussed next. To smooth the piece-wise continuous function, the next step evaluates (\ref{eq:PiecewiseFunction}) on an equipartitioned grid of 400 points and smooths the data using a moving average filter. Specifically, we compute
\begin{equation} \label{eq:PiecewiseFunction}
  \bar{\bar{\bar{\bar{\zeta}}}}(q_k) = \frac{1}{2P+1} \sum_{p=-P}^{P} \bar{\bar{\bar{\zeta}}}(q_{k+p})
\end{equation}
where $P$ is half the filter width, and the filter is applied three times with $P=3,2,$ and then $1$.  

To numerically extract the local maxima, a second order method is used to compute the derivative of the radial Fourier transform, and a combination of a bisection method and interpolation is used to find the roots of the derivative (critical points). To extract the local maxima, the first derivative test is used where the second derivative is computed numerically at the roots.

\section*{References}
\bibliography{films}
\bibliographystyle{elsarticle-num}
\end{document}